
\input phyzzx

\hfuzz=35pt
\titlepage
\line{\hfill astro-ph/9411049}
\line{\hfill BROWN-HET-964}
\line{\hfill October 1994}
\vskip1.0truein
\titlestyle{{MODERN COSMOLOGY AND STRUCTURE FORMATION
}\foot{Invited lectures at TASI-94, Boulder, Colorado, June 1994; to be
published in the proceedings, ed J. Donoghue (World Scientific, Singapore,
1995).}}
\bigskip
\author{Robert H. BRANDENBERGER}
\centerline{{\it Department of Physics}}
\centerline{{\it Brown University, Providence, RI 02912, USA}}
\bigskip
\abstract
An introduction to modern cosmology is given, focusing on theories for the
origin of structure in the Universe. After a brief review of the theory of
growth of cosmological perturbations and a summary of some important
observational results, most of the article is devoted to discussions of the
inflationary Universe scenario and of topological defect models of structure
formation. The final chapter is a summary of the theory of cosmic microwave
background temperature fluctuations and of the current observational status.
\endpage

\chapter{Introduction}

Cosmology has over the past fifteen years emerged as a vibrant and exciting
subfield of physics.  It is based on the marriage of quantum field theory and
particle physics on the one hand with classical general relativity on the
other.  One of the main goals of modern cosmology is to explain the structure
of the Universe on the scale of galaxies and beyond.  Thus, the
experimental/observational basis of the field lies in astronomy, and there is a
lot of interaction between theoretical cosmologists and observational
astronomers and astrophysicists.

The main goal of these lectures is to give an introduction to the two most
developed classes of structure formation theories: those based on inflation and
those based on topological defects.  I will give a brief survey of relevant
observational results from large-scale structure surveys and from searches for
cosmic microwave background (CMB) anisotropies, and I will attempt a
preliminary comparison with theoretical predictions. These notes are intended
as a pedagogical introduction rather than as a comprehensive review. For
comprehehsive discussions of inflation, the reader is referred to Refs. 1-3,
and for detailed reviews of topological defect models to Refs. 4-7. An
introduction to quantum field theory methods used in modern cosmology can be
found in Ref. 8. These notes draw on material presented elsewhere$^{9,10)}$ in
which some of the topics are treated in more detail.

I hope to persuade the reader that cosmology is an exciting area of physics
with close connections to particle and high energy physics, and with a steady
stream of new data from astronomy and astrophysics.

The outline of these lectures is as follows: Section 2 is a review of standard
cosmology, focusing on its basic principless, its observational support and its
problems.

Section 3 is a brief overview of  ``new cosmology."  I argue why, to obtain an
improved cosmological scenario, we need to treat matter using particle physics
and field theory.  Next, I introduce the basic idea of the inflationary
Universe scenario and explain how it leads to a solution of some of the
problems of standard cosmology.  In particular, it provides a mechanism for the
formation of structure in the Universe.  The section continues with a brief
introduction to the topological defect models of structure formation, an
explanation for the need of dark matter, and a survey of the present models.

In Section 4, I present the basics of structure formation, beginning with a
survey of some of the relevant large-scale structure data.  Structure in the
Universe is assumed to grow by gravitational instability.  I summarize the
essentials of the Newtonian theory of cosmological perturbations (valid on
length scales smaller than the apparent horizon (Hubble radius)) and of the
relativistic theory$^{11)}$ (required to study scales beyond the horizon).  I
also discuss free streaming.

Section 5 contains an overview of inflationary Universe models and of the
mechanism for the generation and evolution of perturbations which they provide.

Section 6 presents an overview of topological defect models of structure
formation.  To begin, a classification of defects is given.  Next it is shown
that in models which admit topological defects, they are inevitably formed
during a symmetry breaking phase transition$^{12)}$.  Cosmic string and global
texture models of structure formation are discussed in detail.  In particular,
it is pointed out that if defects are responsible for seeding galaxies, there
must be new physics at a scale of $\eta \sim 10^{16}$ GeV.

Section 7 focuses on CMB anisotropies.  It is shown why theories of structure
formation inevitably produce such anisotropies, the predictions of the various
models are reviewed, and a comparison with recent observations is given.

In this writeup, units in which $c = \hbar = k_B = 1$ are used unless mentioned
otherwise.  The space-time metric $g_{\mu\nu}$ is taken to have signature
$(+,-,-,-)$.  Greek indices run over space and time, latin ones over spatial
indices only.  The Hubble expansion rate is $H (t) = \dot a (t) / a (t)$, with
$a(t)$ the scale factor of a Friedmann-Robertson-Walker (FRW) Universe.  The
present value of $H$ is $100 h$ kms$^{-1}$ Mpc$^{-1}$, where $0.4 < h <1$.
Unless stated otherwise, the value of $h$ is taken to be 0.5. The cosmological
redshift at time $t$ is denoted by $z(t)$.  As usual, the symbols $G$ and
$m_{pl}$ stand for Newton's constant and Planck mass, respectively.  Distances
are measured in pc (``parsec") or Mpc, where 1 pc corresponds to 3.1 light
years.

\chapter{Review of Standard Cosmology}

\section{Principles}

The standard big bang cosmology rests on three theoretical pillars: the
cosmological principle, Einstein's general theory of relativity and a perfect
fluid description of matter.

The cosmological principle$^{13)}$ states that on large distance scales the
Universe is homogeneous.  From an observational point of view this is an
extremely nontrivial statement.  On small scales the Universe looks rather
inhomogeneous.  The inhomogeneities of the solar system are obvious to
everyone, and even by the naked eye it is apparent that stars are not randomly
distributed.  They are bound into galaxies, dynamical entities whose visible
radius is about $10^4$ pc.  Telescopic observations show that galaxies are not
randomly distributed, either.  Dense clumps of galaxies can be identified as
Abell clusters.  In turn, Abell cluster positions are correlated to produce the
large-scale structure dominated by sheets (or filaments), with typical scale
100 Mpc, observed in recent redshift surveys$^{14)}$.  Until recently, every
new survey probing the Universe to greater depth revealed new structures on the
scale of the sample volume.  In terms of the visible distribution of matter
there was no evidence for large-scale homogeneity.  This situation changed in
1992 with the announcement$^{15)}$ that a new redshift survey, complete to a
depth of about $500 h^{-1}$ Mpc, had discovered no prominent structures on
scales larger than $100 h^{-1}$ Mpc.  This is the first observational evidence
from optical measurements in favor of the cosmological principle.  However, to
put this result in perspective we must keep in mind that the observed isotropy
of the CMB temperature$^{16)}$ to better than $10^{-5}$ on large angular scales
has been excellent evidence for the validity of the cosmological principle.

The second theoretical pillar is general relativity, the theory which
determines the dynamics of the Universe.  According to the cosmological
principle, space at any time $t$ is a three dimensional surface with maximal
symmetry (translations and rotations).  There are three families of such
spaces$^{17)}$: flat Euclidean space $R^3$, the three sphere $S^3$, and the
hypersphere $H^3$.  The proper distance $ds^2$ on these three surfaces can be
written in spherical coordinates as
$$
 ds^2 = a(t)^2 \, \left[ {dr^2\over{1-kr^2}} + r^2 (d \vartheta^2 + \sin^2
\vartheta d\varphi^2) \right] \, . \eqno\eq
$$
The constant $k$ is $+1, \, 0$, or $-1$ respectively for $S^3, \, R^3$ and
$H^3$.

The Einstein equations of general relativity imply that $a(t)$ -- called the
scale factor of the Universe -- evolves in time.  The proper distance/time in
space-time is
$$
ds^2 = dt^2 - a(t)^2 \left[ {dr^2\over{1-kr^2}} + r^2 (d \vartheta^2 + \sin^2
\vartheta d\varphi^2) \right] \, . \eqno\eq
$$
By a coordinate transformation, $a(t)$ can be set equal to 1 at the present
time $t_0$.

\midinsert\vskip 4cm
\noindent{\bf Figure 1:} Sketch of the expanding Universe. Concentric circles
indicate space at fixed time, with time increasing as the radius gets larger.
Points at rest have constant comoving coordinates. Their world lines are
straight lines through the origin (e.g. $L$).
\endinsert

  To obtain a simple visualization of an expanding Universe, consider space to
be the surface of a balloon.  We draw a grid on the surface and use it to
define coordinates $\underline{x}^c$ (the superscript $c$ stands for comoving).
 Points at rest on the surface of the balloon have constant comoving
coordinates.  However, if the balloon is being inflated, then the physical
distance $\Delta x^p$ betwen two points at rest with comoving separation
$\Delta x^c$ increases:
$$
\Delta x^p = a(t) \Delta x^c \, . \eqno\eq
$$
The scale factor $a(t)$ is proportional to the radius of the balloon (see Fig.
1).

According to Einstein's equivalence principle, particles in the absence of
external nongravitational forces move on geodesies, curves which extremize
$ds^2$.  The velocity of a particle relative to the expansion of the Universe
is called peculiar velocity $v_p$
$$
v_p = a(t) \, {dx^c\over dt} \eqno\eq
$$
and obeys the equation
$$
\ddot v_p + {\dot a\over a} \, v_p = 0 \, , \eqno\eq
$$
from which it follows that
$$
v_p (t) \sim  a^{-1} (t) \, . \eqno\eq
$$

The dynamics of an expanding Universe  is determined by the Einstein equations,
which relate the expansion  rate to the matter content, specifically to the
energy density $\rho$ and pressure $p$.  For a homogeneous and isotropic
Universe, they reduce to the Friedmann-Robertston-Walker (FRW) equations
$$
\left( {\dot a \over a} \right)^2 - {k\over a^2} = {8 \pi G\over 3 } \rho
\eqno\eq
$$
$${\ddot a\over a} = - {4 \pi G\over 3} \, (\rho + 3 p) \, .\eqno\eq
$$
These equations can be combined to yield the continuity equation (with Hubble
constant $H = \dot a/a$)
$$
\dot \rho = - 3 H (\rho + p) \, . \eqno\eq
$$

The third key assumption of standard cosmology is that matter is described by
an ideal gas with an equation of state
$$
p = w \rho \, . \eqno\eq
$$
For cold matter, pressure is negligible and hence $w = 0$.  From (2.9) it
follows that
$$
\rho_m (t) \sim a^{-3} (t) \, , \eqno\eq
$$
where $\rho_m$ is the energy density in cold matter.  For radiation we have $w
= {1/3}$ and hence it follows from (2.9) that
$$
\rho_r (t) \sim a^{-4} (t) \, , \eqno\eq
$$
$\rho_r (t)$ being the energy density in radiation.

\section{Observational Pillars}

The first observational pillar of standard cosmology is Hubble's
redshift-distance relationship$^{18)}$ (Fig. 2)
$$
z = H d \, , \eqno\eq
$$
where $H$ is the present Hubble expansion constant, $d$ is the distance to a
galaxy, and $z$ is its redshift
$$
z \equiv {\lambda_0\over \lambda_e} - 1 \, , \eqno\eq
$$
$\lambda_e (\lambda_0)$ being the wavelength of light at the time of emission
(detection).

\midinsert\vskip 9cm
\noindent{\bf Figure 2:} A recent redshift-distance plot of galaxies$^{19)}$.
The distances are determined using the Tully-Fisher method. See Ref. 31 for a
detailed discussion of the method and errors.
\endinsert

There is an easy intuitive derivation of this result.  A wave in an expanding
background will have a wavelength which increases as the scale factor $a(t)$.
Hence for light emitted at time $t_e$
$$
z (t_e) = {a (t_0)\over{a (t_e)}} - 1 \, . \eqno\eq
$$
For light emitted close to the present time we can Taylor expand the above
result to obtain
$$
z (t_e) \simeq {\dot a (t_0)\over{a (t_0)}} \, (t_0 - t_e) \simeq H (t_0) \, d
\, . \eqno\eq
$$
Equation (2.15) defines the cosmological redshift, which can be used as a
measure of cosmic time.

The second observational pillar of standard cosmology is the existence and
black body nature of the CMB$^{20, 21)}$

To understand the connection$^{17)}$, consider matter in an expanding Universe.
 As we go backwards in time, the density of matter increases as $a^{-3} (t)$,
and as a consequence the temperature grows.  Above a temperature of 13.6 eV,
atoms are ionized, and a bath of photons in thermal equilibrium must be
present.  Photons still scatter frequently below this temperature.  At some
time $t_{rec}$ the scattering length of a photon becomes longer than the Hubble
radius.  After that, photons travel without scattering.  At $t_{rec}$, the
distribution of photons is of black body type.  A special feature of black body
spectra is that the spectral shape is maintained even after $t_{rec}$.  The
only change is that the temperature redshifts
$$
T(t) = {a (t_{rec})\over{a(t)}} \, T (t_{rec}) \, . \eqno\eq
$$
Hence, the standard Big Bang model predicts a black body spectrum of photons
with temperature
$$
T_o = T_{rec} z (t_{rec})^{-1} \, , \eqno\eq
$$
where $T_{rec}$ is the temperature at $t_{rec}$, determined by comparing the
largest rate of scattering of photons below recombination, that due to Thomson
scattering, with the Hubble expansion rate, yielding the result
$$
T_{rec} \simeq 0.25  \, {\rm eV} \simeq 4000^\circ {\rm K} \, . \eqno\eq
$$
The corresponding redshift is determined by measuring $T_o$.

In 1965, Penzias and Wilson$^{22)}$ discovered this remnant black body
radiation at a temperature of about 3$^\circ$ K.  Since the spectrum peaks in
the microwave region it is now called the CMB (cosmic microwave background).
Recent satellite (COBE)$^{23)}$ and rocket$^{24)}$ experiments have confirmed
the black body nature of the CMB to very high accuracy.  The temperature is
2.73$^\circ$ K$= T_0$ which corresponds to
$$
z (t_{rec}) = z_{rec} \sim 10^3  \, . \eqno\eq
$$

Given the existence of the CMB, we know that matter has two components: dust
(with energy density $\rho_m (t)$) and radiation (with density $\rho_r  (t)$).
At the present time $t_0$, $\rho_m (t) \gg \rho_r (t)$.  The radiation energy
density is determined by $T_0$, and the matter energy density can be estimated
by analyzing the dynamics of galaxies and clusters and using the virial
theorem.  However, since by (2.11) and (2.12) $\rho_m (t) \sim a (t)^{-3}$ and
$\rho_r (t) \sim  a (t)^{-4}$, as we go back in time the fraction of energy
density in radiation increases, and the two components become equal at a time
$t_{eq}$, the time of equal matter and radiation.  The corresponding redshift
is
$$
z_{eq} \simeq \Omega \, h^{-2}_{50} \, 10^4 \eqno\eq
$$
where
$$
\Omega = {\rho\over{\rho_c}} \, (t_0)\, , \eqno\eq
$$
$\rho_c$ being the density for a spatially flat Universe (the critical
density), and $h_{50}$ is the value of $H$ in units of 50 km s$^{-1}$
Mpc$^{-1}$.

The time $t_{eq}$ is important for structure formation.  As we will see in
Section 4, it is only after $t_{eq}$ that perturbations on scales smaller than
the Hubble radius $H^{-1} (t)$ can grow.  Before then, the radiation pressure
prevents growth.  A temperature-time plot of the early Universe is sketched in
Fig. 3,  Note that $t_{eq} < t_{rec}$.

\midinsert \vskip 5.5cm
\noindent{\bf Figure 3:} Temperature-time diagram of standard big bang
cosmology. The present time, time of last scattering and time of equal matter
and radiation are $t_0$, $t_{rec}$ and $t_{eq}$ respectively. The Universe is
radiation-dominated before $t_{eq}$ (Region A) and matter-dominated in Region
B. Before and after $t_{rec}$, respectively, the Universe was opaque and
transparent, respectively, to microwave photons.
\endinsert

The third observational pillar of standard big bang cosmology concerns
nucleosynthesis$^{25, 26)}$ - the production of light elements (heavy elements
are formed in supernovae).  Above a temperature of about $10^9 $ K, the nuclear
interactions are sufficiently fast to prevent neutrons and protons from fusing.
 However, below that temperature, it is thermodynamically favorable for
neutrons and protrons to fuse and form deuterium, helium 3, helium 4 and
lithium 7 through a long and interconnected chain of reactions.  The resulting
light element abundances depend sensitively on the expansion rate of the
Universe and on $\Omega_B$, the fraction of energy density $\rho_B$ at present
in baryons relative to the critical density $\rho_c$.  In Fig. 5, recent
theoretical calculations$^{27)}$ of the abundances are shown and compared with
observations.  Demanding agreement with all abundances leaves only a narrow
window
$$
3 \times 10^{-10} < \eta < 10^{-9} \, , \eqno\eq
$$
where $\eta$ is the ratio of baryon number density $n_B$ to photon number
density $n_\gamma$
$$
\eta = {n_b\over n_\gamma} \, . \eqno\eq
$$
{}From (2.23), it follows that $\Omega_B$ is constrained:
$$
0.01 < \Omega_B h^2 < 0.035 \, . \eqno\eq
$$
In particular, if the Universe is spatially flat and the cosmological constant
is negligible, there must be nonbaryonic dark matter.  We will return to the
dark matter issue in Section 3.

\midinsert \vskip 15.5cm
\noindent{\bf Figure 5:} Light element abundances as a function of the baryon
to entropy ratio $\eta$ (from Ref. 27). The solid curves are the predictions of
homogeneous big bang nucleosynthesis. The observational limits are indicated on
the left vetical axis. Theory and observations are only consistent for a narrow
range of values of $\eta$.
\endinsert

The final pillar of standard cosmology is the near isotropy of the CMB$^{16)}$.
 After subtracting the dipole anisotropy which is presumed to be due to the
motion of the earth relative to the rest frame defined by the CMB, no
anisotropies have been detected to a level of better than $10^{-4}$, i.e., the
temperature difference $\delta T(\vartheta)$ between two beams pointing in
directions in the sky separated by an angle  $\vartheta$ (Fig. 4) satisfies
$$
{\delta T (\vartheta)\over{\bar T}} < 10^{-4} \eqno\eq
$$
on all angular scales $\vartheta$.  Here $\bar T$ is the average temperature.

\midinsert \vskip 4.8cm
\noindent{\bf Figure 4:} Sketch of a CMB anisotropy experiment. Two radio
antennas with beam width $b$ collect microwave radiation from points in the sky
separated by an angle $T$. The difference in beam intensities is measured.
\endinsert

Until recently, the isotropy of the CMB was the only observational support for
the cosmological principle.  Any inhomogeneities of the Universe on length
scales comparable to the comoving Hubble radius at $t_{rec}$ and larger would
generate temperature anisotropies by a mechanism discussed in detail in Section
7.  Hence, the near isotropy of the CMB implies that density fluctuations on
large scales must have been very small in the early Universe.

To summarize, the observational pillars of standard cosmology are Hubble's
redshift-distance relation, the existence and black body nature of the CMB,
primordial nucleosynthesis, and the isotropy of the CMB.  Note, in particular,
that no tests of big bang cosmology say anything about the evolution of the
Universe before the time of nucleosynthesis.  Note, also, that not all
astronomers accept the above observations as support of the Big Bang model.
For a recent criticism see Ref. 28 (and Ref. 29 for a reply to the criticism).

\section{Problems}

Standard Big Bang cosmology is faced with several important problems.  Only one
of these,  the age problem, is a potential conflict with observations.  The
others which I will focus on here -- the homogeneity, flatness and formation of
structure problems (see e.g. Ref. 30) -- are questions which have no answer
within the theory and are therefore the main motivation for the new
cosmological models which will be discussed in the rest of these lectures.

{}From the FRW equations (2.7) and (2.8) it is  easy to calculate the age of
the Universe, given the expansion law (2.11) which holds throughout most of the
history of the Universe.  For a spatially flat Universe, the age $\tau$ depends
on the expansion rate $H$, i.e. on the constant $h$ which is in the range $0.4
< h < 1$:
$$
\tau \simeq {7\over h} 10^9 {\rm yr} \, . \eqno\eq
$$
Globular cluster ages have been estimated to lie in the range $12 - 18 \times
10^9$ yr.  Thus, theory and observations are only consistent if $h < 0.55$ (see
e.g. Ref. 31).  In an open Universe the problem is less severe.  Modern
cosmological models do not add any insight into this problem since they only
modify the evolution of the Universe at very early times $t \ll  t_{eq}$.

The final three problems mentioned above, the homogeneity, flatness and
formation of structure problems, provided a lot of the motivation for the
development of the inflationary Universe scenario$^{30)}$ and will hence be
discussed in detail.

The horizon problem is illustrated in Fig. 6.  As is sketched, the comoving
region $\ell_p (t_{rec})$ over which the CMB is observed to be homogeneous  to
better  than one part in $10^4$ is much larger than the comoving forward light
cone $\ell_f (t_{rec})$ at $t_{rec}$, which is the maximal distance over which
microphysical forces could have caused the homogeneity:
$$
\ell_p (t_{rec}) = \int\limits^{t_0}_{t_{rec}} dt \, a^{-1} (t) \simeq 3 \, t_0
\left(1 - \left({t_{rec}\over t_0} \right)^{1/3} \right) \eqno\eq
$$
$$
\ell_f (t_{rec}) \int\limits^{t_{rec}}_0 dt \, a^{-1} (t) \simeq 3 \, t^{2/3}_0
\, t^{1/3}_{rec} \, . \eqno\eq
$$

{}From the above equations it is obvious that $\ell_p (t_{rec}) \gg \ell_f
(t_{rec})$.  Hence, standard cosmology cannot explain the observed isotropy of
the CMB.

\midinsert \vskip 6.5cm
\noindent
{\bf Figure 6:} A space-time diagram (physical distance $x_p$ versus time $t$)
illustrating the homogeneity problem: the past light cone $\ell_p (t)$ at the
time $t_{rec}$ of last scattering  is much larger than the forward light cone
$\ell_f (t)$ at $t_{rec}$.
\endinsert

In standard cosmology and in an expanding Universe, $\Omega = 1$ is an unstable
fixed point.  This can be seen as follows.  For a spatially flat Universe
$(\Omega = 1)$
$$
H^2 = {8 \pi G\over 3} \, \rho_c \, , \eqno\eq
$$
whereas for a nonflat Universe
$$
H^2 + \varepsilon \, T^2 = {8 \pi G\over 3}  \, \rho \, , \eqno\eq
$$
with
$$
\varepsilon = {k\over{(aT)^2}} \, . \eqno\eq
$$
The quantity $\varepsilon$ is proportional to $s^{-2/3}$, where $s$ is the
entropy density.  Hence, in standard cosmology, $\varepsilon$ is constant.
Combining (2.30) and (2.31) gives
$$
{\rho - \rho_c\over \rho_c} = {3\over{8 \pi G}} \, {\varepsilon T^2\over
\rho_c} \sim T^{-2} \, . \eqno\eq
$$
Thus, as the temperature decreases, $\Omega - 1$ increases.  In fact, in order
to explain the present small value of $\Omega - 1 \sim {\cal O} (1)$, the
initial energy density had to be extremely close to critical density.  For
example, at $T = 10^{15}$ GeV, (2.33) implies
$$
{\rho - \rho_c\over \rho_c} \sim 10^{-50} \, . \eqno\eq
$$
What is the origin of these fine tuned initial conditions?  This is the
flatness problem of standard cosmology.

The last problem of the standard cosmological model I will mention is the
``formation of structure problem."  Observations indicate that galaxies and
even clusters of galaxies have nonrandom correlations on scales larger than 50
Mpc (see e.g. Ref. 14).  This scale is comparable to the comoving horizon at
$t_{eq}$.  Thus, if the initial density perturbations were produced much before
$t_{eq}$, the correlations cannot be explained by a causal mechanism.  Gravity
alone is, in general, too weak to build up correlations on the scale of
clusters after $t_{eq}$ (see, however, the explosion scenario of Ref. 32).
Hence, the two questions of what generates the primordial density perturbations
and what causes the observed correlations, do not have an answer in the context
of standard cosmology.  This problem is illustrated by Fig. 7.

\midinsert \vskip 6.5cm
\noindent
{\bf Figure 7:} A sketch (conformal separation vs. time) of the formation of
structure problem: the comoving separation $d_c$ between two clusters is larger
than the forward light cone at time $t_{eq}$.
\endinsert

Finally, let us address the cosmological constant problem.  All known
symmetries of nature and principles of general relativity allow for the
presence of a term in the Einstein equations which acts like matter with energy
$\Lambda$ and pressure $-\Lambda$, i.e., with an equation of state $p =  -
\rho$.  If it is not to dominate the present expansion rate of the Universe,
the cosmological constant $\Lambda$ must be very small
$$
\Lambda < 3 H^2_0 \sim 10^{-83} {\rm GeV}^2 \, . \eqno\eq
$$
On dimensional grounds, we would expect  $\Lambda$ to be of the order $m_{pl}^2
\sim 10^{38} \, {\rm GeV}^2$.  Thus, the cosmological constant is about 140
orders of magnitude smaller than what we would expect it to be (for recent
reviews of the cosmological constant problem, see Ref. 33).

As we will see, modern cosmology does not address the cosmological constant
problem.  If anything, the problem will manifest itself in a more apparent
manner.

Due to the formation of structure problem, there can be no causal physical
theory for the origin of structure (with nontrivial spatial correlations) in
the Universe in the context of the Standard Big Bang theory.  The main
breakthrough of modern cosmology is that it provides solutions to this problem.
 The key to understanding this breakthrough in cosmology is the realization of
the internal inconsistency of the standard picture when extrapolated to times
much before nucleosynthesis.  Standard cosmology is based on the assumption
that matter continues to be described by an ideal radiation gas to arbitrarily
high temperatures.  This is clearly in contrast to what nuclear and particle
physics tells us.  As we go backwards in time towards the Big Bang, nuclear
physics and eventually particle physics  effects will take over.  To describe
matter correctly, a quantum field theoretic description must be used.

\chapter{New Cosmology and Structure Formation}

The goal of this section is to present an overview of what can be gained if we
go beyond standard cosmology and allow matter to be described in terms of
concepts from particle physics.  Detailed discussions of the models will be
given in later sections.

\section{The Inflationary Unvierse}

The idea of inflation$^{30)}$ is very simple.  We assume there is a time
interval beginning at $t_i$ and ending at $t_R$ (the ``reheating time") during
which the Universe is exponentially expanding, i.e.,
$$
a (t) \sim e^{Ht}, \>\>\>\>\> t \epsilon \, [ t_i , \, t_R] \eqno\eq
$$
with constant Hubble expansion parameter $H$.  Such a period is called  ``de
Sitter" or ``inflationary."  The success of Big Bang nucleosynthesis sets an
upper limit to the time of reheating:
$$
t_R \ll t_{NS} \, , \eqno\eq
$$
$t_{NS}$ being the time of nucleosynthesis.

\midinsert \vskip 3.5cm
\noindent{\bf Figure 8:} The phases of an inflationary Universe. The times
$t_i$ and $t_R$ denote the beginning and end of inflation, respectively.
In some models of inflation, there is no initial radiation domintated FRW
period. Rather, the classical space-time emerges directly in an inflationary
state from some initial quantum gravity state.
\endinsert

The phases of an inflationary Universe are sketched in Fig. 8.  Before the
onset of inflation there are no constraints on the state of the Universe.  In
some models a classical space-time emerges immediately in an inflationary
state, in others there is an initial radiation dominated FRW period.  Our
sketch applies to the second case.  After $t_R$, the Universe is very hot and
dense, and the subsequent evolution is as in standard cosmology.  During the
inflationary phase, the number density of any particles initially in thermal
equilibrium at $t = t_i$ decays exponentially.  Hence, the matter temperature
$T_m (t)$ also decays exponentially.  At $t = t_R$, all of the energy which is
responsible for inflation (see later) is released as thermal energy.  This is a
nonadiabatic process during which the entropy increases by a large factor.  The
temperature-time evolution in an inflationary Universe is depicted in Fig. 9.

\midinsert \vskip 6cm
\noindent{\bf Figure 9:} The time dependence of matter temperature in an
inflationary Universe. During the period of exponential expansion, the
temperature decreases exponentially. At the end of inflation the energy density
of the scalar field responsible for inflation is transferred to ordinary
matter. This leads to reheating. The critical temperature $T_c$ is the
temperature at which the initial matter thermal energy density becomes less
than the scalar field energy density (see Chapter 5).
\endinsert

Fig. 10 is a sketch of how a period of inflation can solve the homogeneity
problem.  $\Delta t = t_R - t_i$  is the period of inflation.  During
inflation, the forward light cone increases exponentially compared to a model
without inflation, whereas the past light cone is not affected for $t \geq
t_R$.  Hence, provided $\Delta t$ is sufficiently large, $\ell_f (t_R)$ will be
greater than $\ell_p (t_R)$.  The condition on $\Delta t$ depends on the
temperature $T_R$ corresponding to time $t_R$, the temperature of reheating.
Demanding that $\ell_f (t_R) > \ell_p (t_R)$ we find, using the analogs of
(2.28) and (2.29), the following criterion
$$
e^{\Delta t H} \geq \, {\ell_p (t_R)\over{\ell_f (t_R) }} \simeq \left(
{t_0\over t_R} \right)^{1/2} = \, \left({T_R\over T_0} \right) \sim 10^{27}
\eqno\eq
$$
for $T_R \sim 10^{14}$GeV and $T_0 \sim 10^{-13}$GeV (the present microwave
background temperature).  Thus, in order to solve the homogeneity problem, a
period of inflation with
$$
\Delta t \gg 50 \, H^{-1} \eqno\eq
$$
is required.

\midinsert \vskip 10cm
\noindent{\bf Figure 10:} Sketch (physical coordinates vs. time) of the
solution of the homogeneity problem. During inflation, the forward light cone
$l_f(t)$ is expanded exponentially when measured in physical coordinates.
Hence, it does not require many e-foldings of inflation in order that $l_f(t)$
becomes larger than the past light cone at the time of last scattering. The
dashed line is the forward light cone without inflation.
\endinsert

Inflation also can solve the flatness problem$^{34, 30)}$  The key point is
that the entropy density $s$ is no longer constant.  As will be explained
later, the temperatures at $t_i$ and $t_R$ are essentially equal.  Hence, the
entropy increases during inflation by a factor $\exp (3 H \Delta t)$.  Thus,
$\epsilon$ decreases by a factor of $\exp (-2 H \Delta t)$.  With the numbers
used in (3.3):
$$
\epsilon_{\rm after} \sim 10^{-54} \, \epsilon_{\rm before} \, . \eqno\eq
$$
Hence, $(\rho - \rho_c) / \rho$ can be of order 1 both at $t_i$ and at the
present time.  In fact, if inflation occurs at all, the theory then predicts
that at the present time $\Omega = 1$ to a high accuracy (now $\Omega < 1$
would require  special initial conditions).

What was said above can be rephrased geometrically: during inflation, the
curvature radius of the Universe -- measured on a fixed physical scale --
increases exponentially.  Thus, a piece of space looks essentially flat after
inflation even if it had measurable curvature before.

Most importantly, inflation provides a mechanism which in a casual way
generates the primordial perturbations required for galaxies, clusters and even
larger objects.  In inflationary Universe models, the Hubble radius
(``apparent" horizon), $3t$, and the ``actual" horizon (the forward light cone)
do not coincide at late times.  Provided (3.3) is satisfied, then (as sketched
in Fig. 11) all scales within our apparent horizon were inside the actual
horizon since $t_i$.  Thus, it is in principle possible to have a casual
generation mechanism for perturbations$^{35-38)}$.

\midinsert \vskip 10cm
\noindent{\bf Figure 11:} A sketch (physical coordinates vs. time of the
solution of the formation of structure problem. Provided that the period of
inflation is sufficiently long, the separation $d_c$ between two galaxy
clusters is at all times smaller than the forward light cone. The dashed line
indicates the Hubble radius. Note that $d_c$ starts out smaller than the Hubble
radius, crosses it during the de Sitter period, and then reenters it at late
times.
\endinsert

The generation of perturbations is supposed to be due to a causal microphysical
process.  Such processes can only act coherently on length scales smaller than
the Hubble radius $\ell_H (t)$ where
$$
\ell_H (t) = H^{-1} (t) \, . \eqno\eq
$$
A heuristic way to understand the meaning of $\ell_H (t)$ is to realize that it
is the distance which light (and hence the maximal distance any causal effects)
can propagate in one expansion time:
$$
\ell_H (t) \sim a (t) \int\limits_{t}^{t+H^{-1} (t)} a (t^\prime)^{-1} \,
dt^\prime \, . \eqno\eq
$$
In Section 5 a more mathematical justification for the definition and role of
$\ell_H (t)$ will be given.

As will be discussed in Section 5, the density perturbations produced during
inflation are due to quantum fluctuations in the matter and gravitational
fields$^{36, 37)}$.  The amplitude of these inhomogeneities corresponds to a
tempertuare $T_H$
$$
T_H \sim H \, , \eqno\eq
$$
the Hawking temperature of the de Sitter phase.  This implies that at all times
$t$ during inflation, perturbations with a fixed physical wavelength $\sim
H^{-1}$ will be produced. Subsequently, the length of the waves is streched
with the expansion of space, and soon becomes larger than the Hubble radius.
The phases of the inhomogeneities are random.  Thus, the inflationary Universe
scenario predicts perturbations on all scales ranging from the comoving Hubble
radius at the beginning of inflation to the corresponding quantity at the time
of reheating.  In particular, provided that inflation lasts sufficiently long
(see (3.4)), perturbations on scales of galaxies and beyond will be generated.

Now that the reader is (hopefully) convinced that inflation is a beautiful
idea, the question arises how to realize this scenario.  The initial hope was
that the same scalar fields (Higgs fields) which particle physicists introduce
in order to spontaneously break the internal symmetries of their field theory
models would lead to inflation.  This hope was based on the fact that the
energy density $\rho$ and pressure $p$ of a real scalar field $\varphi
(\underline{x}, t)$ are given by
$$
\eqalign{\rho (\varphi) & = {1\over 2} \, \dot \varphi^2 + {1\over 2} \,
(\nabla \varphi)^2 + V (\varphi) \cr
p (\varphi) & = {1\over 2} \dot \varphi^2 - {1\over 6} (\nabla \varphi)^2 - V
(\varphi) \, .} \eqno\eq
$$
Thus, provided that at some initial time $t_i$
$$
\dot \varphi (\underline{x}, \, t_i) = \nabla \varphi (\underline{x}_i \, t_i)
= 0 \eqno\eq
$$
and
$$
V (\varphi (\underline{x}_i \, t_i) )  > 0 \, , \eqno\eq
$$
the equation of state of matter will read
$$
p = - \rho \eqno\eq
$$
and, from the FRW equations it will follow that
$$
a (t) = e^{tH} \>  , \> H^2 = \, {8 \pi G\over 3} \, V (\varphi) \, . \eqno\eq
$$

The next question is how to realize the required initial conditions (3.10) and
to maintain the key constraints
$$
\dot \varphi^2 \ll V (\varphi) \> , \> (\nabla \varphi)^2 \ll V (\varphi)
\eqno\eq
$$
for sufficiently long (see (3.4)).  This typically requires potentials which
are very flat near $\varphi (\underline{x}, \, t_i)$.  Worse yet, the
parameters of the potential $V (\varphi)$ must be chosen such that the final
amplitude of density perturbations is sufficiently small to agree with the
limits on CMB anisotropies.  As we will demonstrate in Section 5, these
conditions impose severe constraints on the constants which appear in $V
(\varphi)$.

In light of these difficulties it is important to keep in mind that inflation
can also be generated by modifying gravity at high curvatures (see e.g., Refs.
39-41).  It is also wise to investigate alternative theories of structure
formation which do not rely on inflation.

\section{Topological Defect Models}

According to particle physics theories, matter at high energies and
temperatures must be described in terms of fields.  Gauge symmetries have
proved to be extremely useful in describing the standard model of particle
physics, according to which at high energies the laws of nature are invariant
under a nonabelian group  $G$ of internal symmetry transformations
$$
G = {\rm SU} (3)_c \times {\rm SU} (2)_L \times U(1)_Y \eqno\eq
$$
which at a temperature of about 200 MeV is spontaneously broken down to
$$
G^\prime = {\rm SU} (3)_c \times {\rm U} (1) \, . \eqno\eq
$$
The subscript on the SU(3) subgroup indicates that it is the color symmetry
group of the strong interactions, ${\rm SU} (2)_L \times $ U(1)$_Y$ is the
Glashow-Weinberg-Salam (WS) model of weak and electromagnetic interactions, the
subscripts $L$ and $Y$ denoting left handedness and hypercharge respectively.
At low energies, the WS model spontaneously breaks to the U(1) subgroup of
electromagnetism.

Spontaneous symmetry breaking is induced by an order parameter $\varphi$ taking
on a nontrivial expectation value $< \varphi >$ below a certain temperture
$T_c$.  In some particle physics models, $\varphi$ is a fundamental scalar
field in a nontrivial representation of the gauge group $G$ which is broken.
However, $\varphi$ could also be a fermion condensate.

The transition taking place at $T = T_c$ is a phase transition and $T_c$ is
called the critical temperature.  From condensed matter physics it is well
known that in many cases topological defects form during phase transitions,
particularly if the transition rate is fast on a scale compared to the system
size.  When cooling a metal, defects in the crystal configuration will be
frozen in; during a temperature quench of $^4$He, thin vortex tubes of the
normal phase are trapped in the superfluid; and analogously in a temperature
quench of a superconductor, flux lines are trapped in a surrounding sea of the
superconducting Meissner phase (see Fig. 12 for an example$^{42)}$ of defect
formation).

\midinsert \vskip 10cm
\noindent{\bf Figure 12:} A simulation of defect formation in the $2 + 1$
dimensional Abelian Higgs model$^{42)}$ in an expanding background. The total
energy density is plotted against the two spatial coordinates. The initial
conditions were specified by thermal initial conditions with random phases of
the order parameter on Hubble scales. At later times, the thermal noise has
redshifted away, leaving behind trapped energy density in vortices.
\endinsert

In cosmology, the rate at which the phase transition proceeds is given by the
expansion rate of the Universe.  Hence, topological defects will inevitably be
produced in a cosmological phase transition$^{12)}$, provided the underlying
particle physics model allows such defects.

Topological defects can be point-like (monopoles), string-like (cosmic
strings)$^{43)}$ or planar (domain walls), depending on the particle physics
model (see Section 6).  Also of importance are textures$^{44, 45)}$, point
defects in space-time.

Topological defects represent regions in space with trapped energy density.
These regions of surplus energy can act as seeds for structure  formation as is
illustrated in Fig. 13.  For point-like defects, the force which causes
clustering about the seed can be understood using  Newtonian gravity.  The
process is called gravitational accretion.  For precise calculations (and in
the case of other defects), general relativistic effects must be taken into
account (see Section 6).

\midinsert \vskip 8.5cm
\noindent{\bf Figure 13:} Sketch of the basic gravitational accretion
mechanism. The topological defect (in this case a cosmic string loop) is a
configuration of trapped energy density. This excess density produces a
Newtonian gravitational attractive force on the surrounding matter.
\endinsert

No stable topological defects arise in the breaking of the WS model.  However,
there is good evidence for phase transitions at very high energies.  The
coupling constants of SU(3)$_c$, SU(2) and U(1) are seen to converge at an
energy scale $\eta$ of about
$$
\eta \sim 10^{16} \, {\rm GeV} \, . \eqno\eq
$$
It is therefore not unreasonable to speculate that the standard model results
from the breaking of a larger symmetry group $G_0$ at a scale $\eta$.  A large
class of unified gauge theories$^{46)}$ based on a symmetry breaking
$$
G_0 \longrightarrow G = {\rm SU} (3) \times {\rm SU}(2) \times {\rm U}(1)
\eqno\eq
$$
admit topological defects, many theories have cosmic string solutions.

Topological defect models of structure formation will be discussed in detail in
Section 6.  Here I will briefly point out by which mechanism correlations on
all cosmological scales are induced.  To be concrete, I consider the cosmic
string model$^{47, 48)}$.  Cosmic strings are one-dimensional defects without
ends.  Hence, they must be either infinite in length or else closed loops.  The
fact that at the time of the phase transition $t_c$, the order parameter has
random phases on scales larger than the initial correlation length implies that
at $t_c$ a random walk-like network of infinite strings will form$^{12)}$.
This implies nontrivial correlations of structures seeded by these strings on
all scales larger than the initial correlation length.

\section{Need for Dark Matter}

At this point we have illustrated two classes of mechanisms by which structure
formation in the Universe can be seeded: quantum fluctuations during a period
of inflation, and topological defects.  To completely specify a theory of
structure formation, however, we must also specify what the ``dark matter"
which dominates the energy density of the Universe today consists of (see e.g.
Ref. 49).

The evidence for dark matter has been accumulating over the past decade.
Measurements of galaxy velocity rotation curves indicate that a large fraction
of the mass of a galaxy does not shine.  In Fig. 14, the measured rotation
velocity $v$ is plotted as a function of the distance $r$ from the center of
the galaxy.  This velocity is compared to the velocity $\hat v (r)$ which
results from the virial theorem, assuming that light traces mass (the
luminosity profile is given in the upper frame).  The data is for the spiral
galaxies NGC2403 and NGC3198$^{50)}$.  The comparison shows that the mass of
these galaxies extends significantly beyond the visible radius and that --
assuming Newtonian gravity is applicable -- a large fraction of the mass of a
spiral galaxy must be dark.

\midinsert \vskip 11cm
\noindent{\bf Figure 14:} Velocity rotation curves for two galaxies (from Ref.
50). The upper panel shows the luminosity of the galaxy as a function of the
distance from the center, the lower panel presents the velocity (vertical axis,
in units of $km s^{-1}$) as a function of the same distance. The solid curve is
the velocity inferred from the observed luminosity curve, using the virial
theorem, the dotted curves are the observational results. The fact that the
velocity rotation curves remain constant beyond the visible radius of the
galaxy is strong evidence for galactic dark matter.
\endinsert

The fraction of matter that shines can be expressed as a fraction $\Omega_{\rm
lum}$ of the critical density $\rho_c$.  The present estimates  give
$$
\Omega_{\rm lum} \ll 0.01 \, , \eqno\eq
$$
whereas the fraction $\Omega_g$ of mass in galaxies is much larger$^{51)}$
$$
\Omega_g \sim 0.03 \, . \eqno\eq
$$
It is also possible to estimate the fraction $\Omega_{cl}$ of mass which is
gravitationally bound in clusters.  Current estimates using the virial theorem
give$^{51)}$
$$
\Omega_{cl} \sim 0.1 - 0.2 \eqno\eq
$$
(this comes mainly from studying the infall of galaxies towards the Virgo
cluster).

Finally, the amount $\Omega_{LSS}$ of mass in large-scale structures can be
estimated by measuring the large-scale peculiar velocities  and inferring the
mass required to generate such velocities.  Current estimates give$^{52, 53)}$
$$
\Omega_{LSS} = 0.8 \pm 0.5 \, . \eqno\eq
$$

As mentioned in Section 2, nucleosynthesis provides independent limits on the
fraction $\Omega_B$  of mass in baryons (see (2.25)).  Comparing (3.19-3.22)
with (2.25) we conclude:
\item{i)} There must be baryonic dark matter.  In fact, most of the dark matter
in galaxies and/or clusters could be baryonic, and some of this baryonic dark
matter may have recently been discovered by gravitational microlensing.
\item{ii)} If $\Omega = 1$, then most of the matter in the Universe consists of
nonbaryonic dark matter.  In fact, there is increasing evidence (see (3.22))
that nonbaryonic dark matter must exist independent of the theoretical
prejudice  for $\Omega = 1$.

The dark matter in the Universe is visible only through its gravitational
effects.  Hence, nonbaryonic dark matter candidates can be divided into two
classes, cold dark matter (CDM) and hot dark matter (HDM).

CDM particles are cold, i.e., their peculiar velocity $v$ is negligible at the
time $t_{eq}$ when structure formation begins:
$$
v (t_{eq}) \ll 1 \, . \eqno\eq
$$
Candidates for CDM include the axion (coherent oscillations of a low mass
scalar field) and neutralinos (the lightest stable supersymmetric particle,
which must be neutral).

HDM particles are relativistic at $t_{eq}$:
$$
v (t_{eq}) \sim 1 \, . \eqno\eq
$$
The prime candidate is a $25 h^{+2}_{50}$eV tau neutrino.  Note that this mass
is well within the experimental bounds for the tau neutrino mass, and also that
many particle physics models -- in particular those which lead to neutrino
oscillations -- predict masses of this order of magnitude.

\section{Survey of Models}

Any theory of structure formation must specify both the source of fluctuations
and the composition of the dark matter.  The reader is warned that the model
called the ``CDM Model" is a model with CDM {\bf AND} perturbations generated
by quantum fluctuations during a hypothetical period of inflation.

Inflation-based models were the first to be considered in quantitative detail,
initially assuming a HDM-dominated Universe.  Almost immediately, however,
contradictions with basic observations appeared$^{54)}$.

The problem of HDM-based inflationary models is related to neutrino free
streaming$^{55)}$.  The primordial perturbations in this theory are dark matter
fluctuations, but because of the large velocity of the dark matter particles,
the inhomogeneities are washed out on all scales below the neutrino free
streaming length  $\lambda^c_j (t)$,
$$
\lambda^c_j (t) \sim v (t) z (t) t \, , \eqno\eq
$$
which is the comoving distance the particles move in one Hubble expansion time.
 Since the neutrino velocity $v(t)$ and the redshift $z(t)$ both scale as
$a(t)^{-1}$,  the free streaming length decreases as
$$
\lambda^c_j (t) \sim t^{-1/3} \eqno\eq
$$
after $t_{eq}$ (before $t_{eq}$ the radiation pressure dominates).  Hence, in
an inflationary HDM model all perturbations on scales $\lambda$ smaller than
the maximal value of $\lambda^c_j (t)$ are erased.  The critical scale
$\lambda^{\rm max}_j$ is given by the value of $\lambda^c_j (t)$ at the time
when the neutrinos become non-relativistic which is in turn determined by the
neutrino mass $m_{\nu}$.  The result is
$$
\lambda^{\rm max}_j \simeq 30 \, {\rm Mpc} \, \left({{m_{\nu}} \over {25 {\rm
eV}}} \right)^{-2} \, , \eqno\eq
$$
a scale much larger than the mean separation of galaxies and clusters.  Since
we observe galaxies outside of large-scale structures, this model is in blatant
disagreement with observations.

Inflation-based models are hence only viable if (at least a substantial
fraction of) the dark matter is cold.  Such models have become known as ``CDM
models", and are to a first approximation rather successful at predicting the
clustering properties of galaxies and galaxy clusters$^{56)}$.  There are many
parameters in CDM models: the amplitude of the density perturbations, the power
of the spectrum  (see Section 4), the value of $\Omega$, the fraction
$\Omega_B$ of baryons, to mention some of the main ones.  It is also possible
to add a small fraction $\Omega_v$ of hot dark matter (yielding a class of
so-called ``Mixed Dark Matter" models).

Topological defect models were first developed in the context of CDM.  Theories
based on cosmic strings$^{57-59)}$ or on global textures$^{45, 60)}$ have also
been fairly successful in explaining observations (again to a first
approximation).

It is important to note that if perturbations are seeded by long-lived
topological defects (e.g., cosmic strings), then the above arguments against
hot dark matter disappear$^{61, 62)}$.   The seed perturbations can survive
neutrino free streaming as long as the seeds remain present for many Hubble
expansion times.  If we consider a comoving scale $\lambda$ much smaller than
$\lambda^{\rm max}_j$ of (3.27), then a dark matter perturbation will begin to
grow about the seed fluctuations at a time $t(\lambda)$ when
$$
\lambda^c_j (t (\lambda)) = \lambda \, . \eqno\eq
$$
Cosmic string based hot dark matter models have also been successful at
explaining the qualitative features of observations$^{63)}$.

\chapter{Basics of Structure Formation}

In the structure formation models mentioned in the previous section, small
amplitude seed perturbations are predicted to arise due to particle physics
effects in the very early Universe.  They then grow by gravitational
instability to produce the cosmological structures we observe today.  In order
to be able to make the connection between particle physics and observations, it
is important to understand the gravitational evolution of fluctuations.  This
section will introduce the basic concepts of this topic.  We begin, however,
with an overview of some of the relevant data.

\section{Survey of Data}

It is length scales corresponding to galaxies and larger which are of greatest
interest in cosmology when attempting to find an imprint of the primordial
fluctuations produced by particle physics.  On these scales, gravitational
effects are assumed to be dominant, and the fluctuations are not too far from
the linear regime.  On smaller scales, nonlinear gravitational and
hydrodynamical effects determine the final state and mask the initial
perturbations.

To set the scales, consider the mean separation of galaxies, which is about
5$h^{-1}$ Mpc$^{64)}$, and that of Abell clusters which is around 25$h^{-1}$
Mpc$^{65)}$.  The largest coherent structures seen in current redshift surveys
have a length of about 100$h^{-1}$ Mpc, the recent detections of CMB
anisotropies probe the density field on length scales of about $10^3 h^{-1}$
Mpc, and the present horizon corresponds to a distance of about $3 \cdot 10^3
h^{-1}$ Mpc.

Galaxies are gravitationally bound systems containing billions of stars.  They
are non-randomly distributed in space.  A quantitative measure of this
non-randomness is the ``two-point correlation function" $\xi_2 (r)$ which gives
the excess probability of finding a galaxy at a distance $r$ from a given
galaxy:
$$
\xi_2 (r) = < \, {n (r) - n_0\over n_0} \, >  \, . \eqno\eq
$$
Here, $n_0$ is the average number density of galaxies, and $n(r)$ is the
density of galaxies a distance $r$ from a given one.  The pointed braces stand
for ensemble averaging.

\midinsert \vskip 13cm
\noindent{\bf Figure 15:} Recent observational results for the two point
correlation function of IRAS galaxies, in both a volume limited subsample and a
complete galaxy sample (see Ref. 66 which explains the meaning of the measure
$J_3(r)$).
\endinsert

Recent observational results from a redshift survey of IRAS (infrared) galaxies
yields reasonable agreement$^{66)}$ with a form (see Fig. 15)
$$
\xi_2 (r) \simeq \left({r_0\over r} \right)^\gamma \eqno\eq
$$
with scaling length $r_0 \simeq 5 h^{-1}$ Mpc and power $\gamma \simeq 1.8$.  A
theory of structure formation must explain both the amplitude and the slope of
this correlation function.

Galaxies do not all have the same mass.  There are more smaller galaxies than
large ones (our galaxy is a large one).  The distribution of galaxy masses is
given by the ``galaxy mass function" $n(M)$, where $n (M) dM$ is the number
density of galaxies in the mass range $[M, \, M+dM]$.  Since we can only
measure luminosity but not mass, the observable measure of the galaxy
distribution is $\phi (L)$, the ``galaxy luminosity function."  Now, $\phi (L)
dL$ is the number density of galaxies with luminosity $L$ in the interval $[L,
\, L+dL]$.  Recent results for $\phi (L)$ from the IRAS galaxy survey are
reproduced in Fig. 16$^{67)}$.
\midinsert \vskip 11cm
\noindent{\bf Figure 16:} The luminosity function of IRAS galaxies. The
vertical axis is $\phi (L)$, the horizontal axis is luminosity $L$ in units of
solar luminosity. The solid curve represents the data, the dashed curve is a
fit to a theoretical model (see Ref. 67).
\endinsert

Theories must also be able to explain the internal mass distribution of
galaxies which can be inferred from the galaxy rotation curves (see Fig. 14).
According to the virial theorem, the velocity $v(r)$ at a distance $r$ from the
center of a galaxy is determined by the mass $M(r)$ inside of $r$:
$$
{mv^2\over r} = G \, {m M(r)\over r^2} \eqno\eq
$$
where $m$ is a test mass.  Hence
$$
v (r) = \left(G \, {M (r)\over r} \right)^{1/2} \, . \eqno\eq
$$
A constant rotation velocity implies $M(r) \sim r$ and hence a density profile
$\rho (r)$ of
$$
\rho (r) \sim r^{-2} \, . \eqno\eq
$$
This condition puts constraints on the possible composition of the galactic
dark matter.

An Abell cluster$^{68)}$ is a region in space with greater than fifty bright
galaxies in a sphere of radius $1.5 h^{-1}$ Mpc, i.e., a region with a very
high overdensity of galaxies.  Observations indicate that Abell clusters are
not distributed randomly in space.  The cluster two point correlation function
$\xi_c (r)$ has a form similar to (4.2)$^{65, 69)}$:
$$
\xi_c (r) \simeq \left({r_0\over r} \right)^\gamma \eqno\eq
$$
with $r_0 \simeq 15 h^{-1}$ Mpc and $\gamma \simeq 2$ (see Fig. 17 which is
taken from a recent analysis of rich clusters of galaxies selected from the APM
Galaxy Survey$^{69)}$).

\midinsert \vskip 8cm
\noindent{\bf Figure 17:} The two point correlation function $\xi_c$ (denoted
by $\xi_{cc}$ in the figure) of clusters of galaxies drawn from the APM galaxy
survey as a function of their separation $s$. Results for two different samples
of clusters are given.
\endinsert

There are several remarkable features about the clustering of galaxies and
galaxy clusters.  First, there is evidence for universality of the functional
form of $\xi (r)$; its slope is about 2 for both populations.  Secondly,
relative to their respective mean separations, galaxies are more clustered than
galaxy clusters.  This can be explained by the action of gravity.  Gravity has
had longer to act on the scales of galaxies than on that of clusters, and has
hence amplified the galaxy correlation function relative to that of clusters.

There is a wide spread of cluster masses which can be described by the cluster
multiplicity function $\Phi_c (n)$, where $\Phi_c (n) dn$ is the number density
of clusters containing between $n$ and $n+dn$ galaxies.  Fig. 18 is a sketch of
the observed cluster multiplicity function taken from Ref. 70.  Note that the
cluster mass function inferred from Fig. 18 and the galaxy mass function
deduced from the galaxy luminosity function of Fig. 16 match up quite well at a
mass of about $10^{12} M_\odot$.  Below this mass, the objects are well defined
dynamical entities whereas for larger masses they are composed of fragments.  A
reason for the difference may be due to the fact that clouds of more than
$10^{12} M_\odot$ cannot cool without fragmenting$^{71)}$.

\midinsert \vskip 10.5cm
\noindent{\bf Figure 18:} The cluster multiplicity function $\Phi_c (n)$
(denoted as $n$ on the figure) as a function of the number $n$ (horizontal
axis).
\endinsert

On scales larger than galaxy clusters there is not at present a clear
mathematical description of structure.  Many galaxy redshift surveys have
discovered coherent filamentary and planar structures and voids on scales of up
to $100 h^{-1}$ Mpc$^{14, 72-75)}$.  For example, the astronomers working on
the  ``Center for Astrophysics" redshift survey$^{14)}$ have analyzed many
adjacent slices of the northern celestial sphere.  For all galaxies above a
limiting magnitude of 15.5 they measured the redshifts $z$.  Fig. 19 is a
sketch of redshift versus angle $\alpha$ in the sky for one slice.  The second
direction in the sky has been projected onto the $\alpha -z$ plane.  The most
prominent feature is the band of galaxies at a distance of about $100h^{-1}$
Mpc.  This band also appears in neighboring slices and is therefore presumably
part of a planar density enhancement of comoving planar size of at least $(50
\times 100) \times h^{-2}$ Mpc$^2$.  This structure is often called the ``great
wall."  It is a challenge for theories of structure formation to explain both
the observed scale and topology of the galaxy distribution.

\midinsert \vskip 7.5cm
\noindent{\bf Figure 19:} Results from the CFA redshift survey. Radial distance
gives the redshift of galaxies, the angular distance corresponds to right
ascension. The results from several slices of the sky (at different
declinations) have been projected into the same cone.
\endinsert

Until 1992 there was little evidence for any convergence of the galaxy
distribution towards homogeneity.   Each new survey led to the discovery of new
coherent structures in the Universe on a scale comparable to that of the
survey.  In 1992, preliminary results of a much deeper redshift survey were
announced$^{15)}$ which for the first time found no new coherent structures on
scales larger than $100 h^{-1}$ Mpc.  This is the first direct evidence for the
cosmological principle from optical surveys (the isotropy of the CMB has for a
long time been a strong point in its support).

In summary, a lot of data from optical and infrared galaxies alone are
currently available, and new data are being collected at a rapid rate.  The
observational constraints on theories of structure formation are becoming
tighter.  A lot of theoretical work is needed in order to allow for detailed
comparisons between theory and observations.

\section{Gravitational Instability}

In this article we only discuss theories in which structures grow by
gravitational accretion.  The basic mechanism is illustrated in Fig. 20.
Consider first a flat space-time background.  A density perturbation with
$\delta \rho > 0$ will then give rise to an excess gravitational attractive
force $F$ acting on the surrounding matter.  This force is proportional to
$\delta \rho$, and will hence lead to exponential growth of the perturbation
since
$$
\delta \ddot \rho \sim F \sim \delta \rho \Rightarrow \delta \rho \sim \exp
(\alpha t) \eqno\eq
$$
with some constant $\alpha$.

\midinsert \vskip 8.2cm
\noindent{\bf Figure 20:} Sketch of the gravitational instability mechanism.
The vertical axis is the density perturbation $D$ as a function of a line in
space ($x$). A small initial overdensity ($A$) will cause a gravitational
acceleration $g$ towards it, which will lead to an increase in the perturbation
($B$). Note that in general underdense regions develop in addition to the
growing overdense areas.
\endinsert

In an expanding background space-time, the acceleration is damped by the
expansion.  If $r (t)$ is the physcial distance of a test particle from the
perturbation, then on a scale $r$
$$
\delta \ddot \rho \sim F \sim \, {\delta \rho\over{r^2 (t)}} \, , \eqno\eq
$$
which results in power-law increase of $\delta \rho$.  The goal of this
subsection is to discuss the growth rates of inhomogeneities in more detail
(see e.g. Refs. 76 and 77 for modern reviews).

Because of our assumption that all perturbations start out with  a small
amplitude, we can linearize the equations for gravitational fluctuations.  The
analysis is then greatly simplified by going to Fourier space in which all
modes $\delta (\underline{k})$ decouple.  We expand the fractional density
contrast $\delta (\underline{x})$ as follows:
$$
\delta (\underline{x}) = {\delta \rho (x)\over \rho} = (2 \pi)^{-3/2} V^{1/2}
\int d^3 k \> e^{i \underline{k} \cdot \underline{x}} \delta (\underline{k}) \,
, \eqno\eq
$$
where $V$ is a cutoff volume which disappears from all physical observables.

The ``power spectrum" $P(k)$ is defined by
$$
P (k) = < |\delta (k) |^2 > \, , \eqno\eq
$$
where the braces denote an ensemble average (in most structure formation
models, the generation of perturbations is a stochastic process, and hence
observables can only be calculated by averaging over the ensemble.  For
observations, the braces can be viewed as an angular average).

The physical measure of mass fluctuations on a length scale $\lambda$ is the
r.m.s. mass fluctuation $\delta M/M (\lambda)$ on this scale.  It is determined
by the power spectrum in the following way.  We pick a center $\underline{x}_0$
of a sphere $B_\lambda (x_0)$ of radius $\lambda$ and calculate
$$
\big| {\delta M\over M} \big|^2 \, (\underline{x}_0 , \, \lambda) = \big|
\int\limits_{B_\lambda (\underline{x}_0)} d^3 x \delta (\underline{x}) \,
{1\over{V (B_\lambda)}} \big|^2 \, , \eqno\eq
$$
where $V (B_\lambda)$ is the volume of the sphere.  Inserting the Fourier
decomposition (4.3) we obtain
$$
 \eqalign{
\big| {\delta M\over M} \big|^2 (\underline{x}_0 , \, \lambda) = & {V\over{(2
\pi)^3}} \, {1\over{V(B_\lambda )^2}} \int\limits_{B_\lambda (0)} d^3 x_1 \,
\int\limits_{B_\lambda (0)} d^3 x_2 \, \int d^4 k_1 d^4 k_2 e^{i
(\underline{k}_1 - \underline{k}_2) \cdot \underline{x}_0}\, \cr
& e^{i \underline{k}_1 \cdot \underline{x}_1} e^{-i \underline{k}_2 \cdot
\underline{x}_2} \delta (\underline{k}_1)^{\ast} \, \delta (\underline{k}_2) \,
. } \eqno\eq
$$
Taking the average value of this quantity over all $\underline{x}_0$ yields
$$
< \left( {\delta M\over M} \right)^2 (\lambda) > = \int d^3 k W_k (\lambda) |
\delta (\underline{k} |^2 \eqno\eq
$$
with a window function $W_k (\lambda)$ with the following properties
$$
W_k (\lambda) \cases{\simeq 1 & $k < k_\lambda = 2 \pi / \lambda$ \cr
\simeq 0 & $k > k_\lambda$ .\cr} \eqno\eq
$$
Therefore the r.m.s. mass perturbation on a scale $\lambda$ becomes
$$
< \big| {\delta M\over M} (\lambda) \big|^2 > \sim k_{\lambda}^3 P(k_{\lambda})
\, . \eqno\eq
$$

Astronomers usually assume that $P(k)$ grows as a power of $k$:
$$
P (k) \sim k^n \,  , \eqno\eq
$$
$n$ being called the index of the power spectrum.  For $n = 1$ we get the
so-called Harrison-Zel'dovich scale invariant spectrum$^{78)}$.

Both inflationary Universe and topological defect models of structure formation
predict a roughly scale invariant spectrum.  The distinguishing feature of this
spectrum is that the r.m.s. mass perturbations are independent of the scale $k$
when measured at the time $t_H (k)$ when the associated wavelength is equal to
the Hubble radius, i.e., when the scale ``enters" the Hubble radius.  Let us
derive this fact for the scales entering during the matter dominated epoch.
The time $t_H (k)$ is determined by
$$
k^{-1} a (t_H (k)) = t_H (k) \eqno\eq
$$
which leads to
$$
t_H (k) \sim k^{-3} \, . \eqno\eq
$$
According to the linear theory of cosmological perturbations discussed in the
following subsection, the mass fluctuations increase as $a(t)$ for $t >
t_{eq}$.  Hence
$$
{\delta M\over M} (k, t_H (k)) = \left( {t_H (k)\over t} \right)^{2/3} \,
{\delta M\over M} \, (k, t) \sim {\rm const} \, , \eqno\eq
$$
since the first factor scales as $k^{-2}$ and -- using (4.15) and inserting
$n=1$ -- the second as $k^2$.

\section{Newtonian Theory of Cosmological Perturbations}

The Newtonian theory of cosmological perturbations is an approximate analysis
which is valid on wavelengths $\lambda$ much smaller than the Hubble radius $t$
and for negligible pressure $p$, i.e., $p \ll  \rho$.  It is based on expanding
the hydrodynamical equations about a homogeneous background solution.

The starting points are the continuity, Euler and Poisson equations
$$
\dot \rho + \underline{\nabla} (\rho \underline{v}) = 0 \eqno\eq
$$
$$
\underline{\dot v} + (\underline{v} \cdot \underline{\nabla}) \underline{v} = -
\underline{\nabla} \phi - {1\over \rho} \, \underline{\nabla} p \eqno\eq
$$
$$
\nabla^2  \phi = 4 \pi G \rho \eqno\eq
$$
for a fluid with energy density $\rho$, pressure $p$, velocity $\underline{v}$
and Newtonian gravitational potential $\phi$, written in terms of physical
coordinates $(t, \, \underline{r})$.

The transition to an expanding space is made by introducing comoving
coordinates $\underline{x}$ and peculiar velocity $\underline{u} =
\underline{\dot {x}}$:
$$
\underline{r} = a (t) \underline{x} \eqno\eq
$$
$$
\underline{v} = \dot a (t) \underline{x} + a (t) \underline{u} \, . \eqno\eq
$$
The first term on the right hand side of (4.24) is the expansion velocity.

The perturbation equations are obtained by linearizing Equations (4.20-22)
about a homogeneous background solution $\rho = \bar \rho (t) , \> p = 0$ and
$\underline{u} = 0$.  Using the definition
$$
\delta \equiv {\delta \rho\over \rho} \, ,\eqno\eq
$$
the linearization ansatz can be written
$$
\rho (\underline{x}, t) = \bar \rho (t) (1 + \delta (\underline{x}, t) ) \, .
\eqno\eq
$$
If we consider adiabatic perturbations (no entropy density variations), then
after some algebra the linearized equations become
$$
\dot \delta + \nabla \cdot \underline{u} = 0 \, , \eqno\eq
$$
$$
\underline{\dot {u}} + 2 {\dot a\over a} \underline{u} = - a^2 (\nabla \delta
\phi + c^2_s \nabla \delta) \eqno\eq
$$
and
$$
\nabla^2 \delta \phi = 4 \pi G \bar \rho a^2 \delta \, , \eqno\eq
$$
with the speed of sound $c_s$ given by
$$
c^2_s = {\partial p\over{\partial \rho}} \, . \eqno\eq
$$
The two first order equations (4.27) and (4.28) can be combined to yield a
single second order differential equation for $\delta$.  With the help of
(4.29) this equation reads
$$
\ddot \delta + 2 H \dot \delta -  4 \pi G \bar \rho \delta - {c^2_s\over a^2}
\nabla^2 \delta = 0 \eqno\eq
$$
which in Fourier space becomes
$$
\ddot \delta_{\underline{k}} + 2 H \dot \delta_{\underline{k}} + \left( {c^2_s
k^2 \over a^2} - 4 \pi G \bar \rho \right) \delta_{\underline{k}} = 0 \, .
\eqno\eq
$$
Here, $H(t)$ as usual denotes the expansion rate, and $\delta_{\underline{k}}$
stands for $\delta (\underline{k})$.

Already a quick look at Equation (4.32) reveals the presence of a distinguished
scale for cosmological perturbations, the Jeans length
$$
\lambda_J = {2 \pi\over k_J} \eqno\eq
$$
with
$$
k^2_J = \left( {k\over a} \right)^2 = {4 \pi G \bar \rho\over c^2_s} \, .
\eqno\eq
$$
On length scales larger than $\lambda_J$, the spatial gradient term is
negligible, and the term linear in $\delta$ in (4.32) acts like a negative mass
square quadratic potential with damping due to the expansion of the Universe,
in agreement with the intuitive analysis leading to (4.7) and (4.8).  On length
scales smaller than $\lambda_J$, however, (4.32) becomes a damped harmonic
oscillator equation and perturbations on these scales decay.

For $t > t_{eq}$ and for $\lambda \gg \lambda_J$, Equation (4.32) becomes
$$
\ddot \delta_k + {4\over{3t}} \, \dot \delta_k - {2\over{3t^2}} \, \delta_k = 0
\eqno\eq
$$
and has the general solution
$$
\delta_k (t) = c_1 t^{2/3} + c_2 t^{-1} \, . \eqno\eq
$$
This demonstrates that for $t > t_{eq}$ and $\lambda \gg \lambda_J$, the
dominant mode of perturbations increases as $a(t)$, a result we already used in
the previous subsection (see (4.19)).

For $\lambda \ll \lambda_J$ and $t > t_{eq}$, Equation (4.32) becomes
$$
\ddot \delta_k + 2 H \dot \delta_k + c_s^2 \left({k\over a} \right)^2 \delta_k
= 0 \, , \eqno\eq
$$
and has solutions corresponding to damped oscillations:
$$
\delta_k (t) \sim a^{-1/2} (t) \exp \{ \pm i c_s k \int dt^\prime a
(t^\prime)^{-1} \} \, . \eqno\eq
$$

As an important application of the Newtonian theory of cosmological
perturbations, let us compare sub-horizon scale fluctuations in a
baryon-dominated Universe $(\Omega = \Omega_B = 1)$ and in a CDM-dominated
Universe with $\Omega_{CDM} = 0.9$ and $\Omega = 1$.  We consider scales which
enter the Hubble radius at about $t_{eq}$.

In the initial time interval $t_{eq} < t < t_{rec}$, the baryons are coupled to
the photons.  Hence, the baryonic fluid has a large pressure $p_B$
$$
p_B \simeq p_r = {1\over 3} \, \rho_r \, . \eqno\eq
$$
Hence, the speed of sound is relativistic
$$
c_s \simeq \left( {p_r\over \rho_m} \right)^{1/2} = {1\over{\sqrt{3}}} \,
\left({\rho_r\over \rho_m} \right)^{1/2} \, . \eqno\eq
$$
The value of $c_s$ slowly decreases in this time interval, attaining a value of
about $1/10$ at $t_{rec}$.  From (4.34) it follows that the Jeans mass $M_J$,
the mass inside a sphere of radius $\lambda_J$, increases until $t_{rec}$ when
it reaches its maximal value $M_J^{max}$
$$
M_J^{max} = M_J (t_{rec}) = {4 \pi\over 3} \, \lambda_J (t_{rec})^3 \bar \rho
(t_{rec}) \sim 10^{17} (\Omega h^2)^{-1/2} M_{\odot} \, . \eqno\eq
$$

At the time of recombination, the baryons decouple from the radiation fluid.
Hence, the baryon pressure $p_B$ drops abruptly, as does the Jeans length (see
(4.34)).  The remaining pressure $p_B$ is determined by the temperature and
thus continues to decrease as $t$ increases.  It can be shown that the Jeans
mass continues to decrease after $t_{rec}$, starting from a value
$$
M^-_J (t_{rec}) \sim 10^6 (\Omega h^2)^{-1/2} \, M_\odot \eqno\eq
$$
(where the superscript ``$-$" indicates the mass immediately after $t_{eq}$.

In contrast, CDM has negligible pressure throughout the period $t > t_{eq}$ and
hence experiences no Jeans damping.  A CDM perturbation which enters the Hubble
radius at $t_{eq}$ with amplitude $\delta_i$ has an amplitude at $t_{rec}$
given by
$$
\delta^{CDM}_k (t_{rec}) \simeq \, {a (t_{rec})\over{a (t_{eq})}} \, \delta_i
\, , \eqno\eq
$$
whereas a perturbation with the same scale and initial amplitude in a
baryon-dominated Universe is damped
$$
\delta_k^{BDM} (t_{rec}) \simeq \, \left({a (t_{eq})\over{a (t_{eq})}}
\right)^{-1/2} \, \delta_i \, . \eqno\eq
$$
In order for the perturbations to have the same amplitude today, the initial
size of the inhomogeneity must be much larger in a BDM-dominated Universe than
in a CDM-dominated one:
$$
\delta^{BDM}_k (t_{eq}) \simeq \left( {z (t_{eq})\over{z (t_{eq})}}
\right)^{3/2} \delta_k^{CDM} \, (t_{eq}) \, . \eqno\eq
$$
For $\Omega = 1$ and $h = 1/2$ the enhancement factor is about 30.

In a CDM-dominated Universe the baryons experience Jeans damping, but after
$t_{rec}$ the baryons quickly fall into the potential wells created by the CDM
perturbations, and hence the baryon perturbations are proportional to the CDM
inhomogeneities (see Fig. 21).

\midinsert \vskip 13.7cm
\noindent{\bf Figure 21:} Comparison of the growth of fluctuations in baryon
and CDM dominated Universes. The horizontal axis is time, the vertical axis is
the density perturbation. The curve labelled by $a$ describes the evolution of
fluctuations in a BDM Universe. The growth of CDM perturbations follows the
curve $b$, and curve $c$ is a sketch of the time development of baryon
perturbations in a CDM dominated Universe. To be specific, perturbations on a
scale which enters the Hubble radius at $t_{eq}$ are considered.
\endinsert

The above considerations, coupled with information about CMB anisotropies, can
be used to rule out a model with $\Omega = \Omega_B = 1$.  The argument goes as
follows (see Section 7).  For adiabatic fluctuations, the amplitude of CMB
anisotropies on an angular scale $\vartheta$ is determined by the value of
$\delta \rho/\rho$ on the corresponding length scale $\lambda (\vartheta)$ at
$t_{eq}$:
$$
{\delta T\over T} (\vartheta)  = {1\over 3} \, {{\delta \rho} \over \rho} (
\lambda (\vartheta), \, t_{eq} ) \, . \eqno\eq
$$
 On scales of clusters we know that (for $\Omega = 1$ and $h = 1/2$)
$$
\left({\delta_p \over \rho} \right)_{CDM} \, (\lambda (\vartheta), \, t_{eq})
\simeq z (t_{eq})^{-1} \simeq 10^{-4} \, , \eqno\eq
$$
using the fact that today on cluster scales $\delta \rho/\rho \simeq 1$.  The
bounds on $\delta T/ T$ on small angular scales are
$$
{\delta T\over T} (\vartheta) 10^{-4} \, , \eqno\eq
$$
consistent with the predictions for a CDM model, but inconsistent with those of
a $\Omega = \Omega_B = 1$ model, according to which we would expect
anisotropies of the order of $10^{-3}$.  This is yet another argument in
support of the existence of nonbaryonic dark matter.

To conclude this subsection, let us briefly discuss Newtonian perturbations
during the radiation-dominated epoch.  We consider matter fluctuations with
$c_s = 0$ in a smooth relativistic background.  In this case, Equation (4.32)
becomes
$$
\ddot \delta_k + 2 H \dot \delta_k - 4 \pi G \bar \rho_m \delta_k = 0 \, ,
\eqno\eq
$$
where $\bar \rho_m$ denotes the average matter energy density.  The Hubble
expansion parameter obeys
$$
H^2 = {8 \pi G\over 3} (\bar \rho_m + \bar \rho_r) \, , \eqno\eq
$$
with $\bar \rho_r$ the background radiation energy density.  For $t < t_{eq}$,
$\bar \rho_m$ is negligible in both (4.49) and (4.50), and (4.49) has the
general solution
$$
\delta_k (t) = c_1 \log t + c_2 \, . \eqno\eq
$$
In particular, this result implies that CDM perturbations which enter the
Hubble radius before $t_{eq}$ have an amplitude which grows only
logarithmically in time until $t_{eq}$.  This is sometimes called the Meszaros
effect.

\section{Relativistic Theory of Cosmological Perturbations}

On scales larger than the Hubble radius $(\lambda > t)$ the Newtonian theory of
cosmological perturbations obviously is inapplicable, and a general
relativistic analysis is needed.  On these scales, matter is essentially frozen
in comoving coordinates.  However, space-time fluctuations can still increase
in amplitude.

In principle, it is straightforward to work out the general relativistic theory
of linear fluctuations$^{79)}$.  We linearize the Einstein  equations
$$
G_{\mu\nu} = 8 \pi G T_{\mu\nu} \eqno\eq
$$
(where $G_{\mu\nu}$ is the Einstein tensor associated with the space-time
metric $g_{\mu\nu}$, and $T_{\mu\nu}$ is the energy-momentum tensor of matter)
about an expanding FRW background $(g^{(0)}_{\mu\nu} ,\, \varphi^{(0)})$:
$$
\eqalign{
g_{\mu\nu} (\underline{x}, t) & = g^{(0)}_{\mu\nu} (t) + h_{\mu\nu}
(\underline{x}, t) \cr
\varphi (\underline{x}, t) & = \varphi^{(0)} (t) + \delta \varphi
(\underline{x}, t) \, \cr} \eqno\eq
$$
and pick out the terms linear in $h_{\mu\nu}$ and $\delta \varphi$ to obtain
$$
\delta G_{\mu\nu} \> = \> 8 \pi G \delta T_{\mu\nu} \, . \eqno\eq
$$
In the above, $h_{\mu\nu}$ is the perturbation in the metric and $\delta
\varphi$ is the fluctuation of the matter field $\varphi$.  We have denoted all
matter fields collectively by $\varphi$.

In practice, there are many complications which make this analysis highly
nontrivial.  The first problem is ``gauge invariance"$^{80)}$  Imagine starting
with a homogeneous FRW cosmology and introducing new coordinates which mix
$\underline{x}$ and $t$.  In terms of the new coordinates, the metric now looks
inhomogeneous.  The inhomogeneous piece of the metric, however, must be a pure
coordinate (or "gauge") artefact.  Thus, when analyzing relativistic
perturbations, care must be taken to factor out effects due to coordinate
transformations.

\midinsert \vskip 9.5cm
\noindent{\bf Figure 22:} Sketch of how two choices of the mapping from the
background space-time manifold ${\cal M}_0$ to the physical manifold ${\cal M}$
induce two different coordinate systems on ${\cal M}$.
\endinsert

The issue of gauge dependence is illustrated in Fig. 22.  A coordinate system
on the physical inhomogeneous space-time manifold ${\cal M}$ can be viewed as a
mapping ${\cal D}$ of an unperturbed space-time ${\cal M}_0$ into ${\cal M}$.
A physical quantity $Q$ is a geometrical function defined on ${\cal M}$.  There
is a corresponding physical quantity $^{(0)}Q$ defined on ${\cal M}_0$.  In the
coordinate system given by ${\cal D}$, the perturbation $\delta Q$ of $Q$ at
the space-time point $p \, \epsilon \, {\cal M}$ is
$$
\delta Q (p) = Q (p) - \,^{(0)}Q \, (D^{-1} (p) ) \, . \eqno\eq
$$
However, in a second coordinate system $\tilde {\cal D}$ the perturbation is
given by
$$
\delta \tilde Q (p) = Q (p) - \,^{(0)}Q (\tilde {\cal D}^{-1} (p) ) \, .
\eqno\eq
$$
The difference
$$
\Delta Q (p) = \delta Q (p) - \delta \tilde Q (p) \eqno\eq
$$
is obviously a gauge artefact and carries no physical meaning.

There are various methods of dealing with gauge artefacts.  The simplest and
most physical approach is to focus on gauge invariant variables, i.e.,
combinations of the metric and matter perturbations which are invariant under
linear coordinate transformations.

The gauge invariant theory of cosmological perturbations is in principle
straightforward, although technically rather tedious. In the following I will
summarize the main steps and refer the reader to Ref. 11 for the details and
further references (see also Ref. 81 for a pedagogical introduction and Refs.
82-87 for other approaches).

We consider perturbations about a spatially flat Friedmann-Robertson-Walker
metric
$$
ds^2 = a^2 (\eta) (d\eta^2 - d \underline{x}^2) \eqno\eq
$$
where $\eta$ is conformal time (related to cosmic time $t$ by $a(\eta)  d \eta
= dt$).  A scalar metric perturbation (see Ref. 88 for a precise definition)
can be written in terms of four free functions of space and time:
$$
\delta g_{\mu\nu} = a^2 (\eta) \pmatrix{2 \phi & -B_{,i} \cr
-B_{,i} & 2 (\psi \delta_{ij} + E_{,ij} \cr} \, . \eqno\eq
$$
Scalar metric perturbations are the only perturbations which couple to energy
density and pressure.

The next step is to consider infinitesimal coordinate transformations
$$
x^{\mu^\prime} = x^\mu + \xi^\mu \eqno\eq
$$
which preserve the scalar nature of $\delta g_{\mu\nu}$ and to calculate the
induced transformations of $\phi, \psi, B$ and $E$.  Then we find invariant
combinations to linear order.  (Note that there are in general no combinations
which are invariant to all orders$^{89)}$.)  After some algebra, it follows
that
$$
\eqalign{
\Phi & = \phi + a^{-1} [(B - E^\prime) a]^\prime \cr
\Psi & = \psi - {a^\prime\over a} \, (B - E^\prime) \cr} \eqno\eq
$$
are two invariant combinations.  In the above, a prime denotes differentiation
with respect to $\eta$.

There are various methods to derive the equations of motion for gauge invariant
variables.  Perhaps the simplest way$^{11)}$ is to consider the linearized
Einstein equations (4.54) and to write them out in the longitudinal gauge
defined by
$$
B = E = 0 \eqno\eq
$$
and in which $\Phi = \phi$ and $\Psi = \psi$, to directly obtain gauge
invariant equations.

For several types of matter, in particular for scalar field matter, the
perturbation of $T_{\mu \nu}$ has the special property
$$
\delta T^i_j \sim \delta^i_j \eqno\eq
$$
which imples $\Phi = \Psi$.  Hence, the scalar-type cosmological perturbations
can in this case be described by a single gauge invariant variable.  The
equation of motion takes the form$^{90, 9, 10)}$
$$
\dot \xi = O \left({k\over{aH}} \right)^2 H \xi \eqno\eq
$$
where
$$
\xi = {2\over 3} \, {H^{-1} \dot \Phi + \Phi\over{1 + w}} + \Phi \, . \eqno\eq
$$

The variable $w = p/ \rho$ (with $p$ and $\rho$ background pressure and energy
density respectively) is a measure of the background equation of state.  In
particular, on scales larger than the Hubble radius, the right hand side of
(4.64) is negligible, and hence $\xi$ is constant.

The result that $\dot \xi = 0$ is a very powerful one.  Let us first imagine
that the equation of state of matter is constant, {\it i.e.}, $w = {\rm
const}$.  In this case, $\dot \xi = 0$ implies
$$
\Phi (t) = {\rm const} \, , \eqno\eq
$$
{\it i.e.}, this gauge invariant measure of perturbations remains constant
outside the Hubble radius.

Next, consider the evolution of $\Phi$ during a phase transition from an
initial phase with $w = w_i$  to a phase with $w = w_f$.  Long before and after
 the transition, $\Phi$ is constant because of (4.66), and hence $\dot \xi = 0$
becomes
$$
{\Phi\over{1 + w}} + \Phi = {\rm const} \, , \eqno\eq
$$

In order to make contact with matter perturbations and Newtonian intuition, it
is important to remark that,  as a consequence of the Einstein constraint
equations, at Hubble radius crossing $\Phi$ is a measure of the fractional
density fluctuations:
$$
\Phi (k, t_H (k) ) \sim {\delta \rho\over \rho} \, ( k , \, t_H (k) ) \, .
\eqno\eq
$$
(Note that the latter quantity is approximately gauge invariant on scales
smaller than the Hubble radius).
\chapter{Inflationary Universe Scenarios}
\section{Preliminaries}

Cosmological inflation$^{30)}$ is a period in time during which the Universe is
expanding exponentially, {\it i.e.},
$$
a (t) = e^{tH} \eqno\eq
$$
with constant Hubble expansion rate $H$.  From the FRW equations (2.7) and
(2.9) it follows that the condition for inflation (in the context of Einstein
gravity in a spatially flat Universe) is an equation of state for matter with
$$
p = - \rho \, , \eqno\eq
$$
which neccessitates abandoning a description of matter in terms of an ideal
gas.

As was indicated in Section 3.1, it is possible to achieve inflation if matter
is described in terms of scalar fields, provided that at some period
$$
\eqalign{
\dot \varphi^2 \ll V (\varphi) \cr
(\nabla \varphi)^2 \ll V (\varphi) }\eqno\eq
$$
(see Eqs. (3.9) which give the equation of state for scalar field matter).

\midinsert \vskip 5.5cm
\noindent{\bf Figure 23:} A sketch of two potentials which can give rise to
inflation.
\endinsert

Two examples which can give inflation are shown in Fig. 23.  In (a), inflation
occurs at the stable fixed point $\varphi (\underline{x}, t_i) = 0 = \dot
\varphi (\underline{x}, t_i)$.  However, this model is ruled out by
observation: the inflationary phase has no ending.  $V(0)$ acts as a permanent
nonvanishing cosmological constant.  In (b), a finite period of inflation can
arise if $\varphi (\underline{x})$ is trapped at the local minimum $\varphi =
0$ with $\dot \varphi (\underline{x}) = 0$.  However, in this case $\varphi
(\underline{x})$ can make a sudden transition at some time $t_R > t_i$ through
the potential barrier and move to $\varphi (\underline{x}) = a$.  Thus, for
$t_i < t < t_R$ the Universe expands exponentially, whereas for $t > t_R$ the
contribution of $\varphi$ to the expansion of the Universe vanishes and we get
the usual FRW cosmology.  There are three obvious questions: why does the field
start out at $\varphi = 0$, how does the transition occur and why should the
scalar field have $V(\varphi) = 0$ at the global minimum?  In the following
section the first two questions will be addressed.  The third question is part
of the cosmological constant problem for which there is as yet no convincing
explanation.  Before studying the dynamics of the phase transition, we need to
digress and discuss finite temperature effects.

\section{Finite Temperature Field Theory}
\par
The evolution of particles in vacuum and in a thermal bath are very different.
Similarly, the evolution of fields changes when coupled to a thermal bath.
Under certain conditions, the changes may be absorbed in a temperature
dependent potential, the finite temperature effective
potential.
Here, a heuristic derivation of this potential will be given. The reader is
referred to Ref. 8 or to the original articles$^{91)}$ for the actual
derivation.

        We assume that the scalar field $\varphi (\undertext{x},t)$ is
coupled to a thermal bath which is represented by a second scalar field
$\psi (\undertext{x}, t)$ in thermal equilibrium. The
Lagrangian for $\varphi$ is
$$
{\cal L} = {1\over 2} \partial_\mu \varphi \partial^\mu \varphi \, - \,
V(\varphi) \, - \, {1\over 2} \hat \lambda \varphi^2 \psi^2\, , \eqno\eq
$$
where $\hat \lambda$ is a coupling constant.  The action from which the
equations of motion are derived is
$$
S \, = \, \int d^4 x \, \sqrt{-g} {\cal L}\eqno\eq
$$
where $g$ is the determinant of the metric (2.2).   The resulting equation of
motion for $\varphi (\undertext{x},t)$ is
$$
\ddot \varphi + 3 H \dot \varphi \, - a^{-2} \bigtriangledown^2 \varphi \, =
\,- V^\prime (\varphi) - \hat \lambda \psi^2 \varphi \, . \eqno\eq
$$
If $\psi$ is in thermal equilibrium, we may replace $\psi^2$ by its thermal
expectation value $<\psi^2>_T$.  Now,
$$
<\psi^2>_T \sim T^2\eqno\eq
$$
which can be seen as follows:  in thermal equilibrium, the energy density of
$\psi$ equals that of one degree of freedom in the thermal bath.  In
particular, the potential energy density $V (\psi)$ of $\psi$ is of that order
of magnitude.  Let
$$
V (\psi) \, = \, \lambda_\psi \psi^4\eqno\eq
$$
with a coupling constant $\lambda_\psi$ which we take to be of the order 1
(if $\lambda_\psi$ is too small, $\psi$ will not be in thermal equilibrium).
Since the thermal energy density is proportional to $T^4$, (5.7) follows.
(5.6) can be rewritten as
$$
\ddot \psi + 3H \dot \varphi \, - \, a^{-2} \bigtriangledown^2 \varphi \, = \,
- V_T^\prime (\varphi),\eqno\eq
$$
where
$$
V_T (\varphi) \, = \, V (\varphi) \, + \, {1\over 2} \hat \lambda T^2
\varphi^2\eqno\eq
$$
is called the finite temperature effective potential.  Note that in
(5.10),
$\hat \lambda$ has been rescaled to absorb the constant of proportionality in
(5.7).
\par
These considerations will now be applied to Example A, a scalar field model
with
potential
$$
V (\varphi) \, = \, {1\over 4} \lambda (\varphi^2 - \sigma^2)^2\eqno\eq
$$
($\sigma$ is called the scale of symmetry breaking).  The finite temperature
effective potential becomes (see Fig. 24)
$$
V_T (\varphi) \, = \, {1\over 4} \lambda \varphi^4 - {1\over 2}
\, \left(\lambda \sigma^2 - \hat \lambda T^2\right) \varphi^2 +
\, {1\over 4} \, \lambda \sigma^4 \, . \eqno\eq
$$
For very high temperatures, the effective mass term is positive
and hence the energetically favorable state is $<\varphi> = 0$.
For very low temperatures, on the other hand, the mass term has a
negative sign which leads to spontaneous symmetry breaking.  The
temperature at which the mass term vanishes defines the critical
temperature $T_c$
$$
T_c \, = \, \hat \lambda^{-1/2} \lambda^{1/2} \sigma\,.\eqno\eq
$$

\midinsert \vskip 6cm
\noindent {\bf Figure 24:}  The finite temperature effective potential
for Example A.
\endinsert

As Example B, consider a theory with potential
$$
 V (\varphi) = \, {1\over 4} \varphi^4 - {1\over 3} \, (a + b) \varphi^3 +
{1\over 2} ab \varphi^2\eqno\eq
$$
with ${1\over 2} a > b > 0$.  The finite temperature effective potential is
obtained by adding ${1\over 2} \hat \lambda  T^2 \varphi^2$ to the
right hand side of (5.14).  ${ V_T} (\varphi)$
is sketched in Fig. 25 for various values of $ T$.  The critical temperature
$T_c$ is defined as the temperature when the two minima of ${V_T}(\varphi)$
become degenerate.

\midinsert \vskip 6cm
\noindent {\bf Figure 25:} The finite temperature effective potential
for Example B.
\endinsert

It is important to note that the use of finite temperature effective potential
methods is only legitimate if the system is in thermal equilibrium.  This
point was stressed in Refs. 92 and 93, although the conclusion should be
obvious from the derivation given above.  To be more precise, we require the
$\psi$ field to be in thermal equilibrium and the coupling constant $\hat
\lambda$ of (5.4) which mediates the energy exchange between the $\varphi$
and $\psi$ fields to be large.  However, as shown in Chapter 4, in inflationary
Universe models, the observational
constraints stemming from the amplitude of the primordial energy density
fluctuation spectrum force the self coupling constant $\lambda$ of $\varphi$
to be extremely small.  Since at one loop order, the interaction term ${1\over
2} \hat \lambda \varphi^2 \psi^2$ induces contributions to $\lambda$, it is
unnatural to have $\lambda$ very small and $\hat \lambda$ unsuppressed.
Hence, in many inflationary Universe models - in particular in new
inflation$^{94)}$ and in chaotic inflation$^{92)}$ - finite temperature
effective potential methods are inapplicable.
\section{Phase Transitions}
\par
The temperature dependence of the finite temperature effective potential in
quantum field theory leads to phase transitions in the very early Universe.
These transitions are either first or second order.
\par
Example $A$ of the previous section provides a model in which the transition
is second order (see Fig. 24).  For $T \gg T_c$, the expectation value of the
scalar field $\varphi$ vanishes at all points $\undertext{x}$ in space:
$$
< \varphi (\undertext{x}) > = 0 \, . \eqno\eq
$$
For $T < T_c$, this value of $< \varphi (\undertext{x}) >$ becomes unstable
and $< \varphi (\undertext{x})>$ evolves smoothly in time to a new value $\pm
\sigma$.  The direction is determined by thermal and quantum fluctuations and
is therefore not uniform in space.  There will be domains of average radius
$\xi (t)$ in which $< \varphi (\undertext{x}) >$ is coherent.  By causality,
the coherence length is bounded from above by the horizon.  However, typical
values of $\xi (t)$ are proportional to $\lambda^{-1} \sigma^{-1}$ if
$\varphi$ was in thermal equilibrium before the phase
transition.
\par
In condensed matter physics, a transition of the above type is said to proceed
by spinodal decomposition$^{95)}$, triggered by a rapid quench.
\par
In Example B of the previous section, (see Fig. 25) the phase
transition is first order.  For $T > T_c$, the expectation value
$<\varphi (x) >$ is approximately $0$, the minimum of the high
temperature effective potential.  Provided the zero temperature
potential has a sufficiently high barrier separating the metastable
state $\varphi = 0$ from the global minimum (compared to the energy
density in thermal fluctuations at $T = T_c$), then $\varphi
(\undertext{x})$ will remain trapped at $\varphi = 0$ also for $T <
T_c$.  In the notation of Ref. 96, the field $\varphi$ is trapped in
the false vacuum.  After some time (determined again by the potential
barrier), the false vacuum will decay by quantum tunnelling.
\par
Tunnelling in quantum field theory was discussed in Refs. 96-99 (for
reviews see e.g., Refs. 100 and 8).  The transition proceeds by bubble
nucleation.  There is a probability per unit time and volume that at a
point $\undertext{x}$ in space a bubble of ``true vacuum" $\varphi
(\undertext{x}) = a$ will nucleate.  The nucleation radius is
microscopical.  As long as the potential barrier is large, the bubble
radius will increase with the speed of light after nucleation.  Thus,
a bubble of $\varphi = a$ expands in a surrounding ``sea" of false
vacuum $\varphi = 0$.
\par
To conclude, let us stress the most important differences between the
two types of phase transitions discussed above. In a second order
transition, the dynamics is determined mainly by
classical physics.  The transition occurs homogeneously in space
(apart from the phase boundaries which -- as discussed below -- become
topological defects), and $< \varphi (x) >$ evolves continuously in
time.  In first order transitions, quantum mechanics is essential.
The process is extremely inhomogeneous, and $< \varphi (x) >$ is
discontinuous as a function of time.  As we shall see in the following
sections, the above two types of transitions are the basis of various
classes of inflationary Universe models.

\section{Models of Inflation}

At this stage we have established the formalism to be able to discuss models of
inflation.  I will focus on ``old inflation," ``new inflation"" and ``chaotic
inflation."  There are many other attempts at producing an inflationary
scenario, but there is as of now no convincing realization.

\subsection{Old Inflation}
\par
The old inflationary Universe model$^{30, 101)}$ is based on a scalar field
theory which undergoes a first order phase transition.  As a toy
model, consider a scalar field theory with the potential $V (\varphi)$
of Example B (see Fig. 25).  Note that this potential is fairly general
apart from the requirement that $V (a) = 0$, where $\varphi = a$ is
the global minimum of $V (\varphi)$.  This condition is required to
avoid a large cosmological constant today (no inflationary Universe
model manages to circumvent or solve the cosmological constant problem).
\par
For fairly general initial conditions, $\varphi (x)$ is trapped in the
metastable state $\varphi = 0$ as the Universe cools below the
critical temperature $T_c$.  As the Universe expands further, all
contributions to the energy-momentum tensor $T_{\mu \nu}$ except for
the contribution
$$
T_{\mu \nu} \sim V(\varphi) g_{\mu \nu} \eqno\eq
$$
redshift.  Hence, the equation of state approaches $p = - \rho$, and
inflation sets in.  Inflation lasts until the false vacuum decays.
During inflation, the Hubble constant is given by
$$
H^2 = {8 \pi G\over 3} \, V (0) \, . \eqno\eq
$$

\midinsert \vskip 6cm
\noindent {\bf Figure 26:} A sketch of the spatially inhomogeneous
distribution of $\varphi$ in Old Inflation.
\endinsert

After a period $\Gamma^{-1}$, where $\Gamma$ is the tunnelling decay
rate, bubbles of $\varphi = a$ begin to nucleate in a sea of false
vacuum $\varphi = 0$.  For a sketch of the resulting inhomogeneous
distribution of $\varphi (x)$ see Fig. 26.  Note that inflation stops
after bubble nucleation.
\par
The time evolution in old inflation is summarized in Fig. 27.  We denote the
beginning of inflation by $t_i$ (here $t_i \simeq t_c$), the end by $t_R$
(here $t_R \simeq t_c + \Gamma^{-1}$).

\midinsert \vskip 4cm
\noindent {\bf Figure 27:}  Phases in the old and new inflationary Universe.
\endinsert

It was immediately realized that old inflation has a serious ``graceful exit"
problem$^{102)}$.  The bubbles nucleate after inflation with radius $r \ll
2t_R$ and would today be much smaller than our apparent horizon.  Thus, unless
bubbles percolate, the model predicts extremely large inhomogeneities inside
the Hubble radius, in contradiction with the observed isotropy of the
microwave background radiation.
\par
For bubbles to percolate, a sufficiently large number must be produced so that
they collide and homogenize over a scale larger than the present Hubble
radius.  However, with exponential expansion, the volume between bubbles
expands
exponentially whereas the volume inside bubbles expands only with a low power.
This prevents percolation.

\subsection{New Inflation}

Because of the graceful exit problem, old inflation never was considered to be
a viable cosmological model.  However, soon after the seminal paper by
Guth, Linde and independently Albrecht and Steinhardt put
forwards a modified scenario, the New Inflationary Universe$^{94)}$ (see also
Ref. 103).

\midinsert \vskip 6cm
\noindent {\bf Figure 28:} A sketch of the spatial distribution of
$\varphi$ in New Inflation after the transition. The symbols $+$ and
$-$ indicate regions where $\varphi = + \sigma$ and $\varphi = -
\sigma$ respectively.
\endinsert

The starting point is a scalar field theory with a double well potential which
undergoes a second order phase transition (Fig. 24).  $V(\varphi)$ is
symmetric and $\varphi = 0$ is a local maximum of the zero temperature
potential.  Once again, it was argued that finite temperature effects confine
$\varphi(\undertext{x})$ to values near $\varphi = 0$ at temperatures $T \ge
T_c$.  For $T < T_c$, thermal fluctuations trigger the instability of $\varphi
(\undertext{x}) = 0$ and $\varphi (\undertext{x})$ evolves towards $\varphi =
\pm \sigma$ by the classical equation of motion
$$
\ddot \varphi + 3 H \dot \varphi - a^{-2} \bigtriangledown^2 \varphi = -
V^\prime (\varphi)\, . \eqno\eq
$$
\par
The transition proceeds by spinodal decomposition (see Fig. 28) and hence
$\varphi(\undertext{x})$ will be homogeneous within a correlation length.  The
analysis will be confined to such a small region.  Hence, in Eq.
(5.18)
we can
neglect the spatial gradient terms.  Then, from (3.9) we can read off the
induced equation of state.  The condition for inflation is
$$
\dot \varphi^2 \ll V (\varphi)\, ,\eqno\eq
$$
\ie~ slow rolling.
\par
Often, the  ``slow rolling" approximation is made to find solutions of
(5.18).
This consists of dropping the $\ddot \varphi$ term.  In this case,
(5.18)
becomes
$$
3 H \dot \varphi \, = - \, V^\prime (\varphi)\, . \eqno\eq
$$
As an example, consider a potential which for $|\varphi| < \sigma$ has the
following expansion near $\varphi = 0$
$$
V (\varphi) = V_0 - {1\over 2} m^2 \varphi^2\, . \eqno\eq
$$
With the above $V (\phi)$, (5.20) has the solution
$$
\varphi (t) \, = \, \varphi (0) \exp \left({m^2\over{3H}} t \right)\eqno\eq
$$
(taking $H = \rm const$ which is a good approximation).  Thus, provided $m \ll
\sqrt{3} H \, , \, \ddot \varphi$ is indeed smaller than the other terms in
(5.18) and the slow rolling approximation seems to be satisfied.
\par
However, the above conclusion is premature$^{104)}$.  Equation (5.18) has a
second solution.  For $m > H$ the solution is
$$
\varphi (t) \simeq \varphi (0) e^{mt}\eqno\eq
$$
and dominates over the previous one.  This example shows that the slow rolling
approximation must be used with caution.  Here, however, the conclusion
remains that provided $m \ll H$, then the model produces enough inflation to
solve the cosmological problems.
\par
There is no graceful exit problem in the new inflationary Universe.  Since the
spinodal decomposition domains are established before the onset of inflation,
any boundary walls will be inflated outside the present Hubble radius.
\par
The condition $m^2 \ll 3 H^2$ which must be imposed in order to obtain
inflation, is a fine tuning of the particle physics model -- the first sign of
problems with this scenario.  Consider \eg~ the model (5.11).  By expanding $V
(\varphi)$ about $\varphi = 0$ we can determine both $H$ and $m$ in terms of
$\lambda$ and $\sigma$.  In order that $m^2 < 3 H^2$ be satisfied we need
$$
\sigma > \, \left( {1 \over{6 \pi}}\right)^{1/2} m_{pl} \, , \eqno\eq
$$
which is certainly an unnatural constraint for models motivated by particle
physics.
\par
Let us, for the moment, return to the general features of the new inflationary
Universe scenario.  At the time $t_c$ of the phase transition, $\varphi (t)$
will start to move from near $\varphi = 0$ towards either $\pm \sigma$ as
described by the classical equation of motion, \ie~ (5.22).  At or soon after
$t_c$, the energy-momentum tensor of the Universe will start to be dominated
by $V(\varphi)$, and inflation will commence.  $t_i$ shall denote the time of
the onset of inflation.  Eventually, $\phi (t)$ will reach large values for
which (5.21) is no longer a good approximation to $V (\varphi)$ and for which
nonlinear effects become important.  The time at which this occurs is $t_B$.
For $t > t_B \, , \, \varphi (t)$ rapidly accelerates, reaches $\pm \sigma$,
overshoots and starts oscillating about the global minimum of $V (\varphi)$.
The amplitude of this oscillation is damped by the expansion of the Universe
and (predominantly) by the coupling of $\varphi$ to other fields.  At time
$t_R$,
the energy in $\varphi$ drops below the energy of the thermal bath of
particles produced during the period of oscillation.
\par
The evolution of $\varphi (t)$ is sketched in Fig. 29.  The time period
between $t_B$ and $t_R$ is called the reheating period and is usually short
compared to the Hubble expansion time.  The time evolution of the temperature
$T$ of the thermal radiation bath is also shown in Fig. 29.

\midinsert \vskip 6cm
\noindent {\bf Figure 29:}  Evolution of $\varphi (t)$ and $T (t)$ in the new
inflationary Universe.
\endinsert

Reheating in inflationary Universe models has been considered in Refs.
105-108.
One way to view the process is as follows$^{107, 108)}$.  Consider a second
scalar
field $\psi$ coupled to $\varphi$ via the interaction Lagrangian
$$
{\cal L}_I = {1\over 2} g \varphi^2 \psi^2\, . \eqno\eq
$$
Then, an oscillating $\varphi (t)$ will act as a time dependent mass with
periodic variations in the equation of motion for $\psi$
$$
\ddot \psi_k + 3 H \dot \psi_k + \left(m_\psi^2 + k^2 a^{-2} (t) + g \varphi^2
(t) \right) \psi_k = 0\eqno\eq
$$
where we have neglected nonlinear terms and expanded $\psi$ into Fourier modes
$\psi_k$.  If the expansion of the Universe can be neglected and for periodic
$\varphi^2 (t)$, the above is the well known Mathieu equation$^{109)}$ whose
solutions have instabilities for certain values of $k$.  These instabilities
correspond to the production of $\psi$ particles with well determined
momenta$^{107, 108)}$.  These particles eventually equilibrate and
regenerate a thermal bath.
\par
For $t > t_R$, the Universe is again radiation dominated.  Hence, the stages
of the new inflationary Universe are the same as for old inflation
(Fig. 27).
There is a useful order of magnitude relation between the scale of symmetry
breaking $\sigma$ and $H$.  From
$$
H^2 = \, {8 \pi G\over 3} \, V(0)\eqno\eq
$$
and from the form of the potential (see (5.11)) it follows that
$$
\left({H\over \sigma}\right) \sim \, \lambda^{1/2} \,
\left({\sigma\over{m_{pl}}}\right)\, .\eqno\eq
$$
In particular, for $\sigma \sim 10^{15}$GeV (typical scale of grand
unification) and $\lambda \sim 1$ we obtain $H \sim 10^{11}$GeV.
\par
The new inflationary Universe model -- although it was for a long time
presented as a viable model -- suffers from severe fine tuning and initial
condition problems.  In (5.24) we encountered the first of these problems: in
order to obtain enough inflation, the potential must be fairly flat near
$\varphi = 0$.  A more severe problem will be derived in Chapter 4:
Inflationary Universe models generate energy density perturbations.  The
steeper the potential, the larger the density perturbations.  For a potential
which near $\varphi = 0$ or $\varphi = H$ has the following expansion
$$
V (\varphi) = V (0) - \lambda \, \varphi^4\, ,\eqno\eq
$$
the density perturbations conflict with observations unless (see later)
$$
\lambda < 10^{-12}\, .\eqno\eq
$$
\par
This in itself is an unexplained small number problem.  However, even if we
were
willing to accept this we would run into initial condition
problems$^{92, 93)}$.  For the new inflationary Universe to proceed in the way
outlined above, it is essential that the field $\varphi$ be in thermal
equilibrium with other fields.  This implies that the constant $g$ coupling
$\varphi$ to other fields should not be too small. However,  a coupling term of
the form (5.25) induces
one loop quantum corrections to the self coupling constant $\lambda$
of the order $g^2$.  Hence, the constraint $\lambda < 10^{-12}$ implies a
constraint $g < 10^{-6}$.  Thus, $\varphi$ will not be in thermal equilibrium
at $t_c$, and hence there will be no thermal forces which localize $\varphi$
close to $\varphi = 0$.
\par
Note that the above problem is not an artifact of using quartic potentials
such as (5.29).  Similar constraints would arise in other (\eg~ quadratic)
models.  However, (5.29) was long considered to be the prototypical shape of
$V (\varphi)$ for small values of $\varphi$ since it is the shape which arises
in Coleman-Weinberg$^{110)}$ models.
\par
In the absence of thermal forces which constrain $\varphi$ to start close to
$\varphi = 0$, the only constraints on $\varphi - \,$ at least using classical
physics alone -- come from energetic considerations.  Obviously, it is
unnatural to assume that at the initial time $t_i$ the energy density in
$\varphi$ exceeds the energy density of one degree of freedom of the thermal
bath at time $t_i$ (temperature $T_i$).  This implies
$$
\eqalign{V (\varphi (\undertext{x}, t_i)) & < {\pi^2\over{30}} T_i^4\cr
| \bigtriangledown \varphi (\undertext{x}, t_i) |^2 & < {\pi^2\over{30}} T_i^4
a^2 (t_i)\cr
| \dot \varphi^2 (\undertext{x}, t_i) |^2 &< {\pi^2\over{30}} T_i^4}\eqno\eq
$$
In particular, for the double well potential of (5.11), (5.31) implies that
$\varphi (\undertext{x}, t_i)$ can be of the order
$$
\varphi (\undertext{x}, t_i) \sim \lambda^{-1/4} T_i\eqno\eq
$$
which for $T_i > \sigma$ is much larger than $\sigma$.  In a weakly coupled
model, the only natural time to impose initial conditions on $\varphi
(\undertext{x})$ is the Planck time, \ie~ $T_i \sim m_{pl}$.  Hence, the
initial conditions allow and in fact suggest
$$
\varphi (\undertext{x}, t_i) \sim \lambda^{-1/4} m_{pl} \, \gg \, m_{pl} \quad
\rm{for} \> T_i \sim T_{pl},\eqno\eq
$$
These observations lead to the chaotic inflation scenario, the only of the
original inflationary Universe models which can still be considered as a
viable scenario today.
\subsection{Chaotic Inflation}
\par
Chaotic inflation$^{92)}$ is based on the observation that for weakly coupled
scalar fields, initial conditions which follow from classical considerations
alone lead to very large values of $\varphi (\undertext{x})$ (see
(5.32)).
\par
Consider a region in space where at the initial time $\varphi (\undertext{x})$
is very large, homogeneous (we will make these assumptions quantitative
below) and static.  In this case, the energy-momentum tensor will be
immediately dominated by the large potential energy term and induce an
equation of state $p \simeq - \rho$ which leads to inflation.  Due to the
large Hubble damping term in the scalar field equation of motion, $\varphi
(\undertext{x})$ will only roll very slowly towards $\varphi = 0$.  The
kinetic energy contribution to $T_{\mu \nu}$ will remain small, the spatial
gradient contribution will be exponentially suppressed due to the expansion of
the Universe, and thus inflation persists.  This is a brief survey of the
chaotic inflation scenario.  Note that in contrast to old and new inflation,
no initial thermal bath is required.  Note also that the precise form of
$V(\varphi)$ is irrelevant to the mechanism.  In particular, $V(\varphi)$ need
not be a double well potential.  This is a significant advantage, since for
scalar fields other than Higgs fields used for spontaneous symmetry breaking,
there is no particle physics motivation for assuming a double well potential,
and since the inflaton (the field which gives rise to inflation) cannot be a
conventional Higgs field due to the severe fine tuning constraints.
\par
Let us consider the chaotic inflation scenario in more detail.  For
simplicity, take the potential
$$
V (\varphi) = \, {1\over 2} \, m^2 \varphi^2\eqno\eq
$$
and consider a region in space in which $\varphi (\undertext{x}, t_i)$ is
sufficiently homogeneous.  To be specific, we require
$$
{1\over 2} \, a^{-2} (t_i) | \bigtriangledown \varphi \, (\undertext{x},
t_i)|^2 \, \ll \, V\left(\varphi (\undertext{x}, t_i)\right)\eqno\eq
$$
over a region of size $d_i$
$$
d_i \ge 3 H^{-1} (t_i)\, .\eqno\eq
$$
We also require that the kinetic energy be negligible at the initial time
$t_i$,
$$
\dot \varphi (\undertext{x}, t_i)^2 \, \ll \,
V (\varphi \left(\undertext{x}, t_i) \right)\, ,\eqno\eq
$$
although this assumption can be relaxed without changing the
results$^{111)}$.
{}From (5.35) and (5.37) it follows that at $t_i$ the equation of state is
inflationary, \ie~ $p (t_i) \simeq - \rho (t_i)$.  Condition (5.36) ensures
that no large inhomogeneities can propagate from outside to the center of the
region under consideration.  With these approximations, the equation of motion
for $\varphi$ becomes
$$
\ddot \varphi + 3 H \dot \varphi = - m^2 \varphi\eqno\eq
$$
with
$$
H = \left({4 \pi\over 3}\right)^{1/2} \, {m\over{m_{pl}}} \, \varphi\, .
\eqno\eq
$$
\par
Since we expect $\varphi (\undertext{x}, t)$ to be changing slowly, we make
the slow rolling approximation
$$
3 H \dot \varphi = - m^2 \varphi\eqno\eq
$$
which gives
$$
\dot \varphi = - \left({1\over{12 \pi}}\right)^{1/2} m \, m_{pl}\eqno\eq
$$
and shows that the approximation is self consistent.  In order to get
inflation, we require
$$
{1\over 2} \, \dot \varphi^2 < {1\over 2} m^2 \, \varphi^2\eqno\eq
$$
which (by (5.41)) is satisfied if
$$
\varphi > \left({1\over{12}}\right)^{1/2} m_{pl}\, . \eqno\eq
$$
In order to obtain a period $\tau > 50 H^{-1} (t_i)$ of inflation, a slightly
stronger condition is needed:
$$
\varphi > 3 m_{pl}\, . \eqno\eq
$$
\par
With chaotic inflation, the initial hope that grand unified theories could
provide the answer to the homogeneity and flatness problems has been
abandoned.  The inflaton is introduced as a new scalar field (with no
particular particle physics role) which is very weakly coupled to itself and to
other fields (see Ref. 112 for an attempt to couple the inflaton to non-grand
unified particle physics).  In supergravity and in superstring inspired models
there are scalar fields which are candidates to be the inflaton.  I refer the
reader to Refs. 113-115 for a discussion of this issue.  However, even in such
models the time when the inflaton $\varphi$ decouples from the rest of physics
is the Planck time $t_i = t_{pl}$.  Thus, the chaotic inflation scenario is
often called primordial inflation.
\par
Chaotic inflation is a much more radical departure from standard cosmology
than old and new inflation.  In the latter, the inflationary phase can be
viewed as a short phase of exponential expansion bounded at both ends by
phases of radiation domination.  In chaotic inflation, a piece of the Universe
emerges with an inflationary equation of state immediately after the quantum
gravity epoch.

The chaotic inflationary Universe scenario has been developed in great detail
(see {\it e.g.}, Ref. 116 for a recent review).  One important addition is the
inclusion of stochastic noise$^{117)}$ in the equation of motion for $\varphi$
in order to take into account the effects of quantum fluctuations.  It can in
fact be shown that for sufficiently large values of $| \varphi |$, the
stochastic force terms are more important than the classical relaxation force
$V^\prime (\varphi)$.  There is equal probability for the quantum fluctuations
to lead to an increase or decrease of $| \varphi |$.  Hence, in a substantial
fraction of comoving volume, the field $\varphi$ will climb up the potential.
This leads to the conclusion that chaotic inflation is eternal.  At all times,
a large fraction of the physical space will be inflating.  Another consequence
of including stochastic terms is that on large scales (much larger than the
present Hubble radius), the Universe will look extremely inhomogeneous.

\subsection{General Comments}

Old, new and choatic inflation are all based on the use of new fundamental
scalar fields which cannot be the Higgs field of an ordinary gauge theory.
Instead of introducing new physics via scalar fields -- an approach which makes
the cosmological constant problem more severe -- it is possible to look for
realizations of inflation based on some alternative new physics which do not
invoke fundamental scalar fields.

One possibility is to consider modifications of Einstein gravity which can lead
to inflation.  In fact, the first model of inflation$^{39)}$ was based on
considering an action for gravity of the form
$$
S = \int d^4 x \sqrt{-g} \, (R + \varepsilon R^2) \, , \eqno\eq
$$
where $R$ is the Ricci scalar of the metric $g_{\mu\nu}$.  As shown first by
Whitt$^{118)}$, the equations of motion resulting from this action are the same
as those from Einstein gravity in the presence of a scalar field with a special
potential which allows for chaotic inflation.  The relationship is obtained via
a conformal transformation.

Since perturbative quantum gravity calculations and considerations based on
quantum field theory in curved space-time all point to the presence of higher
derivative terms in the action for gravity, it is not unlikely that successful
inflation will come from the gravity sector.  A recently proposed
theory$^{41)}$, in which the curvature is bounded for all solutions, predicts
that our Universe will have started out in an inflationary period.

Another interesting possibility is that inflation is a result of new physics
associated with a unified theory of all forces such as superstring theory.
Interesting speculations along these lines have recently been made in Ref. 119.
\par
As has hopefully become clear, inflation
is a nice idea which solves many problems of standard big bang
cosmology.  However, no convincing realization of inflation which does
not involve unexplained small numbers has emerged (for a general
discussion of this point see Ref. 120).
\par
It is important to distinguish between models of inflation which are
self consistent and those which are not.  We have shown that new
inflation is not self consistent, whereas chaotic
 inflation is.  One of the key issues involves
initial conditions.  In new inflation, the initial conditions required
can only be obtained if the inflaton field is in thermal equilibrium
above the critical temperature, which however is not possible because
of the density fluctuation constraints on coupling constants.
\par
In chaotic inflation, it can be shown that -- provided the
spatial sections are flat -- a large phase space of initial conditions
(much larger than is apparent from (5.35) and (5.37)) gives chaotic
inflation$^{121-123)}$, whereas the probability to relax dynamically$^{124)}$
to
field configurations which give new inflation (this possibility is
only available in double well potentials) is negligibly small.

\section{Generation and Evolution of Fluctuations}

\subsection{Preliminaries}
\par
In this section, the origin of the primordial density perturbations required
to seed galaxies will be discussed within the context of inflationary Universe
models.  From Chapter 3, we recall the basic reason why in inflationary
cosmology a causal generation mechanism is possible: comoving scales of
cosmological interest today originate inside the Hubble radius early in the de
Sitter period (see Fig. 11).  Hence, it is in principle possible that density
perturbations on these scales can be generated by a causal mechanism at very
early times.
\par
First, let us demonstrate why the Hubble radius $H^{-1} (t)$ is the length
scale of relevance in these considerations.  Consider a scalar field theory
with action
$$
S (\varphi) = \int d^4 x \sqrt{-g} \, \left[ {1\over 2} \partial_\mu \varphi
\partial^\mu \varphi - V (\varphi) \right] \, . \eqno\eq
$$
The resulting equation of motion is
$$
\ddot \varphi + 3 H \dot \varphi - a^{-2} \nabla^2 \varphi = - V^\prime
(\varphi) \, . \eqno\eq
$$
The second term on the left hand side is the Hubble damping term, the third
represents microphysics (spatial gradients).  To simplify the consideration,
assume that $V^\prime (\varphi) = 0$.  Then, the time evolution is influenced
by microphysics and gravity.  For plane wave perturbations with wave number
$k$, the gravitational force is proportional to $H^2 \varphi$ whereas the
microphysical force is $a^{-2} k^2 \varphi$.  Thus, for $ak^{-1} < H^{-1}$,
i.e., on length scales smaller than the Hubble radius, microphysics dominates,
whereas for $ak^{-1} > H^{-1}$, i.e., on length scales larger than the Hubble
radius, the gravitational drag dominates.
\par
Based on the above analysis we can formulate the main idea of the fluctuation
analysis in inflationary cosmology.  In linear order, all Fourier
modes
decouple.  Hence, we fix a mode with wave number $k$.  There are two very
different time intervals to consider.  Let $t_i (k)$ be the time when the
scale crosses the Hubble radius in the de Sitter phase, and $t_f (k)$ the time
when it reenters the Hubble radius after inflation (see Fig. 30).  The first
period lasts until $t = t_i (k)$.  In this time interval microphysics
dominates.  We shall demonstrate that quantum fluctuations generate
perturbations during this period$^{35-38, 125)}$.  The second time interval is
$t_i (k) < t <
t_f (k)$.  Now microphysics is unimportant and the evolution of perturbations
is determined by gravity.  At $t_i (k)$, decoherence sets in; the quantum
mechanical wave functional evolves as a statistical ensemble of
classical configurations after this time.

\midinsert \vskip 7cm
\noindent {\bf Figure 30:} Sketch (physical distance $x_p$ versus time $t$) of
the evolution of two fixed
comoving scales labelled $k_1$ and $k_2$ in the inflationary Universe.
\endinsert

It is possible to give a heuristic derivation of the shape of the spectrum of
cosmological perturbations in an inflationary  Universe based on simple
geometrical arguments$^{35)}$.  Consider first the process of generation of
fluctuations.  If the generation mechanism produces inhomogeneities at all
times on a fixed physical wavelength with an amplitude determined by the
``Hawking temperature" $T_H$ of de Sitter space$^{126)}$,
$$
T_H = {H\over{2 \pi}} \, , \eqno\eq
$$
then the evolution of perturbations on different scales between when they are
formed and when they leave the Hubble radius at times $t_i (k)$ is related by
time translation, and the amplitude of the fluctuations when measured at time
$t_i (k)$ will be independent of $k$:
$$
{\delta M\over M} (k, t_i (k)) = {\cal A}_i = \, {\rm const.} \eqno\eq
$$
In fact, the amplitude ${\cal A}_i$ will be given by the thermal energy
associated with the Hawking temperature (5.49) divided by the ``false vacuum"
energy density $\rho_0$ responsible for inflation
$$
{\cal A}_i \sim {T_H^4\over \rho_0} \, . \eqno\eq
$$

In Chapter 4 it was shown (see (4.68)) that the gauge invariant measure of
density fluctuations evolves trivially on scales outside of the Hubble radius:
it changes by a factor which only depends on the equations of state in the
inflationary and past-inflationary phases.  In particular, this factor is
independent of $k$.  Hence, using (4.68), we arrive at the conclusion that
perturbations are independent of scale when they enter the Hubble radius
$$
{\delta M\over M} (k, t_f (k)) = {\cal A}_f = \, {\rm const} \, . \eqno\eq
$$
Hence, we have demonstrated that a scale invariant spectrum of density
perturbations is a generic feature of an inflationary Universe scenario.

{}From (4.67) and (4.68) if follows that
$$
{\cal A}_f \sim {\cal A}_i \, {1 + w (t_f (k))\over{1 + w (t_i (k))}} \, .
\eqno\eq
$$
On scales which enter the Hubble radius after $t_{eq}$,
$$
1 + w (t_f (k)) = 1 \, . \eqno\eq
$$
The initial value of $w = p /\rho$ during inflation can be determined by using
the expressions (3.9) for energy density and pressure of the scalar field
responsible for inflation.
The result is
$$
1 + w (t_i (k)) = \, {\dot \varphi^2 (t_i (k))\over{\rho (t_i (k))}} \sim
{H^4\over \rho_0} \, , \eqno\eq
$$
where the last proportionality follows by dimensional analysis.  Thus,
combining (5.50-5.54) we obtain
$$
{\delta M\over M} \, (k, t_f (k)) \sim 1 \, , \eqno\eq
$$
which is at least four orders of magnitude too large to conform with the
constraints from CMB anisotropy measurements.

The above problem is known as the ``fluctuation problem"$^{127, 90)}$ and is
common to most inflationary Universe models.  The only known solutions involve
small numbers introduced into the particle physics sector.  This defeats one of
the aims of inflation which is to avoid the need for unnaturally small
constants.  To study this problem in more detail we must turn to a quantitative
analysis of the generation and evolution of fluctuations.

\subsection{Quantum Generation of Fluctuations}

The question of the origin of classical density perturbations from quantum
fluctuations in the de Sitter phase of an inflationary Universe is a rather
subtle issue.  Starting from a homogeneous quantum state ({\it e.g.}, the
vacuum state in the FRW coordinate frame at time $t_i$, the beginning of
inflation), a naive semiclassical anaylsis would predict the absence of
fluctuations since
$$
< \psi | T_{\mu\nu} (x) | \psi > = {\rm const} \, . \eqno\eq
$$

However, as a simple thought experiment shows, such a naive analysis is
inappropriate.  Imagine a local gravitational mass detector $D$ positioned
close to a large mass $M$ which is suspended from a pole (see Fig. 31).  The
decay of an alpha particle will sever the cord (at point $T$) by which the mass
is held to the pole and the mass will drop.  According to the semiclassical
prescription
$$
G_{\mu\nu} = 8 \pi G < \psi | T_{\mu\nu} | \psi > \, , \eqno\eq
$$
the metric ({\it i.e.}, the mass measured) will slowly decrease.  This is
obviously not what happens.  The mass detector shows a signal which corresponds
to one of the classical trajectories which make up the state $| \psi >$, a
trajectory corresponding to a sudden drop in the gravitational force measured.

\midinsert \vskip 4.5cm
\noindent{\bf Figure 31:} Sketch of the thought experiment discussed in the
text.
\endinsert

The origin of classical density perturbations as a consequence of quantum
fluctuations in a homogeneous state $| \psi >$ can be analyzed along similar
lines.  The quantum to classical transition is picking out$^{128-130)}$ one of
the typical classical trajectories which make up the wave function of $| \psi
>$.  We can implement$^{131, 132)}$ the procedure as follows: Define a
classical scalar field
$$
\varphi_{cl} (\underline{x} , t) = \varphi_0 (t) + \delta \varphi
(\underline{x} , t) \eqno\eq
$$
with vanishing spatial average of $\delta \varphi$.  The induced classical
energy momentum tensor $T^{cl}_{\mu\nu} (\underline{x}, t)$ which is the source
for the metric is given by
$$
T^{cl}_{\mu\nu} (\underline{x}, t) = T_{\mu\nu} (\varphi_{cl} (\underline{x},
t) ) \, , \eqno\eq
$$
where $T_{\mu\nu} \, (\varphi_{cl} (\underline{x}, t))$ is defined as the
canonical energy-momentum tensor of the classical scalar field $\varphi_{cl}
(\underline{x}, t)$.  Unless $\delta \varphi$ vanishes, $T^{cl}_{\mu\nu}$ is
inhomogeneous.

For applications to chaotic inflation, we take $| \psi >$ to be a Gaussian
state with mean value $\varphi_0 (t)$
$$
< \psi | \varphi^2 (\underline{x}, t) | \psi > = \varphi_0^0 (t) \, . \eqno\eq
$$
Its width is taken to be the width of the vacuum state of the free scalar field
theory with mass determined by the curvature of $V(\varphi)$ at $\varphi_0$.
This state is used to define the Fourier transform $\delta \tilde \varphi
(k,t)$ by
$$
| \delta \tilde \varphi (k) |^2 = < \psi | \, | \tilde \varphi (k) |^2 \, |
\psi > \, . \eqno\eq
$$
The amplitude of $\delta \tilde \varphi (k)$ is identified with the width of
the ground state wave function of the harmonic osciallator $\tilde \varphi
(k)$.  (Recall that each Fourier mode of a free scalar field is a harmonic
oscillator).  Note that no divergences arise in the above construction.  In
principle, quantum fluctuations contribute a term to $\varphi_0 (t)$, but this
term will be negligible compared to the expectation value (5.60), at least in
chaotic inflation.  The quantum corrections to (5.60) are divergent and must be
regularized and renormalized (see {\it e.g.}, Ref. 133).  They are the source
of the stochastic driving forces in stochastic chaotic inflation.

By linearizing (5.59) about $\varphi_0 (t)$ we obtain the perturbation of the
energy-momentum tensor.  In particular, the energy density fluctuation $\delta
\tilde \rho (k)$ is given by
$$
\delta \tilde \rho (k) = \dot \varphi_0 \delta \dot {\tilde \varphi} (k) +
V^\prime (\varphi_0) \delta \tilde \varphi (k) \, . \eqno\eq
$$
To obtain the initial amplitude ${\cal A}_i$ of (5.49), the above is to to
evaluated at the time $t_i (k)$.

\par
The computation of the spectrum of density perturbations produced in the de
Sitter phase has been reduced to the evaluation of the expectation value
(5.61).  First, we must specify the state $| \psi >$.  (Recall that
in non-Minkowski space-times there is no uniquely defined vacuum state of a
quantum field theory).  We pick the FRW frame of the pre-inflationary
period.  In this frame, the number density of particles decreases
exponentially.  Hence we choose $| \psi >$ to be the ground state in this
frame (see Ref. 134 for a discussion of other choices).  $\psi [ \tilde \varphi
(\undertext{k}), t]$, the wave functional of $|\psi >$, can be calculated
explicitly.  It is basically the superposition of the ground state
wave functions for all oscillators
$$
\psi [ \tilde \varphi (\undertext{k}), t] = N \exp \left\{ - {1\over 2} (2
\pi)^{-3} a^3 (t) \int d^3 k \omega (\undertext{k}, t) | \tilde \varphi
(\undertext{k})|^2 \right\} \, . \eqno\eq
$$
$N$ is a normalization constant and $\omega (\undertext{k}, t) \sim H$ at $t =
t_i (k)$.  Hence
$$
\delta \tilde \varphi (\undertext{k}, t) = (2 \pi)^{3/2}  a^{-3/2}
\omega (\undertext{k}, t)^{-1/2} \sim (2 \pi)^{3/2}  k^{-3/2} H \, , \,
t = t_i (k)\, . \eqno\eq
$$

\subsection{Evolution of Fluctuations}

Given the above determination of the intitial amplitude of density
perturbations at the time when they leave the Hubble radius during the de
Sitter phase, and the general relativistic analysis of the evolution of
fluctuations discussed in Section 4.4, it is easy to evaluate the r.m.s.
inhomogeneities when they reenter the Hubble radius at time $t_f (k)$.

First, we combine (5.62), (5.64), (5.49), (4.10) and (4.15) to obtain
$$
\left( {\delta M\over M} \right)^2 \, (k, t_i (k)) \sim k^3 \left({V^\prime
(\varphi_0) \delta \tilde \varphi (k, t_i (k))\over \rho_0} \right)^2 \sim
\left({V^\prime (\varphi_0) H\over \rho_0 } \right)^2 \, , \eqno\eq
$$
and thus
$$
{\cal A}_i \sim \, {V^\prime (\varphi_0 (t_i (k)) H\over \rho_0} \, . \eqno\eq
$$
If the background scalar field is rolling slowly, then
$$
V^\prime (\varphi_0 (t_i (k))) =  3 H | \dot \varphi_0 (t_i (k)) | \, .
\eqno\eq
$$
Combining (5.66) and (5.67) with (5.52) and (5.54) we get
$$
{\delta M\over M} (k, \, t_f (k)) = {\cal A}_f \sim \, {3 H^2 | \dot \varphi_0
(t_i (k)) |\over{\dot \varphi^2_0 (t_i (k))}} = {3 H^2\over{| \dot \varphi_0
(t_i (k))}} = {3H^2\over{| \dot \varphi_0 (t_i (k))|}} \eqno\eq
$$
This result can now be evaluated for specific models of inflation to find the
conditions on the particle physics parameters which give a value
$$
{\cal A}_f \sim 10^{-5} \eqno\eq
$$
which is required if quantum fluctuations from inflation are to provide the
seeds for galaxy formation and agree with the CMB anisotropy limits.

For chaotic inflation with a potential
$$
V (\varphi) = {1\over 2} m^2 \varphi^2 \, , \eqno\eq
$$
the dynamics of $\varphi$ was analyzed in Section 5b (see in particular (5.43)
and (5.44)).  We have
$$
\varphi (t_i (k)) \sim m_{pl} \eqno\eq
$$
and hence
$$
H (t_i (k)) \sim m^{-1}_{pl} m \, \varphi (t_i (k)) \sim m \, . \eqno\eq
$$
Therefore,
$$
{\delta M\over M} (k, t_f (k)) \sim 3 {H^2\over{| \dot \varphi_0 (t_i (k))|}}
\sim 10 {m\over m_{pl}} \eqno\eq
$$
which implies that
$$
m \sim 10^{13} \, {\rm GeV} \eqno\eq
$$
to agree with (5.70).

Similarly, for a potential of the form
$$
V (\varphi) = {1\over 4} \lambda \varphi^4 \eqno\eq
$$
we obtain
$$
{\delta M\over M} (k, \, t_f (k)) \sim  10 \cdot \lambda^{1/2} \eqno\eq
$$
which requires
$$
\lambda \leq 10^{-12}. \eqno\eq
$$
in order not to conflict with observations.

The conditions (5.74) and (5.77) require the presence of small parameters in
the particle physics model.  It  has been shown quite generally$^{120)}$ that
small parameters are required if inflation is to solve the fluctuation problem.

\subsection{Discussion}
\par
Let us first summarize the main results of the analysis of density
fluctuations in inflationary cosmology:
\item{-} Quantum vacuum fluctuations in the de Sitter phase of an inflationary
Universe are the source of perturbations.
\item{-} The quantum perturbations decohere on scales outside the Hubble
radius and can hence be treated classically.
\item{-} The classical evolution outside the Hubble radius produces a large
amplification of the perturbations.  In fact, unless the particle physics
model contains very small coupling constants, the predicted fluctuations are
in excess of those allowed by the bounds on cosmic microwave anisotropies.
\item{-} Inflationary Universe models generically produce a scale invariant
\hfill \break Harrison-Zel'dovich spectrum
$$
{\delta M\over M} (k , t_f (k) ) = {\rm const.} \eqno\eq
$$

It is not hard to construct models which give a different spectrum.  All that
is required is a significant change in $H$ during the period of
inflation.
\par
I have chosen to present the analysis of fluctuations in inflationary cosmology
in two separate steps in order to highlight the crucial physics issues.
Having done this, it is possible to step back and construct a unified
analysis in which expectation values of gauge invariant variables
are propagated from $t \ll t_i (k)$ to $t_f (k)$ in a consistent way$^{135,
11)}$, and in
which the final values of the expectation values of quadratic operators
are used to construct $T^{cl}_{\mu \nu} (\undertext{x} , t)$.

Once inside the Hubble radius, the evolution of the mass perturbations is
influenced by the damping effects discussed in Section 4.3, which is turn
depend on the composition of the dark matter. The dominant effects are the
Meszaros effect and free streaming.

On scales which enter the Hubble radius before $t_{eq}$, the perturbations can
only grow logarithmically in time between $t_f(k)$ and $t_{eq}$. This implies
that (up to logarithmic corrections), the mass perturbation spectrum is flat
for wavelengths smaller than $\lambda_{eq}$, the comoving Hubble radius at
$t_{eq}$:
$$
{{\delta M} \over M} (\lambda, t) \simeq {\rm const}, \,\,\,\,\,\,\,
t \leq t_{eq},  \, \lambda < \lambda_{eq}, \eqno\eq
$$
whereas on larger scales
$$
{{\delta M} \over M} (\lambda, t) \propto \lambda^{-2}. \eqno\eq
$$
Equations (5.78) and (5.79) give the power spectrum in an $\Omega = 1$
inflationary CDM model.

If the dark matter is hot, then neutrino free streaming cuts off the power
spectrum at $\lambda_J^{max}$ (see (3.27)). The inflationary HDM and CDM
perturbation spectra are compared in Fig. 32.

\midinsert \vskip 9.5cm
\noindent{\bf Figure 32:} The linear theory power spectra for inflationary CDM
(upper curve) and HDM models. The horizontal axis is mass, expressed in units
of solar masses. $M_{eq}$ is the mass inside the comoving horizon at $t_{eq}$,
and $M_{FS}$ is the mass inside the maximal comoving neutrino free streaming
volume.
\endinsert
\chapter{Topological Defects and Structure Formation}
\section{Classification}
\par
In the previous section we have seen that symmetry breaking phase
transitions in unified field theories arising in particle physics
({\it e.g.,} Grand Unified Theories (GUT)$^{46)}$) do not lead, in general, to
inflation.  In most models, the coupling constants which arise in the
effective potential for the scalar field $\varphi$ driving the phase
transition are too large to generate a period of slow rolling which
lasts more than one Hubble time $H^{-1} (t)$.  Nevertheless, there are
interesting remnants of the phase transition: topological
defects.
\par
Consider a single component real scalar field with a typical symmetry breaking
potential
$$
V (\varphi) = {1\over 4} \lambda (\varphi^2 - \eta^2)^2 \eqno\eq
$$
Unless $\lambda \ll 1$ there
will be no inflation.  The phase transition will take place on a short time
scale $\tau < H^{-1}$, and will lead to correlation regions of radius $\xi <
t$ inside of which $\varphi$ is approximately constant, but outside of which
$\varphi$ ranges randomly over the vacuum manifold ${\cal M}$, the set of
values
of $\varphi$ which minimizes $V(\varphi)$ -- in our example $\varphi
= \pm \eta$.  The correlation regions are separated by domain walls, regions in
space where $\varphi$ leaves the vacuum manifold ${\cal M}$ and where,
therefore, potential energy is localized.  Via the usual gravitational
force, this energy density can act as a
seed for structure.

Topological defects are familiar from solid state and condensed matter
systems.  Crystal defects, for example, form when water freezes or
when a metal crystallizes$^{136)}$.  Point defects, line defects and planar
defects are possible.  Defects are also common in liquid crystals.
They arise in a temperature quench from the disordered to the ordered
phase$^{137)}$.  Vortices in $^4$He are analogs of global cosmic strings.
Vortices and other defects are also produced during a quench below the
critical temperature in $^3$He$^{138)}$.  Finally, vortex lines also play an
important role in the theory of superconductivity$^{139)}$.

The analogies between defects in particle physics and condensed matter
physics are quite deep.  Defects form for the same reason: the vacuum
manifold is topologically nontrivial.  The arguments which say that in
a theory which admits defects, such defects will inevitably form, are
applicable both in cosmology and in condensed matter physics.
Different, however, is the defect dynamics.  The motion of defects in
condensed matter systems is friction-dominated, whereas the defects in
cosmology obey relativistic equations, second order in time
derivatives, since they come from a relativistic field theory.

After these general comments we turn to a classification of
topological defects.  We consider theories with an $n$-component order
parameter $\varphi$ and with a potential energy function (free energy
density) of the form (6.1) with
$$
\varphi^2 = \sum\limits^n_{i = 1} \, \varphi^2_i \, . \eqno\eq
$$

There are various types of local and global topological defects
(regions of trapped energy density) depending on the number $n$ of components
of
$\varphi$.  The more rigorous mathematical definition refers to the homotopy
of ${\cal M}$.  The words ``local" and ``global" refer to whether the symmetry
which is broken is a gauge or global symmetry.  In the case of local
symmetries, the topological defects have a well defined core outside of which
$\varphi$ contains no energy density in spite of nonvanishing gradients
$\nabla \varphi$:  the gauge fields $A_\mu$ can absorb the gradient,
{\it i.e.,} $D_\mu \varphi = 0$ when $\partial_\mu \varphi \neq 0$,
where the covariant derivative $D_\mu$ is defined by
$$
D_\mu = \partial_\mu + ie \, A_\mu \, , \eqno\eq
$$
$e$ being the gauge coupling constant.
Global topological defects, however, have long range density fields and
forces.
\par
Table 1 contains a list of topological defects with their topological
characteristic.  A ``v" markes acceptable theories, a ``x" theories which are
in conflict with observations (for $\eta \sim 10^{16}$ GeV).

\midinsert \vskip 5cm
\endinsert

\par
Theories with domain walls are ruled out$^{140)}$ since a single domain wall
stretching
across the Universe today would overclose the Universe.  Local monopoles are
also ruled out$^{141)}$ since they would overclose the Universe.  Local
textures are ineffective at producing structures (see Section 6).

We now describe examples of domain walls, cosmic strings, monopoles and
textures, focussing on configurations with maximal symmetry.

\par
\undertext{Domain walls} arise in theories with a single real order
parameter and free energy density given by (6.1).  The vacuum manifold
of this model consists of two points
$$
{\cal M} = \{ \pm \eta \} \eqno\eq
$$
and hence has nontrivial zeroth homotopy group:
$$
\Pi_0 ({\cal M}) \neq 1 \eqno\eq
$$
(readers not familiar with homotopy groups can simply skip all of the
following statements involving $\Pi_n ({\cal M})$.  They are not
required for an understanding of the physics).

To construct a domain wall configuration with planar symmetry (without
loss of generality the $y-z$ plane can be taken to be the plane of
symmetry), assume that
$$
\eqalign{
\varphi (x) \simeq \eta \>\>\> & x \gg \eta^{-1} \cr
\varphi (x) \simeq - \eta \>\>\> & x \ll - \eta^{-1} } \eqno\eq
$$
By continuity of $\varphi$, there must be an intermediate value of $x$
with $\varphi (x) = 0$.  We can take this point to be $x = 0$, {\it
i.e.},
$$
\varphi (0) = 0 \, . \eqno\eq
$$
The set of points with $\varphi = 0$ constitute the center of the
domain wall.  Physically, the wall is a thin sheet of trapped energy
density.  The width $w$ of the sheet is given by the balance of potential
energy and tension energy.  Assuming that the spatial gradients are
spread out over the thickness $w$ we obtain
$$
w V (0) = w \lambda \eta^4 \sim {1\over w} \, \eta^2 \eqno\eq
$$
and thus
$$
w \sim \lambda^{-1/2} \eta^{-1} \, . \eqno\eq
$$
See Fig. 33 for a sketch of this domain wall configuration.

\midinsert \vskip 4.5cm
\noindent{\bf Figure 33:} A low temperature field configuration containing a
domain wall. The sketch shows a plane $P$ in space. At positions with a $+$ or
$-$, the value of the scalar field is $\eta$ or $- \eta$, respectively. The
solid line is the intersection with $P$ of the plane of points in space with
$\varphi = 0$.
\endinsert

A theory with a complex order parameter $(n = 2)$ admits
\undertext{cosmic strings}.  In this case the vacuum manifold of the
model is
$$
{\cal M} = S^1 \, , \eqno\eq
$$
which has nonvanishing first homotopy group:
$$
\Pi_1 ({\cal M}) = Z \neq 1 \, . \eqno\eq
$$
A cosmic string is a line of trapped energy density which arises
whenever the field $\varphi (x)$ circles ${\cal M}$ along a closed path
in space ({\it e.g.}, along a circle).  In this case, continuity of
$\varphi$ implies that there must be a point with $\varphi = 0$ on any
sheet bounded by the closed path.  The points on different sheets
connect up to form a line overdensity of field energy (see Fig. 34).

\midinsert \vskip 4cm
\noindent{\bf Figure 34:} Sketch of the topological argument for the existence
of cosmic string configurations. Given a field configuration with nontrivial
winding along a circle normal to the plane of this figure, there must be a
point with $\varphi = 0$ on every disk bounded by the circle. Three disks are
depicted: $D$,  $D'$ and $D''$, and the respective points with $\varphi = 0$
are $z$, $z'$ and $z''$. The union of all such points $z$ forms the center
$z(\lambda)$ of the string.
\endinsert

To construct a field configuration with a string along the $z$ axis$^{43)}$,
take $\varphi (x)$ to cover ${\cal M}$ along a circle with radius $r$
about the point $(x,y) = (0,0)$:
$$
\varphi (r, \vartheta ) \simeq \eta e^{i \vartheta} \, , \, r \gg
\eta^{-1} \, . \eqno\eq
$$
This configuration has winding number 1, {\it i.e.}, it covers ${\cal
M}$ exactly once.  Maintaining cylindrical symmetry, we can extend
(6.12) to arbitrary $r$
$$
\varphi (r, \, \vartheta) = f (r) e^{i \vartheta} \, , \eqno\eq
$$
where $f (0) = 0$ and $f (r)$ tends to $\eta$ for large $r$.  The
width $w$ can again be found by balancing potential and tension
energy.  The result is identical to the result (6.9) for domain walls.

For local cosmic strings, {\it i.e.}, strings arising due to the
spontaneous breaking of a gauge symmetry, the energy density decays
exponentially for $r \gg \eta^{-1}$.  In this case, the energy $\mu$
per unit length of a string is finite and depends only on the symmetry
breaking scale $\eta$
$$
\mu \sim \eta^2 \eqno\eq
$$
(independent of the coupling $\lambda$).  The value of $\mu$ is the
only free parameter in a cosmic string model.

To see how the finiteness of the mass per unit length $\mu$ comes
about, consider the simplest theory admitting local strings, the
Abelian Higgs model with Lagrangean
$$
{\cal L} = {1\over 2} \, D_\mu \varphi D^\mu \varphi - V (\varphi) +
{1\over 4} F_{\mu\nu} F^{\mu\nu} \, , \eqno\eq
$$
where $\varphi$ is a complex order parameter with potential (6.1),
$D_\mu$ is the gauge covariant derivative
$$
D_\mu = \partial_\mu + ie \, A_\mu \, , \eqno\eq
$$
the field $A_\mu$ is a U(1) gauge potential with associated field
strength
$$
F_{\mu\nu} = \partial_\mu A_\nu - \partial_\nu A_\mu \, , \eqno\eq
$$
and $e$ is the gauge coupling constant.

For an order parameter configuration (6.12), the gauge fields $A_\mu$
will take on values such that
$$
D_\mu \varphi \simeq 0 \>\>\> r \gg \eta^{-1} \eqno\eq
$$
even though $\partial_\mu \varphi \neq 0$.  Hence, the energy density
decays exponentially for $r \gg \eta^{-1}$.  For strings in a global
theory (no gauge potential), the spatial gradient energy
$(\partial_\mu \varphi)^2$ cannot be cancelled at large $r$, and hence
the mass per unit length is logarithmically divergent as a function of
a large $r$ cutoff.

If the order parameter of the model has three components $(n = 3)$,
then \undertext{monopoles} result as topological defects.  The vacuum
manifold is
$$
{\cal M} = S^2 \eqno\eq
$$
and has topology given by
$$
\Pi_2 ({\cal M}) \neq 1 \, . \eqno\eq
$$
Given a sphere $S$ is space, it is possible that $\varphi$ takes on
values in ${\cal M}$ everywhere on $S$, and that it covers ${\cal M}$
once.  By continuity, there must be a point in space in the interior
of $S$ with $\varphi = 0$.  This is the center of a point-like defect,
the monopole.

To construct a spherically symmetric monopole with the origin as its
center, consider a field configuration $\varphi$ which defines a map
from physical space to field space such that all spheres $S_r$ in
space of radius $r \gg \eta^{-1}$ about the origin are mapped onto
${\cal M}$ (see Fig. 35):
$$
\eqalign{
\varphi: \> & S_r \longrightarrow {\cal M} \cr
& (r, \vartheta,\varphi)  \longrightarrow (\vartheta,
\varphi)  \, . }\eqno\eq
$$
This configuration defines a winding number one magnitude monopole.

Domain walls, cosmic strings and monopoles are examples of
\undertext{topological} \undertext{defects}.  A topological defect has a
well-defined
core, a region in space with $\varphi \notin {\cal M}$ and hence $V
(\varphi) > 0$.  There is an associated winding number which is
quantized, {\it i.e.}, it can take on only integer values.  Since the
winding number can only change continuously, it must be conserved, and
hence topological defects are stable.  Furthermore, topological
defects exist for theories with global and local symmetries.

Now, let us consider a theory with a four-component order parameter
({\it i.e.,} $n = 4$), and a potential given by (6.1).  In this case,
the vacuum manifold is
$$
{\cal M} = S^3 \eqno\eq
$$
and the associated topology is given by
$$
\Pi_3 ({\cal M}) \neq 1 \, . \eqno\eq
$$
The corresponding defects are called ``\undertext{textures}".$^{44, 45)}$

\midinsert \vskip 8.5cm
\noindent {\bf Figure 35:} Construction of a monopole: left is
physical space, right the vacuum manifold.  The field configuration
$\phi$ maps spheres in space onto ${\cal M}$.  However, a core region
of space near the origin is mapped onto field values not in ${\cal
M}$.
\endinsert

Textures, however, are quite different than the previous topological defects.
The texture construction will render this manifest (Fig. 36).  To construct a
radially symmetric texture, we give a field configuration $\varphi (x)$ which
maps physical space onto ${\cal M}$.  The origin 0 in space (an arbitrary point
which will be the center of the texture) is mapped onto the north pole $N$ of
${\cal
M}$.  Spheres surrounding 0 are mapped onto spheres surrounding $N$.  In
particular, some sphere with radius $r_c (t)$ is mapped onto the equator
sphere of ${\cal M}$.  The distance $r_c (t)$ can be defined as the radius of
the texture.  Inside this sphere, $\varphi (x)$ covers half the vacuum
manifold.
Finally, the sphere at infinity is mapped onto the south pole of ${\cal M}$.
The configuration $\varphi (\undertext{x})$ can be parameterized
by$^{60)}$
$$
\varphi (x,y,z) = \left(\cos \chi (r), \> \sin \chi (r) {x\over r}, \>
\sin \chi (r) {y\over r}, \> \sin \chi (r) {z\over r} \right) \eqno\eq
$$
in terms of a function $\chi (r)$ with $\chi (0) = 0$ and $\chi (\infty) =
\pi$.  Note that at all points in space, $\varphi (\undertext{x})$ lies in
${\cal
M}$.  There is no defect core.  All the energy is spatial gradient (and
possibly kinetic) terms.
\par
In a cosmological context, there is infinite energy available in an infinite
space.  Hence, it is not necessary that $\chi (r) \rightarrow \pi$ as $r
\rightarrow \infty$.  We can have
$$
\chi (r) \rightarrow \chi_{\rm max} < \pi \>\> {\rm as} \>\> r \rightarrow
\infty \, . \eqno\eq
$$

\midinsert \vskip 10cm
\noindent{\bf Figure 36:} Construction of a global texture: left is
physical space, right the vacuum manifold. The field configuration
$\phi$ is a map from space to the vacuum manifold (see text).
\endinsert

In this case, only a fraction
$$
n_W = {\chi_{\rm max}\over \pi} - {\sin (2 \chi_{\rm max})\over{2 \pi}}
\eqno\eq
$$
of the vacuum manifold is covered:  the winding number $n_W$ is not quantized.
This is a reflection of the fact that whereas topologically nontrivial maps
from $S^3$ to $S^3$ exist, all maps from $R^3$ to $S^3$ can be deformed to
the trivial map.
\par
Textures in $R^3$ are unstable.  For the configuration described above, the
instability means that $r_c (t) \rightarrow 0$ as $t$ increases: the texture
collapses.  When $r_c (t)$ is microscopical, there will be sufficient energy
inside the core to cause $\varphi (0)$ to leave ${\cal M}$, pass through 0 and
equilibrate at $\chi (0) = \pi$: the texture unwinds.
\par
A further difference compared to topological defects: textures are relevant
only for theories with global symmetry.  Since all the energy is in spatial
gradients, for a local theory the gauge fields can reorient themselves such as
to cancel the energy:
$$
D_\mu \varphi = 0 \, . \eqno\eq
$$
\par
Therefore, it is reasonable to regard textures as an example of a new class of
defects, \undertext{semitopological defects}.  In contrast to topological
defects, there is no core, and $\varphi (\undertext{x}) \in {\cal M}$ for all
$\undertext{x}$.  In particular, there is no potential energy.  In addition,
the
winding number is not quantized, and hence the defects are unstable.  Finally,
they exist only in theories with a global internal symmetry.

\section{Formation}
\par
The Kibble mechanism$^{12)}$ ensures that in theories which admit
topological or semitopological defects, such defects will be produced
during a phase transition in the very early Universe.

Consider a mechanical toy model, first introduced by Mazenko, Unruh
and Wald$^{93)}$ in the context of inflationary Universe models, which
is useful in understanding the scalar field evolution.  Consider (see
Fig. 37) a lattice of points on a flat table.  At each point, a pencil
is pivoted.  It is free to rotate and oscillate.  The tips of nearest
neighbor pencils are connected with springs (to mimic the spatial
gradient terms in the scalar field Lagrangean).  Newtonian gravity
creates a potential energy $V(\varphi)$ for each pencil ($\varphi$ is
the angle relative to the vertical direction).  $V(\varphi)$ is
minimized for $| \varphi | = \eta$ (in our toy model $\eta = \pi /
2$).  Hence, the Lagrangean of this pencil model is analogous to that
of a scalar field with symmetry breaking potential (6.1).

\midinsert \vskip 5.5cm
\noindent{\bf Figure 37:} The pencil model: the potential energy of a
simple pencil has the same form as that of scalar fields used for
spontaneous symmetry breaking.  The springs connecting nearest
neighbor pencils give rise to contributions to the energy which mimic
spatial gradient terms in field theory.
\endinsert

At high temperatures $T \gg T_c$, all pencils undergo large amplitude
high frequency oscillations.  However, by causality, the phases of
oscillation of pencils with large separation $s$ are uncorrelated.
For a system in thermal equilibrium, the length $s$ beyond which
phases are random is the correlation length $\xi (t)$.  However, since
the system is quenched rapidly, there is a causality bound on
$\xi$:
$$
\xi (t) < t \, , \eqno\eq
$$
where $t$ is the causal horizon.

The critical temperature $T_c$ is the temperature at which the
thermal energy is equal to the energy a pencil needs to jump from
horizontal to vertical position.  For $T < T_c$, all pencils want to
lie flat on the table.  However, their orientations are random beyond
a distance of $\xi (t)$ determined by equating the free energy gained by
symmetry breaking (a volume effect) with the gradient energy lost (a surface
effect).  As expected, $\xi (T)$ diverges at $T_c$. Very close to $T_c$, the
thermal energy $T$ is larger than the volume energy gain $E_{corr}$ in a
correlation volume. Hence, these domains are unstable to thermal fluctuations.
As $T$ decreases, the thermal energy decreases more rapidly than $E_{corr}$.
Below the Ginsburg temperature $T_G$, there
is insufficient thermal energy to excite a correlation volume into the
state $\varphi = 0$.  Domains of size
$$
\xi (t_G) \sim \lambda^{-1} \eta^{-1} \eqno\eq
$$
freeze out$^{12, 142)}$.  The boundaries between these domains become
topological defects.

We conclude that in a theory in which a symmetry breaking phase
transitions satisfies the topological criteria for the existence of a
fixed type of defect, a network of such defects will form during the
phase transition and will freeze out at the Ginsburg temperature.  The
correlation length is initially given by (6.29), if the field
$\varphi$ is in thermal equilibrium before the transition.
Independent of this last assumption, the causality bound implies that
$\xi (t_G) < t_G$.

For times $t > t_G$ the evolution of the network of defects may be
complicated (as for cosmic strings) or trivial (as for textures).  In
any case (see the caveats of Refs. 143 and 144), the causality bound
persists at late times and states that even at late times, the mean
separation and length scale of defects is bounded by $\xi (t) \leq t$.

Applied to cosmic strings, the Kibble mechanism implies that at the
time of the phase transition, a network of cosmic strings with typical
step length $\xi (t_G)$ will form.  According to numerical
simulations$^{145)}$, about 80\% of the initial energy is in infinite
strings and 20\% in closed loops.

Note that the Kibble mechanism was discussed above in the context of a
global symmetry breaking scenario.  As pointed out in Ref. 146, the
situation is more complicated in local theories in which gauge field
can cancel spatial gradients in $\varphi$ in the energy functional,
and in which spatial gradients in $\varphi$ can be gauged away.
Nevertheless, as demonstrated numerically (in $2 + 1$ dimensions) in
Refs. 42 and 147 and shown analytically in Ref. 148, the Kibble mechanism also
applies to local symmetries.

\midinsert \vskip 3.5cm
\noindent{\bf Figure 38:} Formation of a loop by a self intersection of an
infinite string. According to the original cosmic string scenario, loops form
with a radius $R$ determined by the instantaneous coherence length of the
infinite string network.
\endinsert

The evolution of the cosmic string network for $t > t_G$ is
complicated (see Section 6.4).  The key processes are loop production
by intersections of infinite strings (see Fig. 38) and loop shrinking
by gravitational radiation.  These two processes combine to create a
mechanism by which the infinite string network loses energy (and
length as measured in comoving coordinates).  It will be shown that as
a consequence, the correlation length of the string network is always
proportional to its causality limit
$$
\xi (t) \sim t \, . \eqno\eq
$$
Hence, the energy density $\rho_\infty (t)$ in long strings is a fixed
fraction of the background energy density $\rho_c (t)$
$$
\rho_\infty (t) \sim \mu \xi (t)^{-2} \sim \mu t^{-2} \eqno\eq
$$
or
$$
{\rho_\infty (t)\over{\rho_c (t)}} \sim G \mu \, . \eqno\eq
$$

We conclude that the cosmic string network approaches a ``scaling
solution" in which the statistical properties of the
network are time independent if all distances are scaled to the
horizon distance.

Applied to textures, the Kibble mechanism implies that on all
scales $r \geq t_G$, field configurations with winding number $n_W
\geq n_{cr}$ are frozen in with a probability $p (n_{cr})$ per volume
$r^3$.  The critical winding number $n_{cr}$ is defined as the winding
number above which field configurations collapse and below which they
expand.  Only collapsing configurations form clumps of energy which
can accrete matter.

The critical winding $n_{cr}$ was determined numerically in Refs. 149
\& 150 and analytically in Ref. 151 (see also Ref. 152).  It is
slightly larger than 0.5.  The probability $p (n_{cr})$ can be
determined using combinatorial arguments$^{153)}$.

For $t > t_G$, any configuration on scale $\sim t$ with winding number
$n_W \ge n_{cr}$ begins to collapse (before $t$, the Hubble damping
term dominates over the spatial gradient forces, and the field
configuration is frozen in comoving coordinates).  After unwinding,
$\varphi (\undertext{x})$ is homogeneous inside the horizon.

The texture model thus also leads to a scaling solution: at all times
$t > t_G$ there is the same probability that a texture configuration
of scale $t$ will enter the horizon, become dynamical and collapse
with a typical time scale $t$.

\section{Topological Defects and Cosmology}

Topological defects are regions in space with trapped energy density.
By Newtonian gravity, these defects can act as seeds about which the
matter in the Universe clusters, and hence they play a very important
role in cosmology.

As indicated in Table 1, theories with domain walls or with local
monopoles are ruled out, and those with only local textures do not give
rise to a structure formation model.  As mentioned earlier, theories with
domain walls are
ruled out since a single wall stretching across the present Universe
would overclose it.  Local monopoles are alos problematic since they
do not interact and come to dominate the energy density of the
Universe.  Local textures do not exist as coherent structures with
nonvanishing gradient energy since the gauge fields can always
compensate scalar field gradients.
\par
Let us demonstrate explicitly why stable domain walls are a
cosmological disaster$^{140)}$.  If domain walls form during a phase transition
in the early Universe, it follows by causality (see however the caveats
of Refs. 143 and 144) that even today there will be at least one wall
per Hubble volume.  Assuming one wall per Hubble volume, the energy
density $\rho_{DW}$ of matter in domain walls is
$$
\rho_{DW} (t) \sim \eta^3 t^{-1} \, , \eqno\eq
$$
whereas the critical density $\rho_c$ is
$$
\rho_c = H^2 \, {3\over{8 \pi G}} \sim m^2_{p\ell} \, t^{-2} \, .
\eqno\eq
$$
Hence, for $\eta \sim 10^{16}$ GeV the ratio of (6.33) and (6.34) is
$$
{\rho_{DW}\over \rho_c} \, (t) \sim \, \left({\eta\over{m_{p\ell}}}
\right)^2 \, (\eta t) \sim 10^{52} \, . \eqno\eq
$$

The above argument depends in an essential way on the dimension of the
defect.  One cosmic string per Hubble volume leads to an energy
density $\rho_{cs}$ in string
$$
\rho_{cs} \sim \eta^2 \, t^{-2} \, . \eqno\eq
$$
Later in this section we shall see that the scaling (6.36) holds in the
cosmic string model.  Hence, cosmic strings do not lead to
cosmological problems.  On the contrary, since for GUT models with
$\eta \sim 10^{16}$ GeV
$$
{\rho_{cs}\over \rho_c} \sim \, \left({\eta\over m_{p \ell}} \right)^2
\sim 10^{-6} \, , \eqno\eq
$$
cosmic strings in such theories could provide the seed perturbations
responsible for structure formation.

Theories with local monopoles are ruled out on cosmological
grounds$^{141)}$ (see again the caveats of Refs. 143 and 144) for
rather different reasons.  Since there are no long range forces
between local monopoles, their number density in comoving coordinates
does not decrease.  Since their contribution to the energy density
scales as $a^{-3} (t)$, they will come to dominate the mass of the
Universe, provided $\eta$ is sufficiently large.

Theories with global monopoles$^{154, 155)}$ are not ruled out, since
there are long range forces between monopoles which lead to a
``scaling solution" with a fixed number of monopoles per Hubble
volume.

In the following we will describe aspects of two of the promising
topological defect models of structure formation, those based on
cosmic strings and on global textures.  The global monopole scenario is in many
aspects similar to the texture theory.

\section{Cosmic String Evolution and Scaling}

If the evolution of the cosmic string network were trivial in the sense that
all strings would only stretch as the universe expands, there would be an
immediate cosmological disaster.  Consider a fixed comoving volume $V$ with a
string passing through.  The energy in radiation decreases as $a^{-1} (t)$
while the energy in string increases as $a (t)$.  Hence trivial evolution
would immediately lead to a string dominated universe, a cosmological
disaster. In order to study the evolution of a cosmic string network, it is
neccessary to know the effective action for a string, and to study what happens
when two strings cross.
\par
The equations of motion of a string are determined by the Nambu action
$$
S = - \mu \int d \sigma d \tau \left(- \det g^{(2)}_{ab} \right)^{1/2} \> \>
a, b = 0, 1\eqno\eq
$$
where $g^{(2)}_{ab}$ is the world sheet metric and $\sigma$ and
$\tau$ are the world sheet coordinates.  In flat space-time, $\tau$
can be taken to be coordinate time, and $\sigma$ is an affine
parameter along the string.  In terms of the string
coordinates $X^\mu (\sigma, \tau)$ and the metric $g^{(4)}_{\mu\nu}$ of
the background space-time,
$$
g^{(2)}_{ab} = X^\mu_{,a} X^v_{,b} g^{(4)}_{\mu\nu} \, . \eqno\eq
$$
{}From
general symmetry considerations,  it is possible to argue that the
Nambu action is the correct action.  However, I shall follow
Foerster$^{156)}$ and
Turok$^{157)}$ and give a direct heuristic derivation.  We start from a
general quantum field theory Lagrangean ${\cal L}_{QFT}$.  The action is
$$
S = \int d^4 y {\cal L}_{QFT} \left(\varphi (y)\right)\eqno\eq
$$
We assume the existence of a linear topological defect at $X^\mu (\sigma,
\tau)$.  The idea now is to change variables such that $\sigma$ and $\tau$ are
two of the new coordinates, and to expand $S$ to lowest order in $w/R$, where
$w$ is the width of the string and $R$ its curvature radius.  As the other new
coordinates we take the coordinates $\rho^2$ and $\rho^3$ in the normal plane
to
$X^\mu (\sigma, \tau)$.  Thus the transformation takes the
old coordinates $y^\mu (\mu = 0, 1, 2, 3)$ to new ones $\sigma^a = (\tau,
\sigma, \rho^2, \rho^3)$:
$$
y^\mu (\sigma^a) = X^\mu (\sigma, \tau) + \rho^i n^\mu_i (\sigma,
\tau)\eqno\eq
$$
where $i = 2,3$ and $n^\mu_i$ are the basis vectors in the normal plane to the
string world sheet.  The measure transforms as
$$
\int d^4 y = \int d \sigma d \tau d \rho^2 d \rho^3 (\det M_a^\mu)\eqno\eq
$$
with
$$
M^\mu_a = \, {\partial y^\mu\over{\partial \sigma^a}} = \, \pmatrix{\partial
X^\mu/\partial (\sigma, \tau)\cr
n^\mu_i\cr} + O (\rho)\, .\eqno\eq
$$
The determinant can easily be evaluated using the following trick
$$
\det M^\mu_a = \, \left( - \det \eta_{\mu \nu} M^\mu_a M^\nu_b \right)^{1/2}
\equiv \sqrt{- \det D_{ab}}\eqno\eq
$$
$$
D = \, \pmatrix{{\partial x^\mu\over{\partial (\sigma, \tau)}} \, {\partial
X^\nu\over{\partial (\sigma, \tau)}} \eta_{\mu \nu} & {\partial X^\mu\over
{\partial (\sigma , \tau)}} n^\nu_b \eta_{\mu\nu}\cr
{\partial X^\mu\over{\partial (\sigma, \tau)}} n^\nu_a \eta_{\mu\nu} & n^\mu_a
n^\nu_b \eta_{\mu\nu}\cr} = \pmatrix{X^\mu_{,a} X^\nu_{,b} \eta_{\mu\nu} & 0\cr
0 & \delta_{ab}\cr} + 0 \, \left({w\over R}\right) \eqno\eq
$$
Hence
$$
\eqalign{S &= \int d \sigma d \tau \left( - \det g^{(2)}_{ab} \right)^{1/2}
\int d \rho^2 d \rho^3 {\cal L} (y (\sigma, \tau, \rho^2, \rho^3)) + O
\left({w\over R} \right)\cr
&= - \mu \int d \sigma d \tau \left( - \det g^{(2)}_{ab} \right)^{1/2} + O
\left({w\over R}\right)\, . }\eqno\eq
$$
$- \mu$ is the integral of {\cal L} in the normal plane of $X$.  To first
order in $w/R$, it equals the integral of $-{\cal H}$; hence it is the mass per
unit length.
\par
This derivation of the Nambu action is instructive as it indicates a method
for calculating corrections to the equations of motion of the string when
extra fields are present, \eg\ for superconducting cosmic strings.  It also
gives a way of calculating the finite thickness corrections to the equations
of motion which will be important at cusps (see below).
\par
In flat space-time we can consistently choose $\tau = t, \dot x \cdot x^\prime
= 0$ and $\dot x^2 + x^{\prime^2} = 0$.  The equations of motion derived from
the Nambu action then become
$$
\ddot {\undertext{x}} - \undertext{x}^{\prime\prime} = 0\, . \eqno\eq
$$
where $\prime$ indicates the derivative with respect to $\sigma$.  The general
solution can be decomposed into a left moving and a right moving mode$^{158)}$
$$
\undertext{x} (t, \sigma) = {1\over 2} \, \left[ \undertext{a} (\sigma - t) +
\undertext{b} (\sigma + t ) \right] \eqno\eq
$$
The gauge conditions imply
$$
\dot {\undertext{a}}^2 = \dot {\undertext{b}}^2 = 1\eqno\eq
$$
For a loop, $\undertext{x} (\sigma, t)$ is periodic and hence the time average
of $\dot {\undertext{a}}$ and $\dot {\undertext{b}}$ vanish.  $\dot
{\undertext{a}}$
and $\dot {\undertext{b}}$ are hence closed curves on the unit sphere with
vanishing
average.  Two such curves generically intersect if they are
continuous.  An intersection corresponds to a point with
$\undertext{x}^\prime = 0$ and $\dot {\undertext{x}} = 1$.  Such a point moving
at the speed of light is called a cusp.  $\dot {\undertext{x}} (\sigma, t)$
need
not be continuous.  Points of discontinuity are called kinks.  Note that both
cusps and kinks will be smoothed out by finite thickness
effects$^{159)}$.
\par
The Nambu action does not describe what happens when two strings hit.  This
process has been studied numerically for both global$^{160)}$ and
local$^{161)}$ strings.  The authors of these papers set up scalar field
configurations
corresponding to two strings approaching one another and evolve the complete
classical scalar field equations.  The result of the analysis is that strings
do not cross but exchange ends, provided the relative velocity is smaller than
0.9.  Thus, by self intersecting, an infinite string will split off a loop
(Fig. 38).  An important open problem is to understand this process
analytically.  For a special value of the coupling constant Ruback$^{162)}$ has
given a mathematical explanation (see also Shellard and
Ruback in Ref. 161).
\par
There are two parts to the nontrivial evolution of the cosmic string network.
Firstly, loops are produced by self intersections of infinite strings.  Loops
oscillate due to the tension and slowly decay by emitting gravitational
radiation.  Combining the two steps we have a process by which energy
is transferred from the cosmic string network to radiation.
\par
There are analytical indications that a stable ``scaling solution"
(already described in Section ) for the
cosmic string network exists.  In the scaling solution, on the order of 1
infinite string segment crosses every Hubble volume.  The correlation length
$\xi (t)$
of an infinite string is thus of the order $t$.  A heuristic argument for the
scaling solution is due to Vilenkin$^{5)}$.  Take
$\tilde \nu (t)$ to be the mean number of infinite string segments per Hubble
volume.  Then the energy density in infinite strings is
$$
\rho_\infty (t) = \mu \tilde \nu (t) t^{-2} \eqno\eq
$$
The number of loops $n(t)$ produced per unit volume is proportional to the
square of $\tilde \nu$, since it takes two string segments to generate a string
intersection. Hence,
$$
{d n (t)\over{dt}} = c \tilde \nu^2 t^{-4} \eqno\eq
$$
where $c$ is a constant of the order $1$.  Conservation of energy in strings
gives
$$
{d \rho_\infty (t)\over{dt}} + {3\over{2 t}} \, \rho_\infty (t) = - c^\prime
\mu t \, {dn\over{dt}} = - c^\prime \mu \tilde \nu^2 t^{-3} \eqno\eq
$$
or, written as an equation for $\tilde \nu (t)$
$$
\tilde {\dot \nu} - \, {\tilde \nu\over{2 t}} = - cc^\prime \tilde \nu^2 t^{-
1}\eqno\eq
$$
Thus if $\tilde \nu \gg 1$ then $\tilde {\dot \nu} < 0$ while if $\tilde \nu
\ll 1$ then $\tilde {\dot \nu} > 0$.  Hence there will be a stable solution
with $\tilde \nu \sim 1$.

The precise value of $\tilde \nu$ must be determined in numerical simulations.
These simulations are rather difficult because of the large dynamic range
required and due to singularities which arise in the evolution equations near
cusps.  In the radiation dominated epoch, $\tilde \nu$ is still uncertain by a
factor of about 10.  The first results were reported in Ref. 163.
More recent results are due three groups.
Bennett and Bouchet$^{164)}$ and Allen and Shellard$^{165)}$ are
converging on
a value $10 < \tilde \nu < 20$, whereas Albrecht and Turok$^{166)}$ obtain a
value which is about 100.
\par
The scaling solution for the infinite strings implies that the network of
strings looks the same at all times when scaled to the Hubble radius.  This
should also imply that the distribution of cosmic string loops is scale
invariant in the same sense.  At present, however, there is no convincing
evidence from numerical simulations that this is really the case.
\par
A scaling solution for loops implies that the distribution of $R_i (t)$, the
radius of loops at the time of formation, is time independent after dividing
by $t$.  To simplify the discussion, I shall assume that the
distribution is monochromatic, \ie\
$$
R_i (t)/t = \alpha\, . \eqno\eq
$$
Based on Fig. 38, we expect $\alpha \sim 1$.  The numerical
simulations$^{164-166)}$,
however, now give $\alpha < 10^{-2}$.
\par
{}From the scaling solution (6.50) for the infinite strings we can derive the
scaling solution for loops.  We assume that the energy density in long strings
-- inasmuch as it is not redshifted -- must go into loops.  $\beta$ shall be a
measure for the mean length $\ell$ in a loop of ``radius" $R$
$$
\ell = \beta R\, . \eqno\eq
$$
If per expansion time and Hubble volume about 1 loop of radius $R_i (t)$ is
produced, then we know that the number density in physical coordinates of loops
of
radius $R_i (t)$ is
$$
n (R_i (t), t) = ct^{-4}\eqno\eq
$$
with a constant $c$ which can be calculated from (6.50), (6.54) and
(6.55).
Neglecting gravitational radiation, this number density simply redshifts
$$
n (R,t) = \, \left({z (t)\over{z (t_f (R))}} \right)^3 n (R, t_f (R))\, ,
\eqno\eq
$$
where $t_f (R)$ is the time when loops of radius $R$ are formed.  Isolating
the $R$ dependence, we obtain
$$
n (R, t) \sim R^{-4} z (R)^{-3}\eqno\eq
$$
where $z (R)$ is the redshift at time $t=R$.  We have the following special
cases:
$$
\eqalign{n (R, t) \sim R^{-5/2} t^{-3/2} \qquad & t < t_{eq}\cr
n (R, t) \sim R^{-5/2} t_{eq}^{1/2} t^{-2} \qquad & t > t_{eq} \, , \, t_f
(R) < t_{eq}\cr
n (R, t) \sim R^{-2} t^{-2} \qquad & t > t_{eq} \, , \, t_f (R) > t_{eq} \,
.}\eqno\eq
$$
\par
The proportionality constant $c$ is
$$
c = {1\over 2} \beta^{-1} \alpha^{-2} \tilde \nu\eqno\eq
$$
(see \eg\ Ref. 167).  In deriving (6.60) it is important to note that $n (R_i
(t), t) dR_i$ is the number density of loops in the radius interval $[R_i ,
R_i + dR_i]$.  Hence, in the radiation dominated epoch
$$
n (R, t) = \nu R^{- 5/2} t^{-3/2} \eqno\eq
$$
with
$$
\nu = {1\over 2} \beta^{-1} \alpha^{1/2} \tilde \nu \, .\eqno\eq
$$
\par
{}From (6.62) we can read off the uncertainties in $\nu$ based on the
uncertainties in the numerical results.  Both $\alpha^{1/2}$ and $\tilde \nu$
are determined only up to one order of magnitude.  Hence, any quantitative
results which depend on the exact value of $\nu$ must be treated with a grain
of salt.
\par
Gravitational radiation leads to a lower cutoff in $n (R, t)$.  Loops with
radius smaller than this cutoff were all formed at essentially the same time
and hence have the same number density.  Thus, $n (R)$ becomes flat.  The
power in gravitational radiation $P_G$ can be estimated using the quadrupole
formula$^{168)}$.  For a loop of radius $R$ and mass $M$
$$
P_G = {1\over 5} G < \dot{\ddot Q} \dot{\ddot Q} > \, , \eqno\eq
$$
where $Q$ is the quadrupole moment, $Q \sim MR^2$, and since the frequency of
oscillation is $\omega = R^{-1}$
$$
P_G \sim G (MR^2)^2 \omega^6 \sim (G \mu) \mu \, . \eqno\eq
$$
\par
Even though the quadrupole approximation breaks down since the loops move
relativistically, (6.64) gives a good order of magnitude of the power of
gravitational radiation.  Improved calculations give$^{169)}$
$$
P_G = \gamma (G \mu) \mu\eqno\eq
$$
with $\gamma \sim 50$.  (6.55) and (6.65) imply that
$$
\dot R = \tilde \gamma G \mu\eqno\eq
$$
with $\tilde \gamma \equiv \gamma/\beta \sim 5$ (using $\beta \simeq
10$).  Note that the rate of decrease is constant.  Hence,
$$
R (t) = R_i - (t - t_i) \tilde \gamma G \mu\eqno\eq
$$
and the cutoff loop radius is
$$
R_c \sim \tilde \gamma G \mu t_i\, . \eqno\eq
$$
\par
Let us briefly summarize the scaling solution
\item{1)} At all times the network of infinite strings looks the same when
scaled by the Hubble radius.  A small number of infinite string segments cross
each Hubble volume and $\rho_\infty (t)$ is given by (6.50).
\item{2)}  There is a distribution of loops of all sizes $0 \le R < t$.
Assuming scaling for loops, then
$$
n (R, t) = \nu R^{-4} \, \left({z (t)\over{z (R)}}\right)^3 \, , \> R
\> \epsilon \> [ \tilde \gamma G \mu t,  \alpha t]\eqno\eq
$$
where $\alpha^{-1} R$ is the time of formation of a loop of radius $R$.  Also
$$
n (R, t) = n ( \tilde \gamma G \mu t,  t) \, , \> R < \tilde \gamma G \mu t\,
. \eqno\eq
$$

Although the qualitative characteristics of the cosmic string scaling
solution are well established, the quantitative details are not.  The
main reason for this is the fact that the Nambu action breaks down at
kinks and cusps.  However, kinks and cusps inevitably form and are
responsible for the small scale structure on strings.  In fact, coarse
graining by integrating out the small scale structure may give an
equation of state for strings which deviates from that of a Nambu
string$^{170)}$.  Attempts at understanding the small scale structure
on strings are at present under way$^{171)}$.

\section{Cosmic Strings and Structure Formation}

The starting point of the structure formation scenario in the cosmic
string theory is the scaling solution for the cosmic string network,
according to which at all times $t$ (in particular at $t_{eq}$, the
time when perturbations can start to grow) there will be a few long
strings crossing each Hubble volume, plus a distribution of loops of
radius $R \ll t$ (see Fig. 39).

\midinsert \vskip 6.5cm
\noindent {\bf Figure 39.}  Sketch of the scaling solution for the
cosmic string network.  The box corresponds to one Hubble volume at
arbitrary time $t$.
\endinsert

The cosmic string model admits three mechanisms for structure
formation:  loops, filaments, and wakes.  Cosmic string loops have the same
time averaged field as a point source with mass$^{172)}$
$$
M (R) = \beta R \mu \, , \eqno\eq
$$
$R$ being the loop radius and $\beta \sim 2 \pi$.  Hence, loops will be seeds
for spherical accretion of dust and radiation.

For loops with $R \leq t_{eq}$, growth of perturbations in a model
dominated by cold dark matter starts at $t_{eq}$.  Hence, the mass at
the present time will be
$$
M (R, \, t_0) = z (t_{eq}) \beta \, R \mu \, . \eqno\eq
$$

In the original cosmic string model$^{47, 48, 57)}$ it was assumed
that loops dominate over wakes.  In this case, the theory could be
normalized ({\it i.e.}, $\mu$ could be determined) by demanding that loops
with the mean separation of clusters $d_{cl}$ (from the discussion in
Section 6.4 it follows that the loop radius $R (d_{cl})$ is determined
by the mean separation) accrete the correct mass, {\it i.e.}, that
$$
M (R (d_{cl}), t_0) = 10^{14} M_{\odot} \, . \eqno\eq
$$
This condition yields$^{57)}$
$$
\mu \simeq 10^{36} {\rm GeV}^2 \eqno\eq
$$
Thus, if cosmic strings are to be relevant for structure formation,
they must arise due to a symmetry breaking at energy scale $\eta
\simeq 10^{16}$GeV.  This scale happens to be the scale of unification (GUT)
of weak, strong and electromagnetic interactions.  It is tantalizing
to speculate that cosmology is telling us that there indeed was new
physics at the GUT scale.

\midinsert \vskip 3.5cm
\noindent{\bf Figure 40.} Sketch of the mechanism by which a long
straight cosmic string $S$ moving with velocity $v$ in transverse
direction through a plasma induces a velocity perturbation $\Delta v$
towards the wake. Shown on the left is the deficit angle, in the
center is a sketch of the string moving in the plasma, and on the
right is the sketch of how the plasma moves in the frame in which the
string is at rest.
\endinsert

The second mechanism involves long strings moving with relativistic
speed in their normal plane which give rise to
velocity perturbations in their wake$^{173)}$.  The mechanism is illustrated in
Fig. 40:
space normal to the string is a cone with deficit angle$^{174)}$
$$
\alpha = 8 \pi G \mu \, . \eqno\eq
$$
If the string is moving with normal velocity $v$ through a bath of dark
matter, a velocity perturbation
$$
\delta v = 4 \pi G \mu v \gamma \eqno\eq
$$
[with $\gamma = (1 - v^2)^{-1/2}$] towards the plane behind the string
results.  At times after $t_{eq}$, this induces planar overdensities,
the most
prominent ({\it i.e.}, thickest at the present time) and numerous of which were
created at $t_{eq}$, the time of equal matter and
radiation$^{58, 59, 63)}$.  The
corresponding planar dimensions are (in comoving coordinates)
$$
t_{eq} z (t_{eq}) \times t_{eq} z (t_{eq}) v \sim (40 \times 40 v) \,
{\rm Mpc}^2
\, . \eqno\eq
$$

The thickness $d$ of these wakes can be calculated using the
Zel'dovich approximation$^{63)}$.  The result is
$$
d \simeq G \mu v \gamma (v) z (t_{eq})^2 \, t_{eq} \simeq 4 v \, {\rm
Mpc} \, . \eqno\eq
$$
\par
Wakes arise if there is little small scale structure on the string.
In this case, the string tension equals the mass density, the string
moves at relativistic speeds, and there is no local gravitational
attraction towards the string.

In contrast, if there is small scale structure on strings,
then the string tension $T$ is smaller$^{170)}$ than the mass per unit
length $\mu$ and the metric of a string in $z$ direction becomes$^{175)}$
$$
ds^2 = (1 + h_{00}) (dt^2 - dz^2 - dr^2 - (1 - 8G \mu) r^2 dy^2 )
\eqno\eq
$$
with
$$
h_{00} = 4G (\mu - T) \ln \, {r\over r_0} \, , \eqno\eq
$$
$r_0$ being the string width.  Since $h_{00}$ does not vanish, there
is a gravitational force towards the string which gives rise to
cylindrical accretion, thus producing filaments.

As is evident from the last term in the metric (6.79), space
perpendicular to the string remains conical, with deficit angle given
by (6.75).  However, since the string is no longer relativistic, the
transverse velocities $v$ of the string network are expected to be
smaller, and hence the induced wakes will be shorter and thinner.

Which of the mechanisms -- filaments or wakes -- dominates is
determined by the competition between the velocity induced by $h_{00}$
and the velocity perturbation of the wake.  The total velocity
is$^{175)}$
$$
u = - {2 \pi G (\mu - T)\over{v \gamma (v)}} - 4 \pi G \mu v \gamma
(v) \, , \eqno\eq
$$
the first term giving filaments, the second producing wakes.  Hence,
for small $v$ the former will dominate, for large $v$ the latter.

By the same argument as for wakes, the most numerous and prominent
filaments will have the distinguished scale
$$
t_{eq} z (t_{eq}) \times d_f \times d_f \eqno\eq
$$
where $d_f$ can be calculated using the Zel'dovich approximation.

The cosmic string model predicts a scale-invariant spectrum of density
perturbations, exactly like inflationary Universe models but for a
rather different reason.  Consider the {\it r.m.s.} mass fluctuations
on a length scale $2 \pi k^{-1}$ at the time $t_H (k)$ when this scale
enters the Hubble radius.  From the cosmic string scaling solution it
follows that a fixed ({\it i.e.}, $t_H (k)$ independent) number
$\tilde v$ of strings of length of the order $t_H (k)$ contribute to
the mass excess $\delta M (k, \, t_H (k))$.  Thus
$$
{\delta M\over M} \, (k, \, t_H (k)) \sim \, {\tilde v \mu t_H
(k)\over{G^{-1} t^{-2}_H (k) t^3_H (k)}} \sim \tilde v \, G \mu \, .
\eqno\eq
$$
Note that the above argument predicting a scale invariant spectrum
will hold for all topological defect models which have a scaling
solution, in particular also for global monopoles and textures.

The amplitude of the {\it r.m.s.} mass fluctuations (equivalently: of
the power spectrum) can be used to normalize $G \mu$.  Since today on
galaxy cluster scales
$$
{\delta M\over M} (k, \, t_0) \sim 1 \, , \eqno\eq
$$
the growth rate of fluctuations linear in $a(t)$ yields
$$
{\delta M\over M} \, (k, \, t_{eq}) \sim 10^{-4} \, , \eqno\eq
$$
and therefore, using $\tilde v \sim 10$,
$$
G \mu \sim 10^{-5} \, . \eqno\eq
$$

A big advantage of the cosmic string model over inflationary Universe
models is that HDM is a viable dark matter candidate.  Cosmic string
loops survive free streaming, as discussed in Section 3.4, and can
generate nonlinear structures on galactic scales, as discussed in
detail in Refs. 61 and 62.  Accretion of hot dark matter by a string wake
was studied in Ref. 63. In this case, nonlinear perturbations
develop only late.  At some time $t_{nl}$, all scales up to a distance
$q_{\rm max}$ from the wake center go nonlinear.  Here
$$
q_{\rm max} \sim G \mu v \gamma (v) z (t_{eq})^2 t_{eq} \sim 4 v \,
{\rm Mpc} \, , \eqno\eq
$$
and it is the comoving thickness of the wake at $t_{nl}$.  Demanding
that $t_{nl}$ corresponds to a redshift greater than 1 leads to the
constraint
$$
G \mu > 5 \cdot 10^{-7} \, . \eqno\eq
$$

The power spectra in the cosmic string models with CDM and HDM are
obviously different on scales smaller than the maximal neutrino free
streaming length (3.27).  Recent calculations$^{176, 177)}$ of the power
spectra are shown in Fig. 41.

\midinsert \vskip 9cm
\noindent{\bf Figure 41:} Power spectra for cosmic string HDM and CDM theories
(dashed curves), compared to those for inflationary HDM and CDM models (solid
curves). In each case, the top curve is for CDM, the bottom one for HDM. Note
that there is substantial power on small scales in the cosmic string HDM
theory.
\endinsert

\section{Global Textures and Structure Formation}

The starting point of the texture scenario of structure formation$^{60)}$ is
the
scaling solution for textures: at any time $t$, there is a fixed
probability $p (n_w) \, dn_w$ that the scalar field configuration over
a Hubble volume covers between $n_w$ and $n_w + dn_w$ of the vacuum
manifold, {\it i.e.}, we have a texture with winding number in the
interval $[n_w, \, n_w + dn_w ]$ entering the Hubble radius.

The dynamics of a texture is easy to understand.  Consider the
spherically symmetric texture configuration of (6.24) with $\chi (r)$
increasing from 0 to $\chi_{\rm max}$ over a distance $d$.  If $d$ is
larger than the Hubble radius, then the Hubble damping term dominates
the equation of motion for $\varphi$ and the field configuration is
frozen in.  Once the Hubble radius $t$ catches up with $d$, the
microphysical forces become dominant and the texture field begins to
evolve.

\midinsert \vskip 5.5cm
\noindent{\bf Figure 42:} A sketch of the forces acting on a spherically
symmetric texture configuration and which cause unwinding in case (a) in which
the winding number is larger than the critical winding, and dissipation if the
winding is smaller than its critical value (case (b)). $r$ is the distance from
the center of the texture, and the vertical axis shows the value of the $\chi$
field.
\endinsert

The evolution of $\varphi$ tends to minimize the field energy.
Consider first large distances from the texture center.  The spatial
gradient energy can be decreased by having $\chi_{\rm max}$ increase
(if $\chi_{\rm max} > \chi_c$) or decrease (if $\chi_{\rm max} <
\chi_c$) (see Figure 42).  The winding associated with $\chi_c$ is called the
critical
winding $n_c$ (see (6.26)).  For a single texture in an infinite
volume we would expect
$$
\chi_c = {\pi\over 2} \, (i.e., \, n_c = 0.5) \, . \eqno\eq
$$
For realistic textures there will be a ``finite volume cutoff"
determined by the separation of textures.  A semi-analytical analysis
and numerical simulations give$^{149-151)}$
$$
n_c \simeq 0.6 \, . \eqno\eq
$$

If $n_W < n_c$, then the field configuration will relax to a trivial
one.  No localized energy concentrations will be generated, and we
cannot speak of a ``texture."  However, if $n_W > n_c$ the field
evolution will be more interesting.  At large $r$, $\chi (r)$ will
increase.  In addition, the radius $r (\chi)$ where $\chi$ takes on a
fixed value $\chi$ tends to decrease, since this leads to a
concentration of gradient energies over a smaller region.  Hence, the
field configuration will contract (see Fig. 42), with increasing total
winding number.  Eventually, close to $r = 0$ there is sufficient
tension energy for $\varphi$ to be able to leave the vacuum manifold
and jump from $\chi = 0$ to $\chi = \pi$.  This is the texture
unwinding event.  After unwinding, energy is radiated radially in the
form of Goldstone bosons.

In the texture model it is the contraction of the field configuration which
leads to density perturbations$^{178)}$.  At the time when the texture enters
the
horizon, an isocurvature perturbation is established:  the energy density in
the scalar field is compensated by a deficit in radiation.  However, the
contraction of the scalar field configuration leads to a clumping of gradient
and kinetic energy at the center of the texture (Fig. 43).  This, in turn,
provides the
seed perturbations which cause dark matter and radiation to collapse in a
spherical manner$^{179, 180)}$.

\midinsert \vskip 6.5cm
\noindent {\bf Figure 43}: A sketch of the density perturbation produced
by a collapsing texture.  The left graph shows the time evolution of
the field $\chi (r)$ as a function of radius $r$ and time (see
(5.18)).  The contraction of $\chi (r)$ leads to a spatial gradient
energy perturbation at the center of the texture, as illustrated on
the right.  The energy is denoted by $\rho$.  Solid lines denote the
initial time, dashed lines are at time $t + \Delta t$, and dotted
lines correspond to time $t + 2 \Delta t$, where $\Delta t$ is a
fraction of the Hubble expansion time (the typical time scale for the
dynamics).
\endinsert

As in the cosmic string model, also in the global texture scenario the
length scale of the dominant structures is the comoving Hubble radius
at $t_{eq}$.  Textures generated at $t_{eq}$ are the most numerous,
and the perturbations induced by them have the most time to grow.

As mentioned in the previous subsection, the texture model predicts a
scale-invariant spectrum of density perturbations.  Hence, in order to
differentiate topological defect models from inflationary scenarios,
and to distinguish between different topological defect theories, we
need statistics which are not determined by the power spectrum alone.
We need statistics which are sensitive to the non-random phases of
topological defect models.  One such statistic is the genus curve$^{181)}$.
For a surface $S$ embedded in $R^3$, the genus $g$ is
$$
g (S) = {\rm \# \, of \, holes \, of} \, S - \, {\rm \# \, of \,
disconnected \, components \, of} \, S + 1 \, . \eqno\eq
$$
The genus $g$ can now be evaluated for the isodensity surface $S
(\rho)$, the surface of points in space with density equal to $\rho$.
The curve
$$
g (\rho) = g (S (\rho)) \eqno\eq
$$
is the genus curve.  To reduce numerical errors, $g$ can also be
evaluated based on a cell decomposition of the volume.  Now, $g (n)$
is the genus of the boundary of the cell complex in which each cell
contains more than $n$ galaxies.  In this case, the genus is simply
$$
g = 1 - {1\over 2} (V-E-F) \eqno\eq
$$
where $V,E,F$ are the number of vertices, edges, and faces of the
polygonal surface, respectively.

\midinsert \vskip 11.5cm
\noindent{\bf Figure 44:} Comparison of the genus curve (genus as a function of
galaxy density) of different toy models of structure formation. Except for the
Gaussian model, all theories have the same linear power spectrum. The
`filament', `wake' and `texture' toy models are based on laying down at random
linear, planar and spherical overdense regions of galaxies. Thus, the figure
demonstrates that the genus statistic is able to distinguish between theories
with different topologies but identical power spectra. The `CDM' model
predictions are computed from linear theory, and the `Poisson' model is
obtained by randomly distributing galaxies. See the senior thesis by
Aguirre$^{182)}$ for further details.
\endinsert

As shown in Fig. 44, the genus statistic is able to distinguish
between models with the same power spectrum but different phase correlations
and topology$^{182)}$. For a texture toy model, the genus curve
is mostly negative, for a cosmic string wake model, it is
predominately positive.  The differences compared to a random phase
inflationary model are statistically significant.

The differences shown in Fig. 44 will only be apparent in large-scale
samples of galaxies, {\it i.e.}, on scales exceeding the comoving
radius at $t_{eq}$.  Such samples should, however, become available in
the near future, and at that point genus curve and other statistics
sensitive to non-random phases should become a powerful tool for
distinguishing the predictions of the different models of structure
formation.

A final word concerning textures: since they are short-lived, only CDM
is a viable dark matter candidate in the context of this structure
formation scenario.

\chapter{Cosmic Microwave Background Anisotropies}

As mentioned in Section 3, the near-isotropy of the CMB is the
strongest evidence in support of the cosmological principle.  By the
same reasoning, any density inhomogeneities in the early Universe will
give rise to CMB anisotropies.  Since our present theories of galaxy
formation are based on the gravitational instability scenario, they
predict such inhomogeneities.  The CMB temperature fluctuations probe
the structure of space at $t_{rec}$, the time of last scattering, a
time when the density perturbations still have a small amplitude and can
be analyzed in linear theory.  Hence, a study of CMB anisotropies will
yield a lot of constraints for structure formation models.  The
information gained will be robust, {\it i.e.}, independent of the
uncertainties of nonlinear gravitational and hydrodynamical effects,
but it will deal only with large scales (comparable or larger than the
comoving horizon at $t_{rec}$).  In this section we shall give a brief overview
of the theory of CMB anisotropies and summarize some recent observational
results.

\section{Basics}

As illustrated in Fig. 45, there are three main sources of CMB
anisotropies.  The first are gravitational potential perturbations at
$t_{rec}$ which lead to fluctuations of the surface of last
scattering.  This produces deviations in the light travel time between
last scattering and detection, and -- given that the photons have the
same temperature on the surface of last scattering -- to temperature
fluctuations for the observer.

\midinsert \vskip 6cm
\noindent{\bf Figure 45:} Space-time plot sketching the origin of CMB
temperature anisotropies. The surface labelled $T_{rec}$ is the last scattering
surface. ${\cal O}$ is the observer at the present time measuring photons
$\gamma$ impinging from directions in the sky separated by an angle $\theta$.
The shaded area labelled $C$ is the world volume of a local overdensity,
leading to distortions of geodesics. Possible velocities of observer and
emitter are indicated by ${\vec v}_o$ and ${\vec v}_e$, respectively.
\endinsert

The second source is due to gravitational perturbations along the line
of sight which lead to deviations of the geodesics and hence to
temperature differences.  A Newtonian way of understanding this effect
is to consider a photon passing through a large mass concentration.
On the way towards the center, the photon is falling into a potential
well and acquires a blueshift, whereas on its way out it is
redshifted.  In an expanding background, this redshift does not
exactly cancel the initial blueshift, and a temperature fluctuation
results.

The third source contributing to CMB anisotropies are peculiar
velocities on the surface of last scattering and of the observer.  The
peculiar motion of the earth gives rise to a dipole anisotropy$^{183)}$
$$
{\delta T\over T} \big|_{\rm dipole} \simeq \, 10^{-3} \eqno\eq
$$
Peculiar velocities induce temperature fluctuations by means of the
Doppler effect.

For linear adiabatic density perturbations in a matter dominated
Universe, the line of sight contributions to $\delta T/T$ can be
written as total time derivative and thus reduces to a contribution
from the surface of last scattering and can be simply combined with
the potential fluctuations at $t_{rec}$.  This is the case first
studied by Sachs and Wolfe, and the combined effect is now called the
Sachs-Wolfe effect$^{184)}$.

An analysis of the Sachs-Wolfe effect reveals a very simple
relationship between temperature fluctuations $\delta T/T \,
(\vartheta)$ on an angular scale $\vartheta$ and the magnitude of
density perturbations on the corresponding lengths scale $\lambda
(\vartheta)$, where at last scattering $\lambda (\vartheta)$ equals
the distance subtended by two light rays with angular separation
$\vartheta$ (see Fig. 45).  A simple derivation$^{11, 185)}$ of this
relationship
makes use of the gauge invariant theory of cosmological perturbations
described in Section 4.4.

The starting point is the phase space distribution function $f (x^\alpha,
\, p_i)$ which would be a function of $p / T$ exclusively in the
absence of inhomogeneities.  In the presence of inhomogeneities, the
deviation of $f$ from homogeneity is associated with temperature
fluctuations:
$$
f (x^\alpha , \, p_i) = \bar f (p / \bar T + \delta T) \, , \eqno\eq
$$
where $\bar T$ is the average temperature and $\bar f (p / T)$ is the
background phase space density.

The phase space distribution function satisfies the collisionless
Boltzmann equation
$$
{d x^\alpha\over{d \eta}} \, {\partial f\over{\partial x^\alpha}} + {d
p_i\over{d \eta}} \, {\partial f\over{\partial p_i}} = 0 \eqno\eq
$$
where, as in Section 4.4, the variable $\eta$ denotes conformal time.
This equation can be integrated along the perturbed geodesics which
are given by
$$
{d p_\alpha\over{d \eta}} = 2p {\partial \Phi\over{\partial x^\alpha}}
\eqno\eq
$$
and
$$
{d x^i\over{d \eta}} = l^i (1 + 2 \Phi), \eqno\eq
$$
with
$$
l^i = - {1\over p} \, p_i \eqno\eq
$$
and
$$
p^2 = p_i p_i \, . \eqno\eq
$$
Inserting these relations into the Boltzman equation gives
$$
\left( {\partial\over{\partial \eta}} + l^i \partial_i \right) \,
{\delta T\over T} = - 2 l^i \partial_i \Phi \, . \eqno\eq
$$
Since in the matter dominated period $\partial_\eta \Phi = 0$ we can
rewrite (7.8) as
$$
\left( {\partial\over{\partial_\eta}} + l^i_i \right) \, \left({\delta
T\over T} + 2 \Phi \right) = 0 \, , \eqno\eq
$$
which implies that
$$
{\delta T\over T} + 2 \Phi = \, {\rm const} \eqno\eq
$$
along the perturbed geodesics.

For isothermal primordial perturbations $( {\delta T\over T} \,
(t_{rec}) = 0)$, the result (7.10) implies that
$$
{\delta T\over T} (\eta_0) = 2 \Phi (\eta_{rec}) + l^i v_i
(\eta_{rec}) \eqno\eq
$$
whereas for primordial adiabatic perturbations (vanishing initial
entropy perturbations)
$$
{\delta T\over T} (\eta_0) = {1\over 3} \Phi (\eta_{rec}) + l^i v_i
(\eta_{rec}) \, , \eqno\eq
$$
{\it i.e.}, the combination of initial curvature fluctuations and line
of sight effects leads to a partial cancellation of the anisotropy.
The second term on the {\rm r.h.s.} of (7.11) and (7.12) is the
Doppler term, and it arises from a determination of the constant in
(7.10) based on considering the initial conditions at $t_{rec}$.

Since $\Phi$ is constant both between $t_{eq}$ and $t_{rec}$ and while
outside the Hubble radius, and since
$$
\Phi (t_H) \sim {\delta \rho\over \rho} \, (t_H) \eqno\eq
$$
at Hubble radius crossing $t_H$, our results imply that (modulo
Doppler terms) for adiabatic perturbations
$$
{\delta T\over T} (\vartheta, \, t_0) = {1\over 3} \Phi (\lambda
(\vartheta), \, t_{eq}) \sim {1\over 3} \, {\delta M\over M} \,
(\lambda (\vartheta), \, t_H)) \, . \eqno\eq
$$
We conclude that the spectrum of primordial mass perturbations can be
normalized by CMB anisotropy detections.  For a scale invariant
spectrum of density perturbations, the {\it r.m.s.} temperature
fluctuations are predicted to be independent of $\vartheta$ on angular
scales larger than the Hubble radius at $t_{rec}$ (between 1 and 2
degrees).

\section{Specific Signatures}

All theories of structure formation give rise to Sachs-Wolfe type
temperature fluctuations given by (7.13) and (7.14).  In topological
defect models there are, in addition, specific signatures which cannot
be described in a linear perturbative analysis.

As described in Section 6.5, space perpendicular to a long straight
cosmic string is conical with deficit angle given by (6.75).  Consider
now CMB radiation approaching an observer in a direction normal to the
plane spanned by the string and its velocity vector (see Fig. 46).
Photons arriving at the observer having passed on different sides of
the string will obtain a relative Doppler shift which translates into
a temperature discontinuity of amplitude$^{186)}$
$$
{\delta T\over T} = 4 \pi G \mu v \gamma (v) \, , \eqno\eq
$$
where $v$ is the velocity of the string.  Thus, the distinctive
signature for cosmic strings in the microwave sky are line
discontinuities in $T$ of the above magnitude.

\midinsert \vskip 7cm
\noindent{\bf Figure 46:} Sketch of the Kaiser-Stebbins effect by
which cosmic strings produce linear discontinuities in the CMB. Photons
$\gamma$ passing on different sides of a moving string $S$ (velocity $v$)
towards the observer ${\cal O}$ receive a relative Doppler shift due to the
conical nature of space perpendicular to the string (deficit angle $\alpha$).
\endinsert

Given ideal maps of the CMB sky it would be easy to detect strings.
However, real experiments have finite beam width.  Taking into account
averaging over a scale corresponding to the beam width will smear out
the discontinuity, and it turns out to be surprisingly hard to
distinguish the predictions of the cosmic string model from that of
inflation-based theories using quantitative statistics which are easy
to evaluate analytically, such as the kurtosis of the spatial gradient
map of the CMB$^{187)}$.

Textures produce a distribution of hot and cold spots on the CMB sky
with typical size of several degrees$^{188)}$.  This signature is much easier
to see in CMB maps.  The mechanism which produces these hot and cold
spots in the CMB is illustrated in Fig. 47.

\midinsert \vskip 5.5cm
\noindent{\bf Figure 47:} Space-time diagram of a collapsing texture. The
unwinding occurs at the point $TX$. The shaded areas correspond to overdense
regions. Photons like $\gamma_1$ are redshifted, those like $\gamma_2$ are
blueshifted.
\endinsert

Photons arriving at the observer having passed through a texture as in the case
of the ray
$\gamma_1$ in Fig. 47 will be redshifted relative to the average
photons since they have to climb out of a potential well, whereas
those in orientation $\gamma_2$ will be blueshifted since they fall
into a potential well.  Taking into account reionization produced by
texture collapse gives an amplitude of $\delta T/T$ of$^{189, 190)}$
$$
{\delta T\over T} \sim 0.06 \times 16 \pi G \eta^2 \, . \eqno\eq
$$
A number of about ten hot and cold spots of angular scale 10$^\circ$
is predicted by the texture model.

Theories of structure formation can now be normalized from CMB
anisotropy data and from large-scale structure considerations.  An
inflationary model with CDM yields agreement between these two
normalizations provided$^{191)}$
$$
b \simeq 1 \, , \eqno\eq
$$
where $b$ is the bias factor determining the ratio of fractional mass
to light perturbations on a scale of 8h$^{-1}$ Mpc.
$$
{\delta L\over L} \Big|_{8 {\rm h}^{-1}{\rm Mpc}} = b \, {\delta
M\over M} \Big|_{8 {\rm h}^{-1} {\rm Mpc}} \, . \eqno\eq
$$
However, agreement between galaxy and cluster correlation properties
seem to require$^{192)}$
$$
b \sim 2 \, . \eqno\eq
$$

Normalizations of the texture model from large-scale structure and CMB
observations$^{189, 190, 193)}$ require a bias
$$
b \sim 3 \, , \eqno\eq
$$
whereas for cosmic strings the two normalizations agree well.  Based
both on numerical simulations and analytical calculations, a
normalization of the cosmic string model from the COBE CMB anisotropy
data gives$^{194, 195)}$
$$
G \mu = (1.3 \pm 0.5) 10^{-6} \, . \eqno\eq
$$

\section{Experimental Results}

Over the past couple of years there has been a spectacular
breakthrough on the observational front.  The DMR experiment on the
COBE satellite$^{196)}$ has produced a temperature map of the entire sky with
beam width of 7$^\circ$, which shows a clear detection of CMB
anisotropies.  Independent confirmation has come from two 5$^\circ$
experiments, FIRAS$^{197)}$ which has mapped 1/4 of the sky, and the Tenerife
experiment$^{198)}$ which surveyed a strip of 70$^\circ$ length in right
ascension at a declination 40$^\circ$.  The FIRAS data cross
correlate very well with the COBE results, and there is even good agreement in
the location of a pronounced feature
in the Tenerife map with a that of a comparable feature in the
two-year COBE maps.

In addition, there are many small angular scale experiments which have
detected anisotropies.  A partial list of observational results is
given in Table 2. In this table, ``Angular Scale" denotes the beam width, the
``results for $\delta
T/T$" stands for the variance of $\delta T$ computed from the CMB
maps, ``cover" indicates the area of the sky mapped.  MAX 1 and MAX 2
denote two separate MAX measurements of $\delta T/T$, one in a region of
the sky $\mu \, {\rm Peg}$, the second near GUM.  OVRO 1 is the first
Owens Valley experiment, a measurement near the North Galactic Cap,
the second is a ring survey. The large numer of anisotropy experiments which
have announced detections of temperature fluctuations since April 1992
indicates the rapid progress in this field.

To a first approximation, the present experimental results are in
agreement with the predictions of a scale invariant spectrum of
density perturbations.  A popular way to show the results is to expand
$T(\undertext{n})$ in spherical harmonics
$$
T (\undertext{n}) = \sum_l \sum_{m = -l}^l a_{lm} Y_{lm} (\undertext{n})
,\eqno\eq
$$
where $\undertext{n}$ is a unit vector on the sky, and to calculate
the temperature correlation function
$$
< T (\undertext{n}_1) T (\undertext{n}_2 ) > = {1\over{4 \pi}}
\sum\limits_l (2 l + 1) C_l P_l (\undertext{n}_1 \cdot \undertext{n}_2
) \eqno\eq
$$
where
$$
< a^\ast_{lm} \, a_{l^\prime m^\prime} > = C_l \delta_{ll^\prime}
\delta_{mm^\prime} \, . \eqno\eq
$$
For a power spectrum of density perturbations
$$
P (k) \sim k^n \eqno\eq
$$
the prediction for the Sachs-Wolfe contribution to $\delta T$ is
$$
l^2 C_l \sim l^{n-1} \eqno\eq
$$
on scales larger than the Hubble radius at $t_{rec}$ ({\it i.e.}, for
small values of $l$).

A direct comparison between theory and experiment is complicated by
two effects:  the Doppler contribution to $\delta T/T$ creates a peak
in the $l^2 C_l$ curve at values of $l$ which correspond to wavelengths
comparable to the Hubble radius at
$t_{rec}$, whose amplitude depends strongly on the ionization history
of the Universe.  Reionization also leads to a decrease in $C_l$ for
large $l$.

The COBE results combined with Tenerife observations favor$^{198)}$ a value of
$n$ larger than what is predicted by simple inflationary models.
However, the error bars are large and the difference is not (yet)
statistically significant.  At present there is the intriguing puzzle
as to why the signal of certain small scale experiments is larger than
the upper limit of other observations at the same angular scale
elsewhere in the sky.  A search for possible non-Gaussian features in
the CMB sky will have high priority in the next years.

\midinsert 

\bigskip
\settabs 5\columns
\centerline{{\bf TABLE 2}:~ CMB Anisotropy Results}
\medskip
\+  Experiment  & Angular Scale   & Result for ${\delta T\over T}$ & Cover
   & Location   \cr\vskip10pt
\+  &  &  &  &  \cr\vskip12pt
\+ COBE-DMR$^{196)}$    &   7$^\circ$  & 1.1 $\pm$ 0.2    & 4$\pi$ & space
\cr\vskip12pt
\+ Tenerife$^{198)}$    &  5.6$^\circ$ & 1.7 $\pm$ 0.4 & 350 deg$^2$ & ground
\cr\vskip12pt
\+ FIRS$^{197)}$    &   4$^\circ$  & 1 - 3  & $\pi$    & balloon
\cr\vskip12pt
\+ SK93$^{199)}$    &   1.45$^\circ$  & 1.4 $\pm$ 0.5    &   & ground
\cr\vskip12pt
\+ SP91$^{200)}$    &   1.4$^\circ$  & 1.1 $\pm$ 0.5    & 13.8 deg$^2$   &
ground    \cr\vskip12pt
\+ ARGO$^{201)}$    &   1$^\circ$  &  2.2 $\pm$ 0.8 &  & balloon
\cr\vskip12pt
\+ Python$^{202)}$    &   0.75$^\circ$  & 3    & 8 deg$^2$    & ground
\cr\vskip12pt
\+ MAX 1$^{203)}$    &   0.5$^\circ$  & $< 3$    &  & balloon   \cr\vskip12pt
\+ MAX 2$^{204)}$    &   0.5$^\circ$  & 4.9 $\pm$ 0.8   & 6 deg$^2$  &
balloon\cr\vskip12pt
\+ MSAM$^{205)}$     &   0.47$^\circ$   & 1.6 $\pm$ 0.4  & 6 deg$^2$  &
balloon\cr\vskip12pt
\+ White Dish$^{206)}$    &   0.2$^\circ$  & $< 2.3$    &  & ground
\cr\vskip12pt
\+ OVRO 1$^{207)}$    &   1.8$^\prime$  & $<1.9$    & 0.03 deg$^2$    & ground
  \cr\vskip12pt
\+ OVRO 2$^{208)}$    &   1.8$^\prime$  & 3.4 $\pm$ 1.1    & 0.1 deg$^2$    &
ground   \cr
\bigskip

\endinsert
\medskip

\chapter{Conclusions}

Modern cosmology has led to the development of several theories of
structure formation, most prominently theories based on inflation, and
topological defect models.  These new theories are all based on the
union between particle physics and general relativity.  The models of
structure formation obey the usual causality principle of relativistic physics.

All of the current theories of structure formation have their
problems.  Most importantly, they do not address the cosmological
constant problem but rather, inasmuch as they make use of scalar matter
fields, make the problem worse.  The inflationary Universe
scenario is still lacking a convincing realization.  Present versions
require very special scalar field potentials.  Topological defect
models, on the other hand, do not explain why the Universe is nearly
homogeneous and spatially flat.  In my opinion, we should regard our
current theories as toy models with which we work and from which we
learn, but which will eventually be replaced by improved and more
convincing theories.

Nevertheless, our present theories are predictive.  To a first
approximation, they all predict a scale invariant spectrum of density
perturbations and induced CMB anisotropies.  Typically, the models
contain one intrinsically free parameter (plus maybe a couple more
parameters with which we can describe our ignorance of the detailed
evolution of the models).  The free parameter can be normalized from
any one of several observables.  It is remarkable that the different
normalizations of the models are consistent (to a first
approximation).  This lets us entertain the hope that we are on the
right track: structure formation proceeds via gravitational
instability (Sections 4.3 and 4.4) with the seed perturbations being
provided by a particle physics theory of the very early Universe.

There is already a wealth of observational data which is fit quite
well by our present toy models.  More and higher accuracy data is
rapidly becoming available.  The data concerns on one hand structure
in the Universe gleamed from optical and infrared galaxy surveys, and
on the other hand from the temperature map of the CMB sky.

With the wealth of data available and steady flow of new observational
results, and given that many important questions remain unresolved,
modern cosmology will remain an exciting area of research for the
forseeable future.

\centerline{\bf Acknowledgments}

I wish to thank John Donoghue for the invitation to lecture at this school, the
TASI-94 students for their interest, and Mark Trodden for proofreading the
text.
I am grateful to all of my research collaborators, on whose work I have
freely drawn. Partial financial support for the preparation of this manuscript
has been provided by the US Department of Energy under Grant DE-FG0291ER40688,
Task A.
\bigskip
\REF\one{A. Linde, `Particle Physics and Inflationary Cosmology'
(Harwood, Chur, 1990).}
\REF\two{S. Blau and A. Guth, `Inflationary Cosmology,' in `300 Years
of Gravitation' ed. by S. Hawking and W. Israel (Cambridge Univ.
Press, Cambridge, 1987).}
\REF\three{K. Olive, {\it Phys. Rep.} {\bf 190}, 307 (1990).}
\REF\four{T.W.G. Kibble, {\it Phys. Rep.} {\bf 67}, 183 (1980).}
\REF\five{A. Vilenkin, {\it Phys. Rep.} {\bf 121}, 263 (1985).}
\REF\six{N. Turok, `Phase Transitions as the Origin of Large-Scale
Structure,' in `Particles, Strings and Supernovae' (TASI-88) ed. by
A. Jevicki and C.-I. Tan (World Scientific, Singapore, 1989).}
\REF\seven{A. Vilenkin and E.P.S. Shellard, `Strings and Other Topological
Defects' (Cambridge Univ. Press, Cambridge, 1994).}
\REF\eight{R. Brandenberger, {\it Rev. Mod. Phys.} {\bf 57}, 1
(1985).}
\REF\nine{R. Brandenberger, in `Physics of the Early Universe,' proc.
of the 1989 Scottish Univ. Summer School in Physics, ed. by J. Peacock, A.
Heavens and A. Davies (SUSSP Publ., Edinburgh, 1990).}
\REF\ten{R. Brandenberger, in proc. of the 1991 Trieste Summer School
in Particle Physics and Cosmology, eds. E. Gava et al. (World
Scientific, Singapore, 1992); \nextline
R. Brandenberger, `Modern Cosmology and Structure Formation', Brown preprint
BROWN-HET-893 (1993), to be publ. in the proc. of the 7th Swieca School in
Particles and Fields (World Scientific, Singapore 1994).}
\REF\eleven{V. Mukhanov, H. Feldman and R. Brandenberger, {\it Phys.
Rep.} {\bf 215}, 203 (1992).}
\REF\twelve{ T.W.B. Kibble, {\it J. Phys.} {\bf A9}, 1387 (1976).}
\REF\thirteen{E. Milne, {\it Zeits. f. Astrophys.} {\bf 6}, 1 (1933).}
\REF\fourteen{V. de Lapparent, M. Geller and J. Huchra, {\it Ap. J.
(Lett)} {\bf 302}, L1 (1986).}
\REF\fifteen{S. Shechtman, P. Schechter, A. Oemler, D. Tucker, R. Kirshner and
H. Lin, Harvard-Smithsonian preprint CFA 3385 (1992), to appear in `Clusters
and Superclusters of Galaxies', ed. by A. Fabian (Kluwer, Dordrecht, 1993);
\nextline
S. Shechtman et al., `The Las Campanas Fiber-Optic Redshift Survey', CFA
preprint (1994), to be publ. in the proc. of the 35th Herstmonceaux Conference
"Wide Field Spectroscopy and the Distant Universe".}
\REF\sixteen{R. Partridge, {\it Rep. Prog. Phys.} {\bf 51}, 647 (1988).}
\REF\seventeen{see e.g., S. Weinberg, `Gravitation and Cosmology'
(Wiley, New York, 1972); \nextline
Ya.B. Zel'dovich and I. Novikov, `The Structure and Evolution of the
Universe' (Univ. of Chicago Press, Chicago, 1983).}
\REF\eighteen{E. Hubble, {\it Proc. Nat. Acad. Sci.} {\bf 15}, 168
(1927).}
\REF\nineteen{J. Mould et al., {\it Ap. J.} {\bf 383}, 467 (1991).}
\REF\twenty{R. Alpher and R. Herman, {\it Rev. Mod. Phys.} {\bf
22}, 153 (1950); \nextline
G. Gamov, {\it Phys. Rev.} {\bf 70}, 572 (1946).}
\REF\twentyone{R. Dicke, P.J.E. Peebles, P. Roll and D. Wilkinson,
{\it Ap. J.} {\bf 142}, 414 (1965).}
\REF\twentytwo{A. Penzias and R. Wilson, {\it Ap. J.} {\bf 142}, 419
(1965).}
\REF\twentythree{J. Mather et al., {\it Ap. J. (Lett.)} {\bf 354}, L37
(1990).}
\REF\twentyfour{H. Gush, M. Halpern and E. Wishnow, {\it Phys. Rev.
Lett.} {\bf 65}, 937 (1990).}
\REF\twentyfive{R. Alpher, H. Bethe and G. Gamov, {\it Phys. Rev.}
{\bf 73}, 803 (1948); \nextline
R. Alpher and R. Herman, {\it Nature} {\bf 162}, 774 (1948).}
\REF\twentysix{For an excellent introduction see S. Weinberg, `The
First Three Minutes' (Basic Books, New York, 1988).}
\REF\twseven{T. Padmanabhan, `Structure Formation in the Universe' (Cambridge
Univ. Press, Cambridge, 1993), Chap. 3 and refs. therein.}
\REF\tweight{H. Arp, G. Burbidge, F. Hoyle, J. Narlikar and N.
Vickramasinghe, {\it Nature} {\bf 346}, 807 (1990).}
\REF\twnine{P.J.E. Peebles, D. Schramm, E. Turner and R. Kron, {\it Nature}
{\bf 352}, 769 (1991).}
\REF\thirty{A. Guth, {\it Phys. Rev.} {\bf D23}, 347 (1981).}
\REF\thone{P.J.E. Peebles, `Principles of Physical Cosmology' (Princeton Univ.
Press, Princeton, 1993).}
\REF\thtwo{J. Ostriker and L. Cowie, {\it Ap. J. (Lett.)} {\bf
243}, L127 (1981).}
\REF\ththree{S. Weinberg, {\it Rev. Mod. Phys.} {\bf 61}, 1 (1989);\nextline
S. Carroll, W. Press and E. Turner, {\it Ann. Rev. Astron. Astrophys.} {\bf
30}, 499 (1992).}
\REF\thfour{D. Kazanas, {\it Ap. J.} {\bf 241}, L59 (1980).}
\REF\thfive{W. Press, {\it Phys. Scr.} {\bf 21}, 702 (1980).}
\REF\thsix{G. Chibisov and V. Mukhanov, `Galaxy Formation and
Phonons,' Lebedev Physical Institute Preprint No. 162 (1980);
\nextline
G. Chibisov and V. Mukhanov, {\it Mon. Not. R. Astron. Soc.} {\bf
200}, 535 (1982).}
\REF\thseven{V. Lukash, {\it Pis'ma Zh. Eksp. Teor. Fiz.} {\bf 31}, 631
(1980).}
\REF\theight{K. Sato, {\it Mon. Not. R. Astron. Soc.} {\bf 195},
467 (1981).}
\REF\thnine{A. Starobinsky, {\it Phys. Lett.} {\bf 91B}, 99 (1980).}
\REF\fourty{M. Mijic, M. Morris and W.-M. Suen, {\it Phys. Rev.} {\bf
D34}, 2934 (1986).}
\REF\foone{V. Mukhanov and R. Brandenberger, {\it Phys. Rev.
Lett.} {\bf 68}, 1969 (1992); \nextline
R. Brandenberger, V. Mukhanov and A. Sornborger, {\it Phys. Rev.} {\bf D48},
1629 (1993).}
\REF\fotwo{J. Ye and R. Brandenberger, {\it Nucl. Phys.} {\bf B346}, 149
(1990).}
\REF\fothree{H. Nielsen and P. Olesen, {\it Nucl. Phys.} {\bf B61}, 45
(1973).}
\REF\fofour{R. Davis, {\it Phys. Rev.} {\bf D35}, 3705 (1987).}
\REF\fofive{N. Turok, {\it Phys. Rev. Lett.} {\bf 63}, 2625 (1989).}
\REF\fosix{see e.g., G. Ross, Grand Unified Theories (Benjamin, Reading,
1985).}
\REF\foseven{Ya.B. Zel'dovich, {\it Mon. Not. R. astron. Soc.} {\bf 192},
663 (1980).}
\REF\foeight{A. Vilenkin, {\it Phys. Rev. Lett.} {\bf 46}, 1169 (1981).}
\REF\fonine{J. Primack, D. Seckel and B. Sadoulet, {\it Ann. Rev. Nucl. Part.
Sci.} {\bf 38}, 751 (1988).}
\REF\fifty{T. van Albada and R. Sancisi, {\it Phil. Trans. R. Soc. London},
{\bf A320}, 447 (1986).}
\REF\fone{V. Trimble, {\it Ann. Rev. Astr. Astrophys.} {\bf 25}, 423 (1987).}
\REF\ftwo{E. Bertschinger and A. Dekel, {\it Ap. J.} {\bf 336},
L5 (1989).}
\REF\fthree{M. Strauss, M. Davis, A. Yahil and J. Huchra, {\it
Ap. J.} {\bf 385}, 421 (1992).}
\REF\ffour{S. White, C. Frenk and M. Davis, {\it Ap. J. (Lett.)} {\bf 274}, L1
(1993).}
\REF\ffive{J. Bond, G. Efstathiou and J. Silk, {\it Phys. Rev. Lett.} {\bf 45},
1980 (1980); \nextline
J. Bond and A. Szalay, {\it Ap. J.} {\bf 274}, 443 (1983).}
\REF\fsix{G. Blumenthal, S. Faber, J. Primack and M. Rees, {\it Nature} {\bf
311}, 517 (1984); \nextline
M. Davis, G. Efstathiou, C. Frenk and S. White, {\it Ap. J.} {\bf 292}, 371
(1985).}
\REF\fseven{N. Turok and R. Brandenberger, {\it Phys. Rev.} {\bf D33},
2175 (1986); \nextline
A. Stebbins, {\it Ap. J. (Lett.)} {\bf 303}, L21 (1986); \nextline
H. Sato, {\it Prog. Theor. Phys.} {\bf 75}, 1342 (1986).}
\REF\feight{T. Vachaspati, {\it Phys. Rev. Lett.} {\bf 57}, 1655 (1986).}
\REF\fnine{A. Stebbins, S. Veeraraghavan, R. Brandenberger, J. Silk and
N. Turok, {\it Ap. J.} {\bf 322}, 1 (1987).}
\REF\sixty{N. Turok, {\it Phys. Scripta} {\bf T36}, 135 (1991).}
\REF\sione{R. Brandenberger, N. Kaiser, D. Schramm and N. Turok, {\it
Phys. Rev. Lett.} {\bf 59}, 2371 (1987).}
\REF\sitwo{R. Brandenberger, N. Kaiser and N. Turok, {\it Phys. Rev.}
{\bf D36}, 2242 (1987).}
\REF\sithree{R. Brandenberger, L. Perivolaropoulos and A. Stebbins, {\it
Int. J. of Mod. Phys.} {\bf A5}, 1633 (1990); \nextline
L. Perivolarapoulos, R. Brandenberger and A. Stebbins, {\it Phys.
Rev.} {\bf D41}, 1764 (1990); \nextline
R. Brandenberger, {\it Phys. Scripta} {\bf T36}, 114 (1991).}
\REF\sifour{M. Davis and J. Huchra, {\it Ap. J.} {\bf 254}, 437 (1982).}
\REF\sifive{N. Bahcall and R. Soneira, {\it Ap. J.} {\bf 270}, 20 (1983);
\nextline
A. Klypin and A. Kopylov, {\it Sov. Astr. Lett.} {\bf 9}, 41 (1983).}
\REF\sisix{M. Strauss et al., {\it Ap. J.} {\bf 385}, 421 (1992).}
\REF\siseven{A. Yahil et al., {\it Ap. J.} {\bf 372}, 380 (1991).}
\REF\sieight{G. Abell, {\it Ap. J. Suppl.} {\bf 3}, 211 (1958).}
\REF\sinine{G. Daulton et al., `The two-point correlation function of rich
clusters of galaxies: results from an extended APM cluster redshift survey',
Oxford Univ. preprint, 1994, MNRAS, in press.}
\REF\seventy{W. Zurek, {\it Ap. J.} {\bf 324}, 19 (1988).}
\REF\sone{M. Rees and J. Ostriker, {\it Mon. Not. R. astr. Soc.} {\bf 179}, 541
(1977).}
\REF\stwo{Ya.B. Zel'dovich, J. Einasto and S. Shandarin, {\it Nature}
{\bf 300}, 407 (1982); \nextline
J. Oort, {\it Ann. Rev. Astron. Astrophys.} {\bf 21}, 373 (1983);
\nextline
R.B. Tully, {\it Ap. J.} {\bf 257}, 389 (1982); \nextline
S. Gregory, L. Thomson and W. Tifft, {\it Ap. J.} {\bf 243}, 411
(1980).}
\REF\sthree{G. Chincarini and H. Rood, {\it Nature} {\bf 257}, 294
(1975); \nextline
J. Einasto, M. Joeveer and E. Saar, {\it Mon. Not. R. astron. Soc.}
{\bf 193}, 353 (1980); \nextline
R. Giovanelli and M. Haynes, {\it Astron. J.} {\bf 87}, 1355 (1982);
\nextline
D. Batuski and J. Burns, {\it Ap. J.} {\bf 299}, 5 (1985).}
\REF\sfour{R. Kirshner, A. Oemler, P. Schechter and S. Shechtman, {\it
Ap. J. (Lett.)} {\bf 248}, L57 (1981).}
\REF\sfive{M. Joeveer, J. Einasto and E. Tago, {\it Mon. Not. R. astron.
Soc.} {\bf 185}, 357 (1978); \nextline
L. da Costa et al., {\it Ap. J.} {\bf 327}, 544 (1988).}
\REF\ssix{see e.g., G. Efstathiou, in `Physics of the Early
Universe,' proc. of the 1989 Scottish Univ. Summer School in Physics,
ed. by J. Peacock, A. Heavens and A. Davies (SUSSP Publ., Edinburgh,
1990).}
\REF\sseven{T. Padmanabhan, `Structure Formation in the Universe' (Cambridge
Univ. Press, Cambridge, 1993).}
\REF\seight{E. Harrison, {\it Phys. Rev.} {\bf D1}, 2726 (1970);
\nextline
Ya.B. Zel'dovich, {\it Mon. Not. R. astron. Soc.} {\bf 160}, 1p
(1972).}
\REF\snine{E. Lifshitz, {\it Zh. Eksp. Teor. Fiz.} {\bf 16}, 587 (1946);
\nextline
E. Lifshitz and I. Khalatnikov, {\it Adv. Phys.} {\bf 12}, 185 (1963).}
\REF\eighty{W. Press and E. Vishniac, {\it Ap. J.} {\bf 239}, 1 (1980).}
\REF\eone{R. Brandenberger, H. Feldman, V. Mukhanov and T. Prokopec,
`Gauge Invariant Cosmological Perturbations: Theory and Applications,'
publ. in ``The Origin of Structure in the Universe," eds. E. Gunzig and P.
Nardone (Kluwer, Dordrecht, 1993).}
\REF\etwo{J. Bardeen, {\it Phys. Rev.} {\bf D22}, 1882 (1980).}
\REF\ethree{R. Brandenberger, R. Kahn and W. Press, {\it Phys. Rev.} {\bf
D28}, 1809 (1983).}
\REF\efour{H. Kodama and M. Sasaki, {\it Prog. Theor. Phys. Suppl.} No.
78, 1 (1984).}
\REF\efive{R. Durrer and N. Straumann, {\it Helvet. Phys. Acta} {\bf
61}, 1027 (1988).}
\REF\esix{D. Lyth and M. Mukherjee, {\it Phys. Rev.} {\bf D38}, 485
(1988).}
\REF\eseven{G.F.R. Ellis and M. Bruni, {\it Phys. Rev.} {\bf D40}, 1804
(1989).}
\REF\eeight{J. Stewart, {\it Class. Quantum Grav.} {\bf 7}, 1169 (1990).}
\REF\enine{J. Stewart and M. Walker, {\it Proc. R. Soc.} {\bf A341}, 49
(1974).}
\REF\ninety{J. Bardeen, P. Steinhardt and M. Turner, {\it Phys. Rev.} {\bf
D28},
679 (1983).}
\REF\none{D. Kirzhnits and A. Linde, {\it Pis'ma Zh. Eksp.
Teor. Fiz.} {\bf 15}, 745 (1972); \nextline
D. Kirzhnits and A. Linde, {\it Zh. Eksp. Teor. Fiz.} {\bf 67}, 1263
(1974);\nextline
C. Bernard, {\it Phys. Rev.} {\bf D9}, 3313 (1974);\nextline
L. Dolan and R. Jackiw, {\it Phys. Rev.} {\bf D9}, 3320 (1974);\nextline
S. Weinberg, {\it Phys. Rev.} {\bf D9}, 3357 (1974).}
\REF\ntwo{A. Linde, {\it Phys. Lett.} {\bf 129B}, 177 (1983).}
\REF\nthree{G. Mazenko, W. Unruh and R. Wald, {\it Phys. Rev.} {\bf
D31}, 273 (1985).}
\REF\nfour{A. Linde, {\it Phys. Lett.} {\bf 108B}, 389 (1982);
\nextline
A. Albrecht and P. Steinhardt, {\it Phys. Rev. Lett.} {\bf 48}, 1220
(1982).}
\REF\nfive{J. Langer, {\it Physica} {\bf 73}, 61 (1974).}
\REF\nsix{S. Coleman, {\it Phys. Rev.} {\bf D15}, 2929 (1977);
\nextline
C. Callan and S. Coleman, {\it Phys. Rev.} {\bf D16}, 1762 (1977).}
\REF\nseven{M. Voloshin, Yu. Kobzarev and L. Okun, {\it Sov. J. Nucl.
Phys.} {\bf 20}, 644 (1975).}
\Ref\neight{M. Stone, {\it Phys. Rev.} {\bf D14}, 3568 (1976);\nextline
M. Stone, {\it Phys. Lett.} {\bf 67B}, 186 (1977).}
\Ref\nnine{P. Frampton, {\it Phys. Rev. Lett.}, {\bf 37}, 1380 (1976).}
\Ref\hundred{S. Coleman, in `The Whys of Subnuclear Physics' (Erice 1977), ed
by
A. Zichichi (Plenum, New York, 1979).}
\REF\hone{A. Guth and S.-H. Tye, {\it Phys. Rev. Lett.} {\bf
44}, 631 (1980).}
\Ref\htwo{A. Guth and E. Weinberg, {\it Nucl. Phys.} {\bf B212}, 321 (1983).}
\Ref\hthree{S. Hawking and I. Moss, {\it Phys. Lett.} {\bf 110B}, 35 (1982).}
\Ref\hfour{R. Matzner, in `Proceedings of the Drexel Workshop on Numerical
Relativity', ed by J. Centrella (Cambridge Univ. Press, Cambridge, 1986).}
\Ref\hfive{A. Albrecht, P. Steinhardt, M. Turner and F. Wilczek, {\it Phys.
Rev.
Lett.} {\bf 48}, 1437 (1982).}
\Ref\hsix{L. Abbott, E. Farhi and M. Wise, {\it Phys. Lett.} {\bf 117B}, 29
(1982).}
\Ref\hseven{J. Traschen and R. Brandenberger, {\it Phys. Rev.} {\bf D42},
2491 (1990).}
\REF\height{L. Kofman, A. Linde and A. Starobinski, {\it Phys. Rev. Lett.}, in
press (1994);\nextline
Y. Shtanov, J. Traschen and R. Brandenberger, {\it Phys. Rev. D.}, in press
(1994).}
\REF\hnine{L. Landau and E. Lifshitz, `Mechanics' (Pergamon, Oxford,
1960); \nextline
V. Arnold, `Mathematical Methods of Classical Mechanics' (Springer,
New York, 1978).}
\REF\hten{S. Coleman and E. Weinberg, {\it Phys. Rev.} {\bf D7}, 1888
(1973).}
\REF\htone{M. Markov and V. Mukhanov, {\it Phys. Lett.} {\bf 104A}, 200
(1984); \nextline
V. Belinsky, L. Grishchuk, I. Khalatnikov and Ya. Zel'dovich, {\it
Phys. Lett.} {\bf 155B}, 232 (1985); \nextline
L. Kofman, A. Linde and A. Starobinsky, {\it Phys. Lett.} {\bf 157B},
36 (1985); \nextline
T. Piran and R. Williams, {\it Phys. Lett.} {\bf 163B}, 331 (1985).}
\REF\httwo{S.-Y. Pi, {\it Phys. Rev. Lett.} {\bf 52}, 1725 (1984).}
\REF\htthree{D. Nanopoulos, K. Olive, M. Srednicki and K. Tamvakis, {\it
Phys. Lett.} {\bf 123B}, 41 (1983); \nextline
J. Ellis, K. Enqvist, D. Nanopoulos, K. Olive and M. Srednicki, {\it
Phys. Lett.} {\bf 152B}, 175 (1985); \nextline
R. Holman, P. Ramond and C. Ross, {\it Phys. Lett.} {\bf 137B}, 343
(1984).}
\REF\htfour{A. Goncharov and A. Linde, {\it JETP} {\bf 59}, 930 (1984);
\nextline
A. Goncharov and A. Linde, {\it Phys. Lett.} {\bf 139B}, 27 (1984);
\nextline
A. Goncharov and A. Linde, {\it Class. Quant. Grav.} {\bf 1}, L75
(1984).}
\REF\htfive{I. Antoniadis, J. Ellis, J. Hagelin and D. Nanopoulos, {\it
Phys. Lett.} {\bf 205B}, 459 (1988); \nextline
I. Antoniadis, J. Ellis, J. Hagelin and D. Nanopoulos, {\it Phys.
Lett.} {\bf 208B}, 209 (1988).}
\REF\htsix{A. Linde, D. Linde and A. Mezhlumian, {\it Phys. Rev.} {\bf D49},
1783 (1994); nextline
A. Linde, `Lectures on Inflationary Cosmology', Stanford preprint SU-ITP-94-36
(1994).}
\REF\htseven{A. Starobinsky, in `Current Trends in Field Theory, Quantum
Gravity, and Strings', Lecture Notes in Physics, ed. by H. de Vega and N.
Sanchez (Springer, Heidelberg, 1986).}
\REF\hteight{B. Whitt, {\it Phys. Lett.} {\bf 145B}, 176 (1984).}
\REF\htnine{M. Gasperini and G. Veneziano, {\it Astropart. Phys.} {\bf 1}, 317
(1993); \nextline
R. Brustein and G. Veneziano, {\it Phys. Lett.} {\bf B329}, 429 (1994).}
\REF\htw{F. Adams, K. Freese and A. Guth, {\it Phys. Rev.} {\bf D43},
965 (1991).}
\REF\htwone{H. Feldman and R. Brandenberger, {\it Phys. Lett.} {\bf
227B}, 359 (1989).}
\REF\htwtwo{J. Kung and R. Brandenberger, {\it Phys. Rev.} {\bf D42},
1008 (1990).}
\REF\htwthree{D. Goldwirth and T. Piran, {\it Phys. Rev. Lett.} {\bf 64},
2852 (1990);\nextline
D. Goldwirth and T. Piran, {\it Phys. Rep.} {\bf 214}, 223 (1992).}
\REF\htwfour{A. Albrecht and R. Brandenberger, {\it Phys. Rev.} {\bf D31},
1225 (1985).}
\REF\htwfive{G. Chibisov and V. Mukhanov, {\it JETP Lett.} {\bf 33}, 532
(1981);\nextline
V. Mukhanov and G. Chibisov, {\it Zh. Eksp. Teor. Fiz.} {\bf 83}, 475
(1982).}
\REF\htwsix{A. Lapedes, {\it J. Math. Phys.} {\bf 19}, 2289 (1978);
\nextline
R. Brandenberger and R. Kahn, {\it Phys. Lett.} {\bf 119B}, 75
(1982).}
\REF\htwseven{A. Guth and S.-Y. Pi, {\it Phys. Rev. Lett.} {\bf 49}, 1110
(1982); \nextline
S. Hawking, {\it Phys. Lett.} {\bf 115B}, 295 (1982); \nextline
A. Starobinsky, {\it Phys. Lett.} {\bf 117B}, 175 (1982); \nextline
R. Brandenberger and R. Kahn, {\it Phys. Rev.} {\bf D28}, 2172 (1984);
\nextline
J. Frieman and M. Turner, {\it Phys. Rev.} {\bf D30}, 265 (1984);
\nextline
V. Mukhanov, {\it JETP Lett.} {\bf 41}, 493 (1985).}
\REF\htweight{W. Zurek, {\it Phys. Rev.} {\bf D24}, 1516 (1982); \nextline
W. Zurek, {\it Phys. Rev.} {\bf D26} 1862 (1982).}
\REF\htwnine{E. Joos and H. Zeh, {\it Z. Phys.} {\bf B59}, 223 (1985);
\nextline
H. Zeh, {\it Phys. Lett.} {\bf 116A}, 9 (1986); \nextline
C. Kiefer, {\it Class. Quantum Grav.} {\bf 4}, 1369 (1987); \nextline
T. Fukuyama and M. Morikawa, {\it Phys. Rev.} {\bf D39}, 462 (1989);
\nextline
J. Halliwell, {\it Phys. Rev.} {\bf D39}, 2912 (1989); \nextline
T. Padmanabhan, {\it Phys. Rev.} {\bf D39}, 2924 (1980); \nextline
W. Unruh and W. Zurek, {\it Phys. Rev.} {\bf D40}, 1071 (1989);
\nextline
E. Calzetta and F. Mazzitelli, {\it Phys. Rev.} {\bf D42}, 4066
(1990); \nextline
S. Habib and R. Laflamme, {\it Phys. Rev.} {\bf D42}, 4056 (1990);
\nextline
H. Feldman and A. Kamenshchik, {\it Class. Quantum Grav.} {\bf 8}, L65
(1991).}
\REF\hthirty{M. Sakagami, {\it Prog. Theor. Phys.} {\bf 79}, 443
(1988);\nextline
R. Brandenberger, R. Laflamme and M. Mijic, {\it Mod. Phys.
Lett.} {\bf A5}, 2311 (1990).}
\REF\hthone{J. Bardeen, unpublished (1984).}
\REF\hthtwo{R. Brandenberger, {\it Nucl. Phys.} {\bf B245}, 328 (1984).}
\REF\hththree{N. Birrell and P. Davies, `Quantum Fields in Curved Space'
(Cambridge Univ. Press, Cambridge, 1982).}
\REF\hthfour{R. Brandenberger and C. Hill, {\it Phys. Lett.} {\bf 179B},
30 (1986).}
\REF\hthfive{W. Fischler, B. Ratra and L. Susskind, {\it Nucl. Phys.} {\bf
B259}, 730 (1985).}
\REF\hthsix{D. Mermin, {\it Rev. Mod. Phys.} {\bf 51}, 591 (1979).}
\REF\hthseven{P. de Gennes, `The Physics of Liquid Crystals' (Clarendon Press,
Oxford, 1974); \nextline
I. Chuang, R. Durrer, N. Turok and B. Yurke, {\it Science} {\bf 251}, 1336
(1991); \nextline
M. Bowick, L. Chandar, E. Schiff and A. Srivastava, {\it Science} {\bf 263},
943 (1994).}
\REF\htheight{M. Salomaa and G. Volovik, {\it Rev. Mod. Phys.} {\bf 59}, 533
(1987).}
\REF\hthnine{A. Abrikosov, {\it JETP} {\bf 5}, 1174 (1957).}
\REF\hfourty{Ya.B. Zel'dovich, I. Kobzarev and L. Okun, {\it Zh. Eksp.
Teor. Fiz.} {\bf 67}, 3 (1974).}
\REF\hfoone{Ya.B. Zel'dovich and M. Khlopov, {\it Phys. Lett.} {\bf 79B},
239 (1978); \nextline
J. Preskill, {\it Phys. Rev. Lett.} {\bf 43}, 1365 (1979).}
\REF\hfotwo{T.W.B. Kibble, {\it Acta Physica Polonica} {\bf B13}, 723
(1982).}
\REF\hfothree{P. Langacker and S.-Y. Pi, {\it Phys. Rev. Lett.} {\bf 45}, 1
(1980).}
\REF\hfofour{T.W.B. Kibble and E. Weinberg, {\it Phys. Rev.} {\bf D43},
3188 (1991).}
\REF\hfoseven{T. Vachaspati and A. Vilenkin, {\it Phys. Rev.} {\bf D30},
2036 (1984).}
\REF\hfoeight{S. Rudaz and A. Srivastava, {\it Mod. Phys. Lett.} {\bf A8}, 1443
(1993).}
\REF\hfnine{F. Liu, M. Mondello and N. Goldenfeld, {\it Phys. Rev. Lett.} {\bf
66}, 3071 (1991).}
\REF\hsixty{M. Hindmarsh, A.-C. Davis and R. Brandenberger, {\it Phys. Rev.}
{\bf D49}, 1944 (1994); \nextline
R. Brandenberger and A.-C. Davis, {\it Phys. Lett.} {\bf B332}, 305 (1994).}
\REF\hsione{T. Prokopec, A. Sornborger and R. Brandenberger, {\it Phys.
Rev.} {\bf D45}, 1971 (1992).}
\REF\hsitwo{J. Borrill, E. Copeland and A. Liddle, {\it Phys. Lett.} {\bf
258B}, 310 (1991).}
\REF\hsithree{A. Sornborger, {\it Phys. Rev.} {\bf D48}, 3517 (1993).}
\REF\hsifour{L. Perivolaropoulos, {\it Phys. Rev.} {\bf D46}, 1858
(1992).}
\REF\hsifive{T. Prokopec, {\it Phys. Lett.} {\bf 262B}, 215 (1991);\nextline
R. Leese and T. Prokopec, {\it Phys. Rev.} {\bf D44}, 3749
(1991).}
\REF\hfofive{M. Barriola and A. Vilenkin, {\it Phys. Rev. Lett.} {\bf 63},
341 (1989).}
\REF\hfosix{S. Rhie and D. Bennett, {\it Phys. Rev. Lett.} {\bf 65}, 1709
(1990).}
\REF\hsisix{D. Foerster, {\it Nucl. Phys.} {\bf B81}, 84 (1974).}
\REF\hsiseven{N. Turok, in `Proceedings of the 1987 CERN/ESO Winter School
on Cosmology and Particle Physics' (World Scientific, Singapore,
1988).}
\REF\hsieight{T.W.B. Kibble and N. Turok, {\it Phys. Lett.} {\bf 116B}, 141
(1982).}
\REF\hsinine{R. Brandenberger, {\it Nucl. Phys.} {\bf B293}, 812 (1987).}
\REF\hseventy{E.P.S. Shellard, {\it Nucl. Phys.} {\bf B283}, 624 (1987).}
\REF\hsone{R. Matzner, {\it Computers in Physics} {\bf 1}, 51 (1988);
\nextline
K. Moriarty, E. Myers and C. Rebbi, {\it Phys. Lett.} {\bf 207B}, 411
(1988); \nextline
E.P.S. Shellard and P. Ruback, {\it Phys. Lett.} {\bf 209B}, 262
(1988).}
\REF\hstwo{P. Ruback, {\it Nucl. Phys.} {\bf B296}, 669 (1988).}
\REF\hsthree{A. Albrecht and N. Turok, {\it Phys. Rev. Lett.} {\bf 54},
1868 (1985).}
\REF\hsfour{D. Bennett and F. Bouchet, {\it Phys. Rev. Lett.} {\bf 60},
257 (1988).}
\REF\hsfive{B. Allen and E.P.S. Shellard, {\it Phys. Rev. Lett.} {\bf
64}, 119 (1990).}
\REF\hssix{A. Albrecht and N. Turok, {\it Phys. Rev. } {\bf D40}, 973
(1989).}
\REF\hsseven{R. Brandenberger and J. Kung, in `The Formation and Evolution
of Cosmic Strings' eds. G. Gibbons, S. Hawking and T. Vachaspati
(Cambridge Univ. Press, Cambridge, 1990).}
\REF\hseight{see e.g., C. Misner, K. Thorne and J. Wheeler, `Gravitation'
(Freeman, San Francisco, 1973).}
\REF\hsnine{T. Vachaspati and A. Vilenkin, {\it Phys. Rev.} {\bf D31},
3052 (1985); \nextline
N. Turok, {\it Nucl. Phys.} {\bf B242}, 520 (1984); \nextline
C. Burden, {\it Phys. Lett.} {\bf 164B}, 277 (1985).}
\REF\heighty{B. Carter, {\it Phys. Rev.} {\bf D41}, 3869 (1990).}
\REF\heone{E. Copeland, T.W.B. Kibble and D. Austin, {\it Phys. Rev.}
{\bf D45}, 1000 (1992).}
\REF\hetwo{N. Turok, {\it Nucl. Phys.} {\bf B242}, 520 (1984).}
\REF\hethree{J. Silk and V. Vilenkin, {\it Phys. Rev. Lett.} {\bf 53},
1700 (1984).}
\REF\hefour{A. Vilenkin, {\it Phys. Rev.} {\bf D23}, 852 (1981);
\nextline
J. Gott, {\it Ap. J.} {\bf 288}, 422 (1985); \nextline
W. Hiscock, {\it Phys. Rev.} {\bf D31}, 3288 (1985); \nextline
B. Linet, {\it Gen. Rel. Grav.} {\bf 17}, 1109 (1985); \nextline
D. Garfinkle, {\it Phys. Rev.} {\bf D32}, 1323 (1985); \nextline
R. Gregory, {\it Phys. Rev. Lett.} {\bf 59}, 740 (1987).}
\REF\hefive{D. Vollick, {\it Phys. Rev.} {\bf D45}, 1884 (1992);
\nextline
T. Vachaspati and A. Vilenkin, {\it Phys. Rev. Lett.} {\bf 67}, 1057 (1991).}
\REF\hesix{A. Albrecht and A. Stebbins, {\it Phys. Rev. Lett.} {\bf 68}, 2121
(1992).}
\REF\heseven{A. Albrecht and A. Stebbins, {\it Phys. Rev. Lett.} {\bf 69}, 2615
(1992).}
\REF\heeight{D. Spergel, N. Turok, W. Press and B. Ryden, {\it Phys. Rev.}
{\bf D43}, 1038 (1991).}
\REF\henine{A. Gooding, D. Spergel and N. Turok, {\it Ap. J. (Lett.)}
{\bf 372}, L5 (1991); \nextline
C. Park, D. Spergel and N. Turok, {\it Ap. J. (Lett.)} {\bf 373}, L53 (1991).}
\REF\hninety{R. Cen, J. Ostriker, D. Spergel and N. Turok, {\it Ap. J.}
{\bf 383}, 1 (1991).}
\REF\hnone{J. Gott, A. Melott and M. Dickinson, {\it Ap. J.} {\bf 306},
341 (1986).}
\REF\hntwo{S. Ramsey, Senior thesis, Brown Univ. (1992); \nextline
D. Kaplan, Senior thesis, Brown Univ. (1993); \nextline
R. Brandenberger, D. Kaplan and S. Ramsey, `Some statistics for measuring
large-scale structure', Brown preprint BROWN-HET-922 (1993);
A. Aguirre, Senior thesis, Brown Univ. (1994).}
\REF\hnthree{D. Fixen, E. Cheng and D. Wilkinson, {\it Phys. Rev. Lett.} {\bf
50}, 620 (1983).}
\REF\hnfour{R. Sachs and A. Wolfe, {\it Ap. J.} {\bf 147}, 73 (1967).}
\REF\hnfive{V. Mukhanov and G. Chibisov, {\it Sov. Astron. Lett.} {\bf 10}, 890
(1984).}
\REF\hnsix{N. Kaiser and A. Stebbins, {\it Nature} {\bf 310}, 391
(1984).}
\REF\hnseven{R. Moessner, L. Perivolaropoulos and R. Brandenberger, {\it Ap.
J.} {\bf 425}, 365 (1994).}
\REF\hneight{N. Turok and D. Spergel, {\it Phys. Rev. Lett.} {\bf 64}, 2736
(1990).}
\REF\thundred{R. Durrer, A. Howard and Z.-H. Zhou, {\it Phys. Rev.} {\bf D49},
681 (1994).}
\REF\twhone{U.-L. Pen, D. Spergel and N. Turok, {\it Phys. Rev.} {\bf D49}, 692
(1994).}
\REF\twhtwo{E. Wright et al., {\it Ap. J. (Lett.)} {\bf 396}, L13 (1992).}
\REF\twohthree{S. White, C. Frenk, M. Davis and G. Efstathiou, {\it Ap. J.}
{\bf 313}, 505 (1987).}
\REF\twohfive{D. Bennett and S. Rhie, {\it Ap. J. (Lett.)} {\bf 406}, L7
(1993).}
\REF\twohfour{D. Bennett, A. Stebbins and F. Bouchet, {\it Ap. J. (Lett.)}
{\bf 399}, L5 (1992).}
\REF\twohsix{L. Perivolaropoulos, {\it Phys. Lett.} {\bf 298B}, 305 (1993).}
\REF\twohseven{G. Smoot et al., {\it Ap. J. (Lett.)} {\bf 396}, L1
(1992).}
\REF\twoheight{S. Meyer, E. Cheng and L. Page, {\it Ap. J. (Lett.)} {\bf 371},
L7 (1991);\nextline
K. Ganga, E. Cheng, S. Meyer and L. Page, {\it Ap. J. (Lett.)} {\bf 410}, L57
(1993).}
\REF\twohnine{S. Hancock et al., {\it Nature} {\bf 317}, 333 (1994).}
\REF\twohten{E. Wollack et al., {\it Ap. J. (Lett.)} {\bf 419}, L49 (1993).}
\REF\twoheleven{T. Gaier et al., {\it Ap. J. (Lett.)} {\bf 398}, L1 (1992);
\nextline
J. Schuster et al., {\it Ap. J. (Lett.)} {\bf 412}, L47 (1993).}
\REF\twohtwelve{P. de Bernardis et al., {\it Ap. J. (Lett.)} {\bf 422}, L33
(1994).}
\REF\twohthteen{M. Dragovan et al., Princeton preprint (1993).}
\REF\twohfourteen{P. Meinhold et al., {\it Ap. J. (Lett.)} {\bf 409}, L1
(1993).}
\REF\twohfifteen{J. Gunderson et al., {\it Ap. J. (Lett.)} {\bf 413}, L1
(1993).}
\REF\twohsixteen{E. Cheng et al., {\it Ap. J. (Lett.)} {\bf 422}, L37 (1994).}
\REF\twohseventeen{G. Tucker, G. Griffin, H. Nguyen and J. Peterson, {\it Ap.
J. (Lett.)} {\bf 419}, L45 (1993).}
\REF\twoheighteen{A. Readhead et al., {\it Ap. J.} {\bf 346}, 566 (1989).}
\REF\twohnineteen{S. Myers, A. Readhead and A. Lawrence, {\it Ap. J.} {\bf
405}, 8 (1993).}
\refout

\end